\documentclass[sigconf]{acmart}

\AtBeginDocument{%
  }

\copyrightyear{2026}
\acmYear{2026}
\setcopyright{cc}
\setcctype{by}
\acmConference[CHI '26]{Proceedings of the 2026 CHI Conference on Human Factors in Computing Systems}{April 13--17, 2026}{Barcelona, Spain}
\acmBooktitle{Proceedings of the 2026 CHI Conference on Human Factors in Computing Systems (CHI '26), April 13--17, 2026, Barcelona, Spain}
\acmPrice{}
\acmDOI{10.1145/3772318.3791431}
\acmISBN{979-8-4007-2278-3/2026/04}

\usepackage{acmart-taps}
\usepackage{makecell}
\usepackage{multirow}
\usepackage{cleveref}

\definecolor{revbg}{HTML}{FFF3CD} 
\definecolor{aibg}{HTML}{E7F1FF}  

\definecolor{revbg}{HTML}{FFF7E6}     
\definecolor{revline}{HTML}{FFB300}   
\definecolor{aibg}{HTML}{E9F3FF}      
\definecolor{ailine}{HTML}{1E88E5}    

\newcommand{\revlabel}{\textsc{Reviewer comment}} 
\newcommand{\ailabel}{\textsc{Feedback to the reviewer}} 

\newsavebox{\tapsbox}

\newenvironment{revcomment}{%
  \par\vspace{3pt}\noindent
  \begin{lrbox}{\tapsbox}%
  \begin{minipage}{\dimexpr\linewidth-14pt}%
  \vspace{3pt}\raggedright
}{%
  \vspace{3pt}%
  \end{minipage}%
  \end{lrbox}%
  \textcolor{revline}{\rule[-\dimexpr\dp\tapsbox+\fboxsep\relax]{1.5pt}{\dimexpr\ht\tapsbox+\dp\tapsbox+2\fboxsep\relax}}%
  \colorbox{revbg}{\usebox{\tapsbox}}%
  \par\vspace{3pt}%
}

\newenvironment{aifeedback}{%
  \par\vspace{3pt}\noindent
  \begin{lrbox}{\tapsbox}%
  \begin{minipage}{\dimexpr\linewidth-14pt}%
  \vspace{3pt}\raggedright
}{%
  \vspace{3pt}%
  \end{minipage}%
  \end{lrbox}%
  \textcolor{ailine}{\rule[-\dimexpr\dp\tapsbox+\fboxsep\relax]{1.5pt}{\dimexpr\ht\tapsbox+\dp\tapsbox+2\fboxsep\relax}}%
  \colorbox{aibg}{\usebox{\tapsbox}}%
  \par\vspace{3pt}%
}

\newenvironment{emailboxfancy}{%
  \par\vspace{5pt}\noindent
  \begin{lrbox}{\tapsbox}%
  \begin{minipage}{\dimexpr\linewidth-14pt}%
  \vspace{5pt}
}{%
  \vspace{6pt}%
  \end{minipage}%
  \end{lrbox}%
  \fbox{\usebox{\tapsbox}}%
  \par\vspace{5pt}%
}
\newcommand{\emailline}[2]{\textbf{#1}:~#2\\}

\begin{document}

\title[What happens when reviewers receive AI feedback in their reviews?]{What Happens When Reviewers Receive \\ AI Feedback in Their Reviews?}


\author{Shiping Chen}
\authornote{Both authors contributed equally to this paper.}
\orcid{0000-0001-9477-2420}
\affiliation{%
  \institution{UCL Interaction Centre,\\University College London}
  \city{London}
  \country{United Kingdom}
} 
\email{shiping.chen.20@ucl.ac.uk}

\author{Shu Zhong}
\authornotemark[1]
\orcid{0000-0002-1820-6424}
\affiliation{%
  \institution{Department of Computer Science,\\University College London}
  \city{London}
  \country{United Kingdom}
}
\email{shu.zhong.21@ucl.ac.uk}

\author{Duncan P. Brumby}
\affiliation{%
 \institution{UCL Interaction Centre,\\University College London}
  \city{London}
  \country{United Kingdom}}
\email{d.brumby@ucl.ac.uk}
\orcid{0000-0003-2846-2592}

\author{Anna L. Cox}
\affiliation{%
 \institution{UCL Interaction Centre,\\University College London}
  \city{London}
  \country{United Kingdom}}
\email{anna.cox@ucl.ac.uk}
\orcid{0000-0003-2231-2964}

\renewcommand{\shortauthors}{Chen et al.}

\begin{abstract}

AI is reshaping academic research, yet its role in peer review remains polarising and contentious. Advocates see its potential to reduce reviewer burden and improve quality, while critics warn of risks to fairness, accountability, and trust. At ICLR 2025, an official AI feedback tool was deployed to provide reviewers with post-review suggestions. We studied this deployment through surveys and interviews, investigating how reviewers engaged with the tool and perceived its usability and impact. Our findings surface both opportunities and tensions when AI augments in peer review. This work contributes the first empirical evidence of such an AI tool in a live review process, documenting how reviewers respond to AI-generated feedback in a high-stakes review context. We further offer design implications for AI-assisted reviewing that aim to enhance quality while safeguarding human expertise, agency, and responsibility. 
\end{abstract}

\ccsdesc[500]{Human-centered computing}
\ccsdesc[500]{Computing methodologies~Artificial intelligence}

\keywords{Peer Review, Human-AI interaction, AI for Work, AI Review, Large Language Models}

\maketitle

\section{Introduction}
Artificial intelligence is changing how research is imagined, produced, and communicated~\cite{posner2020ai, chubb2022speeding, kapania2025m, hwang2025human}. Generative AI now can support every stage of academic work, from ideation and creative writing~\cite{lee2022coauthor, clark2018creative, chakrabarty2022help, ippolito2022creative, buschek2021impact}, to code generation~\cite{vaithilingam2022expectation,barke2022grounded} and data analysis~\cite{chew2023llm, dai2023llm}, as well as the polishing of full manuscripts \cite{liang2024mapping}. These tools lower the barrier to produce near publication-ready work and are contributing to rapid growth in scholarly output~\cite{khalifa2024using, hanson2024strain}.

This surge places heavy pressure on the volunteer review system \cite{hanson2024strain}. Submission volume is growing faster than the community’s capacity to review it \cite{hanson2024strain, kim2025position}. CHI, for example, grew from 3{,}182 submissions in 2023 to 6{,}730 in 2026, more than doubling in three years \cite{chi2026stats}. NeurIPS received over 9{,}000 submissions in 2020, a 40\% year-on-year increase, with totals now above 10{,}000 annually \cite{kim2025position}. Reviewers must handle more papers within the same short cycles.
Maintaining a fair, sustainable review ecosystem that protects reviewer wellbeing, supports sound judgement, and ensures fairness has become increasingly challenging \cite{uncoveringlee, snell2005reviewersperception, nobarany2014rethinking, shah2021overview, aczel2025present}.

As pressure grows, conferences and publishers are exploring AI support. Some treat AI assistance as a way to ease reviewer workload and strengthen reviews, while others remain cautious about introducing such tools into high-stakes peer review system. Critics argue that using AI in peer review is anti-scientific, raising concerns about fairness, transparency, and responsibility \cite{naddaf2025ai}. Publication venues and communities have responded differently to these debates. The Conference on Computer Vision and Pattern Recognition (CVPR), for example, introduced strict rules in 2025 prohibiting the use of large language models (LLMs) in any part of review writing or translation \cite{CVPR202556online}. Reviewers are prohibited from using LLMs to generate review content, regardless of whether the model is accessed locally or through an API. Reviews suspected of AI generation could be flagged as irresponsible and lead to desk rejections of the reviewer’s own submissions. In sharp contrast, other venues have adopted more exploratory stances, experimenting with ways AI might ease the growing pressures of review.

\begin{figure*}[t]
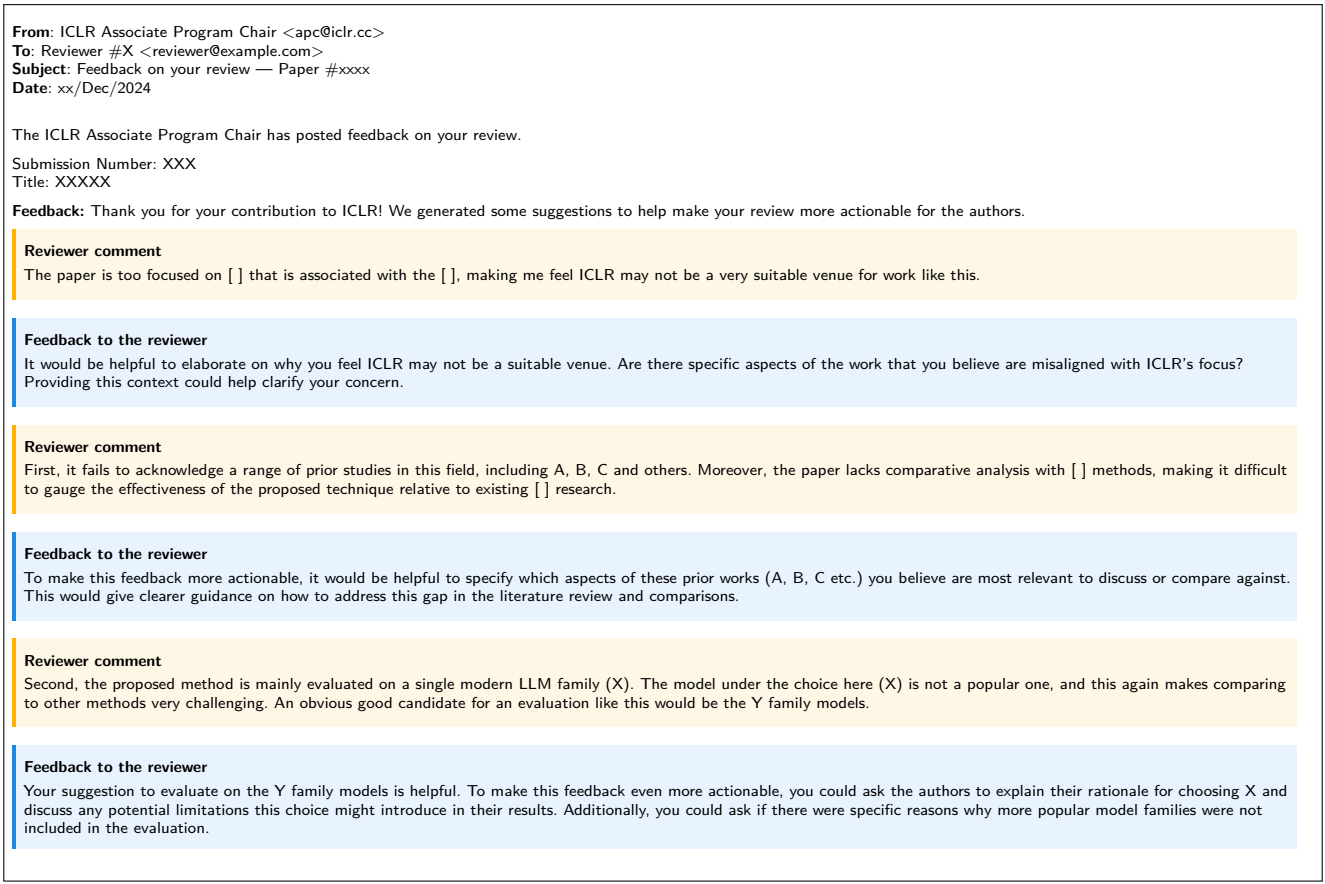

\centering
\aptLtoX[graphic=no, type=html]{
\noindent\emailline{From}{ICLR Associate Program Chair $<$apc@iclr.cc$>$}
\emailline{To}{Reviewer \#X $<$reviewer@example.com$>$}
\emailline{Subject}{Feedback on your review --- Paper \#xxxx}
\emailline{Date}{xx/Dec/2024}

\medskip
\noindent The ICLR Associate Program Chair has posted feedback on your review.

\medskip
\noindent Submission Number: XXX\\
Title: XXXXX

\medskip
\noindent\textbf{Feedback:} Thank you for your contribution to ICLR! We generated some suggestions to help make your
review more actionable for the authors.

\begin{quote}
\textbf{\revlabel}\\
The paper is too focused on [ ] that is associated with the [ ], making me feel ICLR may not be a very
suitable venue for work like this.
\end{quote}

\begin{quote}
\textbf{\ailabel}\\
It would be helpful to elaborate on why you feel ICLR may not be a suitable venue. Are there specific
aspects of the work that you believe are misaligned with ICLR's focus? Providing this context could help
clarify your concern.
\end{quote}

\begin{quote}
\textbf{\revlabel}\\
First, it fails to acknowledge a range of prior studies in this field, including A, B, C and others.
Moreover, the paper lacks comparative analysis with [ ] methods, making it difficult to gauge the
effectiveness of the proposed technique relative to existing [ ] research.
\end{quote}

\begin{quote}
\textbf{\ailabel}\\
To make this feedback more actionable, it would be helpful to specify which aspects of these prior works
(A, B, C etc.) you believe are most relevant to discuss or compare against. This would give clearer guidance
on how to address this gap in the literature review and comparisons.
\end{quote}

\begin{quote}
\textbf{\revlabel}\\
Second, the proposed method is mainly evaluated on a single modern LLM family (X). The model under the
choice here (X) is not a popular one, and this again makes comparing to other methods very challenging.
An obvious good candidate for an evaluation like this would be the Y family models.
\end{quote}

\begin{quote}
\textbf{\ailabel}\\
Your suggestion to evaluate on the Y family models is helpful. To make this feedback even more actionable,
you could ask the authors to explain their rationale for choosing X and discuss any potential limitations
this choice might introduce in their results. Additionally, you could ask if there were specific reasons
why more popular model families were not included in the evaluation.
\end{quote}
}{

\begin{emailboxfancy}
{\scriptsize\fontfamily{lmss}\selectfont 
\emailline{From}{ICLR Associate Program Chair <apc@iclr.cc>}
\emailline{To}{Reviewer \#X <reviewer@example.com>}
\emailline{Subject}{Feedback on your review — Paper \#xxxx}
\emailline{Date}{xx/Dec/2024}

\medskip
The ICLR Associate Program Chair has posted feedback on your review.

\medskip
Submission Number: XXX\\
Title: XXXXX

\medskip
\textbf{Feedback:} Thank you for your contribution to ICLR! We generated some suggestions to help make your
review more actionable for the authors.

\begin{revcomment}
\textbf{\revlabel}\\[2pt]
The paper is too focused on [ ] that is associated with the [ ], making me feel ICLR may not be a very
suitable venue for work like this.
\end{revcomment}

\begin{aifeedback}
\textbf{\ailabel}\\[2pt]
It would be helpful to elaborate on why you feel ICLR may not be a suitable venue. Are there specific
aspects of the work that you believe are misaligned with ICLR's focus? Providing this context could help
clarify your concern.
\end{aifeedback}

\begin{revcomment}
\textbf{\revlabel}\\[2pt]
First, it fails to acknowledge a range of prior studies in this field, including A, B, C and others.
Moreover, the paper lacks comparative analysis with [ ] methods, making it difficult to gauge the
effectiveness of the proposed technique relative to existing [ ] research.
\end{revcomment}

\begin{aifeedback}
\textbf{\ailabel}\\[2pt]
To make this feedback more actionable, it would be helpful to specify which aspects of these prior works
(A, B, C etc.) you believe are most relevant to discuss or compare against. This would give clearer guidance
on how to address this gap in the literature review and comparisons.
\end{aifeedback}

\begin{revcomment}
\textbf{\revlabel}\\[2pt]
Second, the proposed method is mainly evaluated on a single modern LLM family (X). The model under the
choice here (X) is not a popular one, and this again makes comparing to other methods very challenging.
An obvious good candidate for an evaluation like this would be the Y family models.
\end{revcomment}

\begin{aifeedback}
\textbf{\ailabel}\\[2pt]
Your suggestion to evaluate on the Y family models is helpful. To make this feedback even more actionable,
you could ask the authors to explain their rationale for choosing X and discuss any potential limitations
this choice might introduce in their results. Additionally, you could ask if there were specific reasons
why more popular model families were not included in the evaluation.
\end{aifeedback}
}
\end{emailboxfancy}
}
\caption[Example AI feedback]{Example of ICLR AI feedback from a participant. Bracketed items \texttt{[ ]} are omitted for confidentiality. Coloured labels are used here to distinguish reviewer comments from AI feedback; the original feedback email was presented in plain text.}
\label{fig:iclr_email}
\Description{Example email showing how AI-generated feedback was integrated with reviewer comments at ICLR. Reviewer comments appear in plain text, and AI feedback is displayed in blue boxes below each comment, suggesting ways to make the review clearer and more actionable.}
\end{figure*}

A notable case is the International Conference on Learning Representations (ICLR) \footnote{\url{https://iclr.cc/}}, one of the top three Machine Learning conferences (h5-index 362). ICLR is facing explosive growth in submissions, increasing from 3,422 in 2022 to 11,672 in 2025 \cite{ICLRStat13:online}. To address this, the conference introduced an AI feedback tool during the ICLR 2025 review cycle, aimed at enhancing review quality at scale \cite{Assistin40:online}. The system sent \textit{post-submission} feedback to reviewers, offering optional suggestions to improve clarity, tone and specificity after their initial reviews were submitted. Reviewers received this feedback via email (\Cref{fig:iclr_email}), and were given the option, but not the obligation, to revise their reviews accordingly. This deployment provides a rare opportunity to study how reviewers experience AI support in a live, high-stakes reviewing environment.

In practice, AI use in peer review remains informal and highly individual. Some reviewers already use AI tools privately and share personal experience of using AI in reviewing~\cite{UseAIforReviewNature}, while others avoid AI entirely. Reviewers carry the core responsibility for rigour and fairness in decisions that shape research fields~\cite{uncoveringlee, mason2023bringing}. Their adoption of AI support influences whether such tools become trusted aids, remain unused or create new risks in the review process. Yet little is known about how reviewers make sense of AI support, how they decide whether to act on it and how such tools affect them. If conferences move towards standardising or mandating AI usage, understanding reviewers’ perceptions and reactions is essential for designing AI systems that support peer review.

In response to this, we investigate how reviewers experienced this timely form of AI feedback in peer review. We conducted a mixed-methods study with reviewers from ICLR 2025 who experienced the AI feedback system firsthand. Prior work on AI support for reviewing has relied largely on hypothetical scenarios or Wizard-of-Oz studies, such as automated reviewing or real-time writing assistants \cite{chen2025envisioning, sun2024reviewflow}. Our work captures interactions with an officially deployed AI tool during an active review cycle. This unique setting allows us to move beyond speculation and document how reviewers engaged with AI-generated feedback in practice, and how it shaped their judgements of ownership, authority and accountability.

We studied this large-scale deployment through surveys and interviews with reviewers. We examine how reviewers perceived the feedback, whether and how they acted on it, and how it influenced their judgements, decisions, and sense of ownership over their reviews. Specifically, we asked:

\begin{itemize}
    \item RQ1: How do reviewers perceive the AI feedback tool?
    \item RQ2: How do reviewers respond to the AI-generated feedback, and what actions do they take in revising their reviews?
    \item RQ3: What benefits and drawbacks do reviewers identify in such AI-assisted peer review, and what forms of AI support do they envision for the future?
\end{itemize}

Our study offers an early empirical window into how AI is reshaping peer review in practice. Reviewers engaged seriously with the tool but revealed tensions between recognising its potential and questioning its authority. These dynamics highlight both the promise and the limits of post-review assistance as a form of augmentation. Our contributions are threefold:
\begin{enumerate}
    \item \textbf{An empirical account of AI assistance in a real, high-stakes review process.} Unlike prior speculative or Wizard-of-Oz studies, our work examines how actual reviewers at ICLR 2025 interacted with post-review AI feedback during a live review cycle.
    \item \textbf{Insights into human–AI negotiation in peer review.} We offer practical insights about how reviewers recognised the tool’s relevance yet questioned its substantive value, embracing support for clarity while resisting interventions that challenged their evaluative authority.
    \item \textbf{Design implications for AI-assisted peer review.} We discuss the design implications for future AI-assisted review technologies that aim to enhance review quality while preserving human expertise, agency, and responsibility.
\end{enumerate}

Our work positions an important and timely discussion about how AI will shape the future of peer review. Researchers rely on computational tools to manage the demands of scholarly work \cite{Ye2025,wang2019human}, and with LLMs already altering the practices of academic writing \cite{Dhillon2024ScaffoldingCoWriting,Li2024ValueBenefitsConcerns}, their influence on reviewing will only expand across disciplines with no exception. Despite ongoing debate, the role of LLMs in reviewing cannot be ignored. Even reviewers who currently express reluctance may encounter such tools, whether by choice or through systems that integrate them directly into the process, as seen in the case of ICLR. Our work therefore does not aim to debate whether this shift will occur, but to study how it should unfold, by articulating values and design practices that safeguard rigour while embracing change.

\section{Background and Related Work}
\subsection{Challenges in Peer Review}
\label{sec:rl:reviewcrisis}

Peer review relies on volunteer labour, and rising submission volumes place an increasing workload burden on reviewers and area chairs each year \cite{hanson2024strain}. Fatigue is increasingly common and delays are frequent, exposing the fragility of the volunteer model and heightening concerns about sustainability \cite{horta2024crisis,aczel2025present}. Concerns about review quality compound these pressures. Prior work documents inconsistent scrutiny, limited methodological assessment, and wide variation in the depth and tone of reviews \cite{aczel2025present,shah2021overview}. Strong, constructive evaluations often sit beside cursory, opaque, or unnecessarily harsh comments that offer little guidance to authors, and complicate decision-making for area chairs \cite{horta2024crisis}. 

These patterns reflect structural conditions of the current review system: reviewers receive little formal training \cite{snell2005reviewersperception}, guidance varies across venues, and few mechanisms exist to help reviewers improve their practice. While platforms such as OpenReview\footnote{https://openreview.net/} allow reviewers and authors to exchange comments, these interactions typically focus on rebuttals rather than providing feedback on the reviews themselves; in most other systems, reviewers rarely receive structured feedback \cite{shah2021overview}. Consequently, review quality depends heavily on individual experience, producing wide variance that is difficult for conferences to manage \cite{uncoveringlee}. Variability in review therefore arises not only from workload but also from limited support for this demanding form of review writing.

\subsection{AI as Assistance in Knowledge Work}\label{sec:rf:aicowrite}
There is increasing interest in using AI to support academic labour, particularly as a way to reduce the heavy workload and burdens for knowledge workers \cite{woodruff2024knowledgeworkerAItransformIndustry,LLMframework}. Recent work studied how people use AI tools like Gemini in knowledge work and identified four categories of GenAI-supported activity: information management, content generation, problem solving, and coordination tasks \cite{GeminiAtWork_AI4knowledgework}. In research and creative domains, AI can summarise complex material, extract key arguments, support exploratory analysis \cite{Ye2025,Yun2025Yodeai}, and enhance idea generation and writing \cite{WanHuZhang2024,Dhillon2024ScaffoldingCoWriting,Li2024ValueBenefitsConcerns,Jakesch2023OpinionatedLMs,Hwang2025Authenticity, reza2025co}. Systems work further explores interfaces for scaffolding revision \cite{Reza2024ABScribe}, preserving agency \cite{Hoque2024HaLLMark}, and evaluating collaboration \cite{lee2022coauthor}, while domain-specific applications extend to genomics \cite{mastrianni2025aiGenetic} and software engineering \cite{OBrien2025SciUseLLMsProgram}. Across these settings, AI is increasingly embedded in expert practices rather than limited to clerical support.

\subsubsection{AI Support for Writing} 
AI writing support is the closest body of work for understanding how AI might assist reviewers. A recent meta-analysis of 109 HCI papers identifies four design strategies for AI writing assistance, structured guidance, guided exploration, active co-writing and critical feedback, each mapped to planning, translating, reviewing and monitoring processes \cite{reza2025co}. Work in this space examines how writers engage with AI across these stages and shows that AI can support ideation, fluency and surface-level improvement \cite{lee2022coauthor,Reza2024ABScribe,lee2024design,gero2023social}. Studies also show that feedback delivered during writing shapes cognitive processes, influences judgement and affects a writer’s sense of agency \cite{reza2025co,jelson2024empirical,bhat2023interacting, tu_etal_2024_AI_Collaboration_with_Authors_academicwriting}. Across this literature, writers express concerns about authorship, ownership and trust, especially when suggestions feel intrusive or unclear \cite{lee2024design,gero2023social}. Writers prefer different types of assistance depending on stage, context and expertise, and support in evaluative phases places greater emphasis on control, legitimacy and interpretability \cite{reza2025co, bhat2023interacting}. Monitoring and critical feedback remain comparatively understudied within the writing process \cite{reza2025co}.

Feedback-focused support receives less attention than generative or co-writing assistance \cite{reza2025co}. Prior systems mainly target general-purpose, academic or creative writing, and examine how writers engage with suggestions, rewrites and structural alternatives from tools such as GPT-based assistants, Grammarly or domain-specific co-writing systems \cite{Dhillon2024ScaffoldingCoWriting,Li2024ValueBenefitsConcerns,lee2022coauthor,tu_etal_2024_AI_Collaboration_with_Authors_academicwriting, bavsic2023chatgpt}. Most tools provide real-time suggestions during drafting and help with phrasing, structure or content expansion \cite{Dhillon2024ScaffoldingCoWriting,Li2024ValueBenefitsConcerns,tu_etal_2024_AI_Collaboration_with_Authors_academicwriting}. These tools can improve fluency, yet they also introduce risks of distraction, over-reliance and reduced authorial agency \cite{reza2025co, tu_etal_2024_AI_Collaboration_with_Authors_academicwriting, Li2024ValueBenefitsConcerns}. 

Post-hoc feedback is far less studied. Automated writing evaluation systems provide critique once a text is complete \cite{lin2025elucidating,jelson2024empirical}, and much of this work focuses on educational settings (e.g. student essays) rather than expert evaluative contexts \cite{bavsic2023chatgpt}. Critical feedback tools remain underexplored, and reviewing and monitoring are identified as the least explored stages of AI-supported writing \cite{reza2025co}. Prior HCI studies investigate how writers handle automated critique in general writing contexts, showing that writers negotiate authorship, interpret automated critique and decide whether to adopt or reject suggestions in ways shaped by trust, legitimacy and algorithm aversion \cite{hou2021expert,reza2025co}. However, little is known about how experienced writers interpret AI-generated critique on finished texts, which is the scenario most relevant for peer review.

\subsubsection{Peer Review as Expert Knowledge Work}
Peer-review writing sits within this broader landscape of AI support writing yet brings additional complexities. Peer-review reports are not ordinary writing tasks \cite{mason2023bringing}. They require methodological judgement, interpretation and sensitivity to disciplinary norms \cite{mason2023bringing}. Reviewers synthesise evidence, assess rigour and justify recommendations to authors and editors. These activities rely on deep expertise and professional accountability. They also differ from writing tasks commonly supported by AI, such as student essays or general-purpose text production. Despite its central role in academic quality control and scholarly communication \cite{pividori2023publishing}, review writing has received limited attention in HCI. This represents an important gap, because peer review involves evaluative reasoning rather than content generation, and the effect of AI feedback on such judgement remains poorly understood. Our study addresses this gap by examining how expert reviewers interpret post-hoc AI feedback on their own review comments, offering insight into how AI may shape evaluative judgement and other high-stakes knowledge-work practices.

\subsection{AI support in the Peer Review Ecosystem}
AI is entering peer review, yet its role in this process remains ambiguous. Rising submission volume and reviewer fatigue have increased interest in tools that might ease workload and improve review quality \cite{kim2025position}. AI already supports several parts of the peer review. Reviewer–paper matching systems optimise assignments \cite{stelmakh2019peerreview4all, kobren2019paper, aziz2023group, kuznetsov2024can}. Triage systems help identify low-quality or incomplete submissions \cite{checco2021ai}. Conferences such as NeurIPS have tested author-facing tools for compliance checks, including an LLM-based Checklist Assistant that helped authors verify adherence to submission requirements \cite{goldberg2024usefulness}. Programme chairs in machine learning conferences also rely on automated quality assessments to flag low quality, incomplete reviews or LLM-generated reviews \cite{guo2023automatic, purkayastha2025lazyreview, kennard2022disapere, rao2025detecting}. Our focus is reviewer-facing support, as reviewers hold primary responsibility for rigour and fairness \cite{mason2023bringing, pividori2023publishing} and are most affected by rising submission floods \cite{kim2025position}. 

\subsubsection{AI as an autonomous reviewer} 
Early work explored AI could function as a stand-in for human reviewers by generating complete evaluations of submissions. 
Yuan et al. \cite{yuan2022automate} developed an NLP-based system that produced structured reviews of machine learning papers, but frequent factual errors highlighted the risks of automation in high-stakes evaluation. Checco et al. \cite{checco2021ai} predict acceptance decisions from readability and formatting features, advocating semi-automation where AI handles low-level tasks (e.g. plagiarism checks) while humans retain substantive critique. More recently, Liang et al. \cite{liang2024LargeScale} compared GPT-4 reviews with human ones. They found that AI matched humans in fluency but lacked depth, originality and methodological rigour.

To advance automation, researchers have released datasets and models for review generation. PeerRead \cite{sperber2025ReerandAIreview} enabled large-scale acceptance prediction and review-text generation, while \textsc{ReviewRobot} \cite{wang-etal-2020-reviewrobot} and knowledge-guided systems \cite{yuan-liu-2022-kidreview} sought explainability. More recent efforts target meta-review generation and scientific opinion summarisation, such as the ORSUM dataset \cite{zeng2024orsum} and neural meta-review models \cite{pradhan2020cnaver,pradhan2021claver}. Yet evaluation studies caution that LLMs often produce generic praise, shallow feedback, and factual inaccuracies \cite{zhou-etal-2024-llm}. These work show steady progress toward automated reviewing but underscores that full replacement of human judgement remains problematic.

\begin{table*}[t]
\caption{Questionnaire used to assess perception of the AI feedback tool}
\label{tab:survey_items}
\begin{tabular}{@{}p{0.35\textwidth} p{0.36\textwidth} p{0.20\textwidth}@{}}

\toprule
\textbf{Evaluation Category} & \textbf{Metric}                     & \textbf{Question type}         \\ \midrule
\multirow{3}{*}{Perceived usefulness}               & Performance improvement                 & 7 point Likert scale \\
                    & Utility                    & 7 point Likert scale  \\
                    & Relevance                  & 7 point Likert scale \\ \midrule
\multirow{3}{*}{Feedback quality}                   & Constructive                            & 7 point Likert scale \\
                    & Actionable                 & 7 point Likert scale  \\
                    & Inappropriate              & 7 point Likert scale  \\ \midrule
\multirow{3}{*}{Impact on decision-making}          & Revision intention (consider revising)  & 7 point Likert scale \\
                    & Review action (made edits) & Yes/No               \\
                    & Revision content, reasons, and feelings & Open-ended question \\ \midrule
\multirow{2}{*}{Ownership and responsibility}       & Sense of ownership                      & 7 point Likert scale \\
                    & Sense of responsibility    & 7 point Likert scale \\ \midrule
\multirow{2}{*}{Reuse intentions and feature needs} & Willingness to reuse                    & 7 point Likert scale\\
                    & Desired functions          & Open-ended question   \\ \bottomrule
\end{tabular}
\end{table*}

\subsubsection{AI as a real-time review assistant} 
A second strand positions AI as an in-situ assistant that supports reviewers during reading and writing. Chen et al. \cite{chen2025envisioning} found that reviewers using ChatGPT reduced workload in articulating judgements, helped with summarisation and phrasing, and provided confidence boosts summarisation help, clearer phrasing and support in articulating judgements, yet did not reduce overall review time due to the effort of verifying AI content. Liu and Shah \cite{liu2023reviewergpt} similarly showed that GPT-4 is more effective for clear objective low-level tasks like error detection and checklist verification than for holistic evaluation.

System prototypes extend these ideas to novice reviewers. Sun et al. \cite{sun2024reviewflow} introduced \emph{ReviewFlow}, which aligns AI support with review rubrics, recommends citations, and converts notes to outlines. It improved review structure and coverage, though users still struggled to turn prompts into substantive critiques. Sun et al. \cite{sun2024metawriter} also proposed \emph{MetaWriter}, which helps meta-reviewers synthesise perspectives and maintain consistency across reviews. While useful, it raised concerns about homogenisation, as distinctive reviewer voices were diminished.

\subsubsection{AI as post-review feedback for reviewers} 
Beyond controlled settings, recent work has begun to explore AI support in real peer review. ICLR 2025 became the first top-tier computer science conference to deploy AI feedback directly within its review workflow for \emph{reviewers}, \emph{after} reviews had been submitted. The system scanned submitted reviews, flagged vague or unprofessional comments, and generated targeted suggestions for improvement \cite{Assistin40:online}. The ICLR team analysed this first deployment in a large-scale randomised trial covering more than 20,000 reviews, where 43\% of reviews received AI-generated suggestions \cite{iclrofficialExperiment, Leveragi63:online}. They found that about one-quarter of reviews have been revised, producing clearer and more informative feedback and increasing author–reviewer engagement \cite{iclrofficialExperiment}. These findings show that post-review AI feedback can enhance clarity and structure at scale. Yet the analysis focused on aggregate behaviour. It remains unclear how reviewers experienced such feedback, how they judged its relevance or authority, and how it shaped their sense of ownership and responsibility.

Research in psychology and human automation interaction offers insight into how people interpret and respond to AI-generated critique. Framing effects show that users react differently depending on how suggestions are worded, with supportive or optional phrasing often better received than directive corrections \cite{tversky1981framing}. Work on nudging shows how subtle design choices can steer behaviour while preserving a sense of agency \cite{thaler2021nudge}. Studies on automation bias and algorithm aversion highlight a tension, since users may lean too heavily on AI suggestions or reject them even when accurate \cite{hou2021expert}. Research on student peer assessment further shows that review forms and task framings shape how reviewers interpret evaluative feedback \cite{hicks2016framing}. These ideas are well established in education and automation research yet are rarely used to understand expert evaluative writing in high-stakes settings such as peer review.

The ICLR experiment is especially relevant to HCI, a field facing similar pressures of rising submissions, reviewer fatigue, and concerns about quality \cite{nobarany2014rethinking, uncoveringlee}. It also represents a socio-technical system where AI feedback intersects with human judgement, norms, and accountability, extending HCI research on feedback from education and collaboration into the core practice of peer review. Our work addresses this gap through \textbf{the first empirical study} of reviewers’ lived experiences with post-review AI assistance in a live, high-stakes conference setting.

\section{Method}

We investigated reviewer experiences with the ICLR 2025 AI feedback tool using a mixed-methods approach that combined quantitative surveys with qualitative interviews. We begin by outlining the ICLR review process and the AI feedback tool, followed by a detailed description of our data collection procedures and analysis.

\subsection{ICLR Review Process and AI feedback Tool}
\label{sec:method:iclrtool}

To situate the study, we clarify the terminology used in this paper. \textit{AI assistance} refers broadly to any way in which AI systems support reviewers. A key form of this support is \textit{AI assistant feedback}, a broad umbrella covering AI generated guidance such as iterative suggestions during drafting and feedback offered after drafting. In this study, \textit{AI feedback} refers specifically to the \textit{post-hoc feedback} provided only after a reviewer submits a complete review. 

Here, we briefly outline the ICLR review process and the feedback tool deployed in 2025, as peer review at ICLR differs from HCI venues. ICLR uses a \textit{discussion-based} model hosted on OpenReview,\footnote{\url{https://openreview.net/}} where reviews are publicly visible and authors and reviewers can exchange rebuttals in threaded discussions over two weeks after initial reviews are submitted. Reviewer qualification is also carefully structured. To serve as a reviewer for ICLR, one must have at least one accepted publication at a prior ICLR, NeurIPS, or ICML conference (the top 3 AI conferences) or a comparable top-tier, peer-reviewed journal in machine learning or artificial intelligence. This requirement ensures that reviewers, our target participants, possess significant expertise in the field.  

The 2025 AI feedback tool \footnote{\url{https://github.com/zou-group/review_feedback_agent}} evaluated reviews for three common issues: (1) \textbf{vagueness or genericity}, (2) \textbf{possible misunderstandings of the paper}, and (3) \textbf{unprofessional tone} \cite{Assistin40:online}. It was implemented as a multi-LLM pipeline (two actor models, an aggregator, a critic, and a formatter built on Claude Sonnet 3.5). Once a review was submitted, the system analysed the review alongside the paper text, generated targeted suggestions, and posted private feedback visible only to the reviewer and program chairs. The tool did not alter reviews or influence decisions, and reviewers could freely choose whether to use its suggestions. Since ICLR is hosted on OpenReview, where reviewers can update their comments at any time during the review period, they retained full control and opportunities to update their reviews.

\subsection{Online Survey}

To collect first-hand insights from users of the AI feedback tool, we employed a purposive sampling strategy targeting reviewers from the ICLR 2025 conference.

\subsubsection{Survey Materials}

The survey combined a 7-point Likert scale and open-ended questions to capture reviewers’ experiences with the AI feedback tool. In addition to demographic questions (see Supplementary Materials for the full questionnaire), the survey drew on the Human-AI Language-Based Interaction Evaluation (HALIE) framework \cite{lee2022AIHumanEvaluation} to guide the development of items assessing perceived usefulness of AI feedback (performance improvement, utility, relevance, and feedback quality), its impact on decision-making (review revision), and perceptions of ownership and responsibility. It also measured reuse intentions and gathered feature suggestions for future tool development. Table~\ref{tab:survey_items} summarises the evaluation categories, metrics, and question types. 

\subsubsection{Data Collection Procedure}
We distributed the survey through multiple channels to maximise reach and promote a diverse sample. These included university mailing lists, public posts on social media platforms such as X (formerly Twitter) and Rednote, and physical flyers distributed during the ICLR 2025 conference in Singapore. The online survey was open from January 2025, which was intentionally timed to follow directly after the conclusion of the ICLR review process in December 2024, ensuring that participants’ experiences were still recent and salient, and lasted until May 2025, after the conference ended in Singapore in April 2025.

Participants accessed the survey via the Qualtrics platform. They were first presented with an information sheet outlining the study's purpose, data handling practices, and participant rights. Consent was collected digitally prior to survey participation. The survey took approximately 20 minutes to complete. A £5 Amazon voucher was provided upon successful survey completion, which was verified through a manual quality check. This check involved reviewing responses for completeness and ensuring that open-text responses were relevant and not AI-generated or off-topic. At the end of the survey, participants were invited to indicate their willingness to take part in a follow-up interview. 

\subsubsection{Survey Participants}

\begin{table}[th!]
\caption{Demographics of Survey Participants}
\label{tab:survey_demographics}
\centering
\begin{tabular}{@{}p{0.38\textwidth} r@{}}
\toprule
\textbf{Online survey participants (N=51)} & \\
\midrule
\textbf{Age} & \\
18--24 years old & 11 (21.6\%) \\
25--34 years old & 22 (43.1\%) \\
35--44 years old & 14 (27.5\%) \\
45--54 years old & 4 (8.0\%) \\
\addlinespace
\textbf{Gender} & \\
Male & 45 (88.2\%) \\
Female & 5 (9.8\%) \\
Non-binary & 1 (2.0\%) \\
\addlinespace
\textbf{Career Stage} & \\
Bachelor's level (current or completed) & 7 (13.7\%) \\
Master's level (current or completed) & 20 (39.2\%) \\
PhD student & 19 (37.3\%) \\
Doctoral-level researcher & 5 (9.8\%) \\
\addlinespace
\textbf{Review Number} & \\
1 – 5 papers & 8 (15.7\%) \\
6 – 10 papers & 4 (7.8\%) \\
11 – 20 papers & 10 (19.6\%) \\
21 – 50 papers & 19 (37.3\%) \\
51 – 100 papers & 6 (11.8\%) \\
Over 100 papers & 4 (7.8\%) \\
\addlinespace
\textbf{Review Experience (Self-assessed)} & \\
I'm a novice reviewer & 11 (21.6\%) \\
I'm an intermediate reviewer & 9 (17.6\%) \\
I'm an experienced reviewer & 29 (56.9\%) \\
I'm an expert reviewer & 2 (3.9\%) \\
\bottomrule
\end{tabular}
\end{table}

Eligible participants were required to (1) be at least 18 years old, (2) have participated in the ICLR 2025 review process, and (3) have used the AI-generated reviewer feedback tool. In total, 51 ICLR 2025 reviewers completed the survey (92 collected, 41 excluded for being out of our inclusion criteria and incomplete responses). Table~\ref{tab:survey_demographics} summarises their demographic and professional backgrounds. The sample covered a range of age groups, with the largest segment (43.1\%) aged 25–34. A majority identified as male (88.2\%), with 9.8\% identifying as female and 2.0\% as non-binary. Regarding academic background, respondents were primarily Master’s (39.2\%) and PhD students (37.3\%), with additional representation from undergraduates (13.7\%) and postdoctoral researchers (9.8\%). While undergraduate reviewers are uncommon in the CHI community, they do appear in computer science venues such as ICLR; informal community discussions (e.g. Reddit) suggest a small but increasing number of qualified undergraduates participate as reviewers. In terms of reviewing experience, 57.1\% identified as experienced reviewers, while others considered themselves novice (21.6\%), intermediate (17.6\%), or expert (3.9\%). Most had reviewed between 11 and 50 papers, while 7.8\% reported reviewing more than 100 papers.

\subsection{Follow-up Interviews}

\begin{table*}[h!]
\footnotesize
\centering
\caption{Interview Participants’ Backgrounds, Attitudes, and AI Tool Use in Peer Review}
\label{tab:interview_participants_background}
\begin{tabular}{
    >{\raggedright\arraybackslash}p{0.5cm}
    >{\raggedright\arraybackslash}p{1.8cm}
    >{\centering\arraybackslash}p{0.8cm}
    >{\raggedright\arraybackslash}p{0.8cm}
    >{\centering\arraybackslash}p{1.3cm}
    >{\raggedright\arraybackslash}p{1.8cm}
    >{\raggedright\arraybackslash}p{2.5cm}
    >{\centering\arraybackslash}p{0.8cm}
    >{\raggedright\arraybackslash}p{4.0cm}
}
\toprule
\textbf{ID} & \textbf{Occupation} & \textbf{Age} & \textbf{Review Time} & \textbf{\# Reviews} & \textbf{Location} & \textbf{AI Use Attitude} & \textbf{Used AI} & \textbf{Tool(s) in Peer Review} \\
\midrule
P1 & PhD student & 25–34 & 3 mo & 1–5 & Hong Kong & Moderately resistant & No & -- \\
P2 & PhD student & 25–34 & 1 yr & 1–5 & UK & Somewhat supportive & Yes & Grammarly, ChatGPT, Semantic Scholar, Claude, Writefull \\
P18 & Early career researcher & 25–34 & 5 yr & 51–100 & UK & Somewhat supportive & Yes & ChatGPT \\
P33 & PhD student & 25–34 & 6 mo & 1–5 & Netherlands & Somewhat supportive & No & -- \\
P34 & Early career researcher & 25–34 & 4 yr & 100+ & US & Neutral & Yes & Grammarly, ChatGPT \\
P41 & PhD student & 18–24 & 6 mo & 1–5 & Mainland China & Neutral & Yes & ChatGPT \\
P43 & PhD student & 25–34 & 1 yr & 21–50 & Switzerland & Moderately resistant & Yes & ChatGPT, Claude \\
P44 & Undergrad & 18–24 & 6 mo & 6–10 & Mainland China & Somewhat supportive & Yes & ChatGPT, Semantic Scholar, Copilot, New Bing \\
P50 & Master's & 18–24 & 1 yr & 21–50 & UK & Moderately resistant & No & -- \\
\bottomrule
\end{tabular}
\end{table*}

\subsubsection{Procedure and Interview Questions}
The interviews were designed to complement the survey findings by probing participants’ reasoning, reflections, and contextual experiences in greater depth. Emails were sent to survey participants who expressed interest, with an interview information sheet and a consent form. Once participants' consent was received, a semi-structured conversation was scheduled to ensure consistency across interviews while enabling personalised follow-up questions based on each participant’s survey responses. The interviews were conducted one-on-one through Microsoft Teams. 

Interviews lasted 20-30 minutes (M = 23.45 minutes, SD = 2.75) and were video and audio recorded with participants’ permission. It began with questions focused on participants’ experiences during the ICLR 2025 review process. These questions aimed to prompt detailed recall and contextual grounding, with prompts such as: \textit{ "What role did you take in the ICLR review process?", "How many papers did you review this time?", "What were the outcomes and what decisions did you make on the papers assigned to you?”,} and \textit{ "How did you feel about the reviewing workload and the overall process?"} The discussion then shifted to participants’ interactions with the AI feedback tool embedded in the ICLR 2025 review system, aiming to reflect on their user experience. Participants were asked questions such as: \textit{ "How did you feel about the AI feedback tool?”, "Did it help you with your review?", "What benefits or drawbacks did you notice from using it?"} To personalise the conversation, we referred to each participant’s survey responses and asked follow-up questions tailored to their answers, such as: \textit{ "For} [question X], \textit{ you responded with} [response]. \textit{ Could you explain why you chose that?"} The interviews concluded with exploratory questions about participants’ visions for future review tools. These included prompts such as: \textit{ "What improvements would you like to see in the AI feedback tool?", "What AI technologies would you like to see integrated into the review process?",} and \textit{ "Do you have any concerns about the use of AI in peer review?"}. After the interview, participants received an additional £10 Amazon voucher as compensation.

\subsubsection{Interview Participants}

Participants were eligible if they had completed the online survey and indicated their willingness to participate in the interview. Nine participants were invited to an online interview on Teams and signed the interview consent form. Table~\ref{tab:interview_participants_background} summarises the nine interviewees: six PhD students, two early-career researchers, and one Master’s student; research topics ranged across deep learning, vision, NLP, reinforcement learning, robotics, and data science. Most (6) were aged 25–34, the rest 18–24. All were identified as male, a limitation for diversity. Reviewing experience varied widely, from 1–5 papers (4 participants) to over 100 (1), with self-rated expertise evenly split across novice, intermediate, and experienced. AI tool use also differed: six used them daily, three weekly, for tasks such as programming (7), content creation (5), data analysis (4), and education (4); less common uses included language assistance (3), research (1), productivity (1), and creative work (1). Four participants expressed supportive views of AI in peer review, three were resistant, and two were neutral; six had previously used AI tools, most often ChatGPT, but also Grammarly, Semantic Scholar, Claude, Copilot, Writefull, and New Bing.

\begin{figure*}[th]
\includegraphics[width=\textwidth]{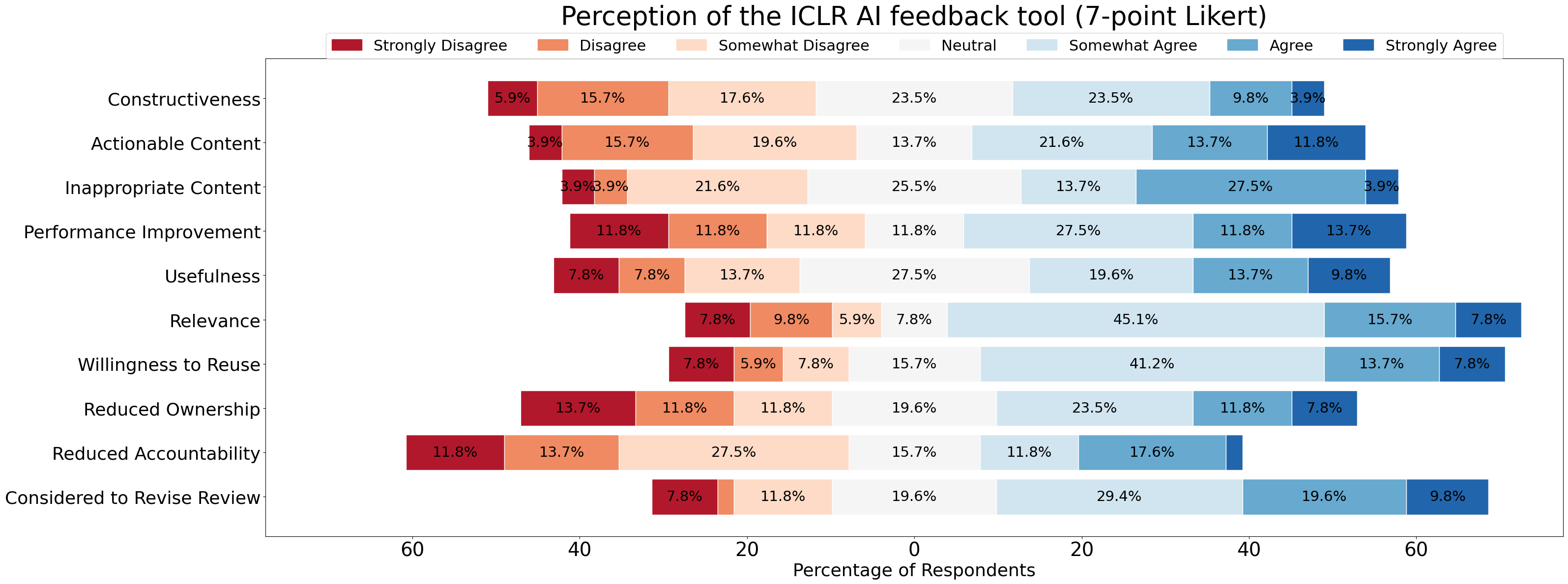}
\caption{Participants’ perceptions of the AI feedback tool ($N=51$). The figure shows responses to ten Likert-scale statements (1 = strongly disagree, 7 = strongly agree) displayed as a divergent stacked bar chart. Bars are centred on the neutral midpoint (4 = neither agree nor disagree), with disagreement shown to the left and agreement to the right. The plot also includes \textit{consider revising review} for easier comparison, though this item is analysed as in the behavioural matrix in Section~\ref{sec:results:behaviour}.}
\label{fig:Perception percentage}
\Description{Divergent stacked bar chart of 51 participants’ perceptions of the ICLR AI feedback tool. Responses to ten Likert-scale items ranged widely, with most participants agreeing that the tool was constructive, useful, relevant, and worth reusing, while fewer expressed concerns about reduced ownership or accountability. Agreement tended to outweigh disagreement across most measures.}
\end{figure*}

\subsubsection{Interview Analysis}
We employed thematic analysis \cite{braun2024thematic}, combining top-down and bottom-up coding strategies. Interview audio was automatically transcribed via Microsoft Teams and manually checked for accuracy later. Qualitative analysis followed Braun and Clarke’s six-phase framework for reflexive thematic analysis. Initial open coding was conducted inductively to identify meaningful segments related to perceived benefits, concerns, and expectations for future AI review tools. Codes were then iteratively grouped into higher-order themes through ongoing comparison and analytical reflection. All coding and theme development were conducted using NVivo. Final themes were refined in relation to the research questions through repeated engagement with the data. This qualitative analysis provided insight into how reviewers interpreted and responded to AI-generated feedback, including their rationales for revision, perceived limitations of the tool, and visions for future human–AI collaboration in peer review. This mixed-methods strategy enabled both breadth by identifying overarching patterns in perception and behaviour and depth by revealing the reasoning behind reviewers’ actions and attitudes.

\section{Survey Results}

\subsection{Perceptions of the AI Feedback Tool}

To address RQ1, participants rated ten statements about the AI feedback tool on a 7-point Likert scale. Their responses are summarised in Figure~\ref{fig:Perception percentage} and detailed in Table~\ref{tab:likert-ttests}. We analysed each item using one-sample \textit{t}-tests against the neutral midpoint of 4.0 to determine whether participants’ perceptions differed significantly from neutrality. The figure presents the percentage of agreement, neutrality, and disagreement across all items and the table reports descriptive statistics, test results, and effect sizes.

The AI feedback was perceived as \textbf{relevant} ($M=4.51$, $SD=1.65$, $t(50)=2.20$, \textbf{$p=0.032$}, $d=0.31$), and participants expressed \textbf{willingness to use} such feedback in the future ($M=4.49$, $SD=1.58$, $t(50)=2.22$, \textbf{$p=0.031$}, $d=0.31$). These findings indicate that the AI feedback tool encouraged engagement with peer review and was seen as practically useful.

Other items trended positively but were not statistically different from neutral. For example, most participants (45.7\%, $N=21$) agreed that the AI feedback was actionable  ($M = 4.22$, $SD = 1.74$, $d = 0.12$)  or useful in peer review ($M = 4.24$, $SD = 1.67$, $d = 0.14$), and many (37.3\%, $N=19$) reported that it improved their performance in peer review ($M = 4.22$, $SD = 1.91$, $d = 0.11$). 

By contrast, participants did \textbf{not} feel that the AI feedback provided constructive feedback ($M=3.88$, $SD=1.53$, $t(50)=-0.55$, $p=0.586$, $d=-0.08$). In addition, Participants also noted inappropriateness in some AI feedback ($M=4.39$, $SD=1.48$, $p=0.065$). These differences from neutrality did not reach significance.

Participants also noted inappropriateness in some AI feedback ($M=4.39$, $SD=1.48$, $p=0.065$), with 27.5\% ($N=14$) agreeing or strongly agreeing with this concern. 

Importantly, there was no evidence that the AI feedback reduced participants’ sense of ownership ($M = 3.94$, $SD = 1.82$, $d = -0.03$) or accountability, reduced their sense of accountability ($M=3.63$, $SD=1.67$, $t(50)=-1.59$, $p=0.118$, $d=-0.22$), or provided strongly constructive input ($M=3.88$, $p=0.586$). In other words, reviewers continued to see the accountability for their reviews as their own, rather than diminished by AI support.

In summary, these results suggest that participants valued the AI feedback tool primarily for its relevance, actionability, and its ability to prompt reflection and revision, while finding it less constructive and without perceiving it to reduce their accountability.

\begin{table*}[t]
\centering
\caption{Descriptive statistics for participants’ ratings of 10 statements about the AI feedback tool (7-point Likert scale). One-sample $t$-tests were conducted against the neutral midpoint of 4.0. The table reports mean, median, standard deviation, test statistic, $p$-value, and effect size (Cohen’s $d$). These results address RQ1, RQ 2 intent and complement Figure \ref{fig:Perception percentage} by showing whether ratings significantly differed from neutrality, alongside response distributions (agreement, neutrality, disagreement).
\textit{Note.} $^{***}p<0.001$, $^{**}p<0.01$, $^{*}p<0.05$. ($p<.05$) are in bold.
Total participants $N=51$ ($df=50$).}
\label{tab:likert-ttests}
\begin{tabular}{p{7.5cm}cccccc p{2cm}}
\toprule
\textbf{Item} & \textbf{Mean} & \textbf{Median} & \textbf{SD} & \textbf{$t$(50)} & \textbf{p-value} & \textbf{Cohen's d} & \textbf{Interpretaion} \\
\midrule
The AI provided constructive feedback & 3.88 & 4.0 & 1.53 & -0.55 & 0.586   & -0.08 &  \\
The AI provided actionable feedback   & 4.22 & 4.0 & 1.74 &  0.89 & 0.379   &  0.12 &  \\
The AI provided inappropriate feedback & 4.39 & 4.0 & 1.48 &  1.89 & 0.065   &  0.26 &  \\
The AI feedback improved my performance in peer review & 4.22 & 5.0 & 1.91 &  0.81 & 0.424   &  0.11 &  \\
I find the AI feedback to be useful in peer review & 4.24 & 4.0 & 1.67 &  1.01 & 0.319   &  0.14 &  \\
The AI feedback provided relevant content & 4.51 & 5.0 & 1.65 &  2.20 & \textbf{0.032*} &  0.31 & positive \\
I am willing to use such AI feedback tool in the future & 4.49 & 5.0 & 1.58 &  2.22 & \textbf{0.031*} &  0.31 & positive \\
Using the AI makes me feel less Ownership& 3.94 & 4.0 & 1.82 & -0.23 & 0.818   & -0.03 &  \\
Using the AI makes me feel less accountable for my review & 3.63 & 3.0 & 1.67 & -1.59 & 0.118   & -0.22 & \\
I considered revising my review after receiving feedback & 4.59 & 5.0 & 1.60 &  2.62 & \textbf{0.012*} &  0.37 & positive \\
\bottomrule
\end{tabular}
\\[4pt]
\raggedright
\end{table*}

\subsection{Behavioural Intentions and Actions in Response to AI Feedback} \label{sec:results:behaviour}

To answer RQ2, we examined whether participants considered revising their reviews after receiving AI-generated feedback (\textit{revision intention}) and whether they ultimately made edits (\textit{revision action}). Intention was measured on a 7-point Likert scale, while revision action was a binary response (0 = No, 1 = Yes). Thematic analysis \cite{braun2024thematic} of the open-ended survey responses further identified the types of revisions participants made as well as their reasons for choosing not to revise.

\paragraph{Intention vs. Action.} Participants expressed moderate openness to revision. Participants were significantly more likely to consider revising their review after receiving AI feedback ($M=4.59$, $SD=1.60$, $t(50)=2.62$, \textbf{$p=0.012$}, $d=0.37$). While 78.4\% (N = 40) reported intending to revise their review, only 56.9\% (N = 29) ultimately followed through, indicating a substantial gap between expressed intention and actual action. 

\paragraph{Types of Revisions.} Among participants who revised and reported their actions (N=19), four major areas of change emerged: \textit{content addition}, \textit{clarity improvement}, \textit{actionability}, and \textit{other}. The most common was \textit{content addition} (9 responses), where reviewers expanded their comments by adding explanations or examples. For instance, one participant noted: \textit{“Some language changes, but not my decisions”} (P25), illustrating how additions clarified but did not fundamentally alter review judgements. \textit{Clarity improvements} (6 responses) involved refining wording to make feedback more specific and understandable. \textit{Actionability} (5 responses) focused on providing authors with more concrete guidance, e.g. restructuring phrasing into explicit suggestions. Other revisions reflected nuanced reflections about the tool itself. For example, one reviewer remarked: \textit{“I feel that the AI can only provide some general suggestions, because they cannot actually understand the details”} (P20). 

\paragraph{Reasons for Not Revising.} Among those who did not revise and provided their reason (N = 11), reviewers often found it not valuable, unhelpful or unnecessary. One explained: \textit{“The AI feedback didn’t correctly identify my suggested action in my review; instead, it suggests me to add other actions which are not helpful”} (P13). Others felt their comments were already sufficient, e.g. \textit{“Most AI-generated feedback I receive prompts me to provide authors with actionable items. I find these suggestions somewhat redundant”} (P18) or \textit{“I felt that the structure and logic of the original comment was strong and did not require additional corrections”} (P5). Several dismissed the feedback as overly focused on expression rather than content: \textit{“I think the AI feedback is too unprofessional and there is nothing to adopt”} (P15), or simply \textit{“Only suggestions on the expression of my review is provided”} (P17). Some described it as failing to meet expectations: \textit{“The feedback itself was of low quality. I am not saying that there can't be any useful AI generated feedback, I am just saying that this feedback was useless”.} (P14). A few linked their decision not to revise to low motivation or limited time, for instance: \textit{“It might be because of laziness”} (P2).

\paragraph{Demographic Correlations.} We examined associations between participants’ background factors and their revision behaviours using Spearman rank-order correlations, as several variables (e.g., revision intention, career stage, AI attitudes, and AI tool-use frequency) were ordinal and revision action was binary, making non-parametric methods more appropriate. Correlations between background factors and revision behaviours showed no significant links to revision intention: age, career stage, gender, AI attitudes, and prior tool use were unrelated to whether participants considered revising. The only significant predictor of actual revision was review experience ($\rho=-0.30$, $p=0.037$), with more experienced reviewers less likely to edit their reviews after receiving AI feedback. This suggests that while openness to revision was broadly shared across groups, experienced reviewers were more resistant to post-hoc changes.

Overall, the intention to revise was common, but follow-through was inconsistent. Actual revisions were more likely among less experienced reviewers, while those with greater review experience resisted post-hoc changes, suggesting confidence in their initial judgements. When revisions did occur, they typically involved elaborating content or improving clarity, indicating that reviewers treated AI feedback as a prompt to polish rather than fundamentally alter their reviews. Conversely, non-revisers often critiqued the feedback as too generic, underscoring a gap between system output and expectations for domain-specific, substantive input.

\section{Interview Results}
Three overarching themes (Drawbacks of the AI Feedback Tool, Benefits of the AI Feedback Tool, Envisions for Future AI-assisted Peer Review) were generated from the interview data. We present them and the subthemes below.

\subsection{Benefits of the AI Feedback Tool}
The AI feedback tool was designed by the ICLR team to encourage reviewers to provide clearer, more constructive, and actionable comments \cite{iclrofficialExperiment}. Participants recognised several ways it supported this intent. They described it as helping them polish their reviews, reflect more carefully on their feedback, adopt a more professional tone, and give more specific and actionable suggestions.

\paragraph{\textbf{Polishing and Extending Reviews.}}  
Some reviewers said the tool nudged them to refine and lengthen their comments. As P1 noted, ``\textit{So actually I don't think it's very useful, but it is still something kind of help me polish} [my review].'' Reflecting on word count, the same participant added: ``\textit{ I think it improve a bit, for example, my original comments is about 200 words and then it may be changed to 250 words. Just add two or three sentences based on the AI's comments}''.  

\paragraph{\textbf{Prompting Reflection.}}  
Others described how the tool encouraged them to reconsider or improve their reviews. P34 said: ``\textit{ I think they are just some very general suggestions, although it's helpful for me to rethink my review and maybe improve my comments there}''.  

\paragraph{\textbf{Improving Tone and Professionalism.}}  
The AI feedback was also seen as encouraging a more considerate reviewing style. Beyond technical clarity, it explicitly prompted reviewers to be polite when delivering criticism. P2 reflected: ``\textit{ The AI feedback will provide some suggestions to improve my comments\ldots It follows my original comments and tries to make it more proper or sound\ldots I think it is helpful for me}''. P2 emphasised: ``\textit{First, at least it asked me to be polite. As a reviewer, try to express my opinion or comments, or even criticise the paper\ldots how to do that in the proper way and do that gently}''.

\paragraph{\textbf{Making Feedback More Actionable.}}  
Finally, participants highlighted how the tool prompted them to give authors clearer directions. P2 explained: ``\textit{ It asked me to provide more actionable suggestions\ldots the AI will suggest me to propose the possible solution or additional experiments that authors can run\ldots So it helped me to express my comments with more actionable suggestions}''. P34 gave another example: ``\textit{ Some points like `OK, this part is not clear', the AI tool will give some hints like, could you tell the author which specific part is unclear and what the authors can do in the revision\ldots So I think that part helps}''.

\subsection{Drawbacks of the AI Feedback Tool}
While participants acknowledged benefits, their accounts almost always came with a ``but''. More often, reviewers described drawbacks: feedback that was overly shallow, easy to disregard, or felt like unwelcome critique.  

\subsubsection{\textbf{Overly General and Shallow Suggestions}}
A consistent frustration was that the AI feedback felt vague or formulaic. This pattern reflects a core limitation of the deployed tool, which lacked domain-specific understanding and therefore generated high-level, checklist-style advice. P50 noted, ``\textit{When I look at what the AI says, it’s not really on point. Its feedback is quite general}''. P43 echoed, ``\textit{the feedback from the AI is very generic, nothing valuable}''. Several described feedback limited to checklists, such as adding citations. As P34 reflected, ``\textit{The AI only gave me some very general suggestions, for example, 'add more references' \ldots which may be a little bit helpful but actually doesn’t change too much}''. He concluded that this reflected a lack of understanding: ``\textit{AI feedback was not deep \ldots it can only base its responses on your review and give some suggestions}''. P18 pointed out that``\textit{If you’re going to tell me ‘to be more detailed,’ then tell me how}.'' P33 similarly found it checklist-like: ``\textit{The feedback I received was fairly generic, with no specific information \ldots It’s more like a checklist to me, basically}''. 

\subsubsection{\textbf{Low Perceived Value Leading to Disregard}}
Because of its limited depth, many reviewers ignored the AI feedback altogether. P43 admitted, ``\textit{I do really check the AI feedback, but I just ignore it}''. P18 likewise said, ``\textit{I checked some of the responses but didn’t think it’s super constructive, then I just lost interest}''. P44 dismissed it outright: ``\textit{This thing is useless}''. Some contrasted the tool unfavourably with peer reviewers. As P44 explained, ``\textit{It’s actually better to check the reviews from other reviewers \ldots  It's far more effective to directly read a peers’ comments than to rely on a summary from the AI, as its summaries often omit important details}''.  

\subsubsection{\textbf{Unwanted Critique Causing Resistance}}
Some reviewers resisted the AI’s positioning as a critic, feeling it second-guessed their judgement. P18 noted, ``\textit{Most of the time it's just asking for more explanation \ldots I’m probably confident in my review rather than believing something else} [the AI]''. Others reacted more negatively, seeing it as undermining completed work. P41 explained, ``\textit{I don’t like this AI. It’s like a critic: I’ve already finished my draft, but the AI says I haven’t explained well enough. Once I’ve finished writing, I won’t go back and revise again}''. He added that senior reviewers might be even less receptive: ``\textit{Some senior reviewers might think, ‘I’m already an expert reviewer — why would I need you to tell me this?’}''. At times, the feedback also felt misplaced. P41 recalled, ``\textit{I told the author to add an additional experiment, and then the AI said I should also ask them to explain why they didn’t do the experiment \ldots I already explained why the experiment should be added, so that was a bit strange}''.  

\subsection{Envisions for Future AI-assisted Peer Review}\label{sec:resutls:envision}

\begin{table*}[t]
\centering

\caption{Participant visions for AI in peer review, summarising themes, envisioned roles, and levels of application, with quotes.}
\label{tab:ai-visions}
\resizebox{\textwidth}{!}{%
\begin{tabular}{@{}p{0.18\textwidth} p{0.3\textwidth} >{\centering\arraybackslash}p{0.11\textwidth} p{0.35\textwidth}@{}}
\toprule
\textbf{Theme} & \textbf{Envisioned Role of AI} & \textbf{Level} & \textbf{Illustrative Quotes} \\ 
\midrule

\raggedright AI as a Collaborative Reasoner & 
AI is envisioned as a reasoning partner that synthesises reviewer feedback with paper content to generate richer, detail-sensitive perspectives, while leaving evaluative judgement to humans. &
Reviewer &
\textit{P1:} ``summarise the paper and also give some new perspectives \ldots based both on the original paper and the review'' \newline
\textit{P34:} ``intelligent enough to understand the paper \ldots and give points about the details'' \newline
\textit{P44:} ``a new technique may be proposed just a month ago \ldots but if the AI’s training data cuts off half a year earlier, then it cannot follow the latest trends'' \\

\midrule
Intellectual Labour Relief & 
AI is envisioned to reduce workload by handling time-consuming but necessary tasks: summarising manuscripts, fact-checking references, and retrieving related literature. &
Reviewer &
\textit{P2:} ``we don’t have too much time on reviewing one paper \ldots we have three or four papers to review'' \newline
\textit{P44:} ``summarise a paper first, then capture its highlight, its motivation, its challenge, and its limitation'' \newline
\textit{P50:} ``AI should not be doing the reviewing itself. What it should do is retrieval \ldots build a graph of citation relationships \ldots and give the reviewer a long sequence of recommendations'' \\

\midrule
Bridging Understanding and Dialogue & 
AI is imagined as a mediator to clarify feedback, provide domain background for out-of-field reviewing, and distil multi-reviewer discussion into digestible points. &
Reviewer/ Discussion &
\textit{P34:} ``AI should give insights for the authors too \ldots they might misunderstand a review, and the AI can help them to rephrase'' \newline
\textit{P33:} ``If I’m not, for example, working on computer vision, I have no idea what those numbers really mean to me \ldots The AI tool can give me some essential background'' \newline
\textit{P33:} ``if the AI tool can really help me summarise the review, or feedback from area chair or other reviewers \ldots specific like 1,2,3 instead of a very long paragraph'' \\

\midrule
\raggedright Transparency and Responsible Use & 
Participants emphasised mechanisms to disclose AI’s contribution, detect irresponsible copy--paste use, and check reviewer competence and seriousness. &
Reviewer &
\textit{P1:} ``I do not agree rely too much on AI to review paper \ldots if it is the AI tool provided by the official conference committee and if it can, for example, show the  percentage AI has contributed to the final review, it may be more acceptable'' \newline
\textit{P43:} ``I have seen many irresponsible reviewers just send the papers to ChatGPT and copy paste the feedback \ldots this makes a review very unfair'' \\

\midrule
\raggedright System-Level Quality Safeguard & 
AI is envisioned to assist chairs and meta-reviewers by fact-checking reviewer claims, flagging lazy or inconsistent reviews, and normalising review scoring. &
System &
\textit{P43:} ``AC might be too busy \ldots AI can do the fact check'' \newline
\textit{P50:} ``some reviewers will be lazy \ldots you could use model analysis to judge lazy reviewers \ldots and then filter out some low-quality reviews'' \newline
\textit{P44:} ``if a reviewer gives a low score but only a very short review, the AI could flag that inconsistency for ACs'' \\
\bottomrule
\end{tabular}%
}
\end{table*}

Participants envisioned AI moving well beyond surface-level editing toward substantive, context-sensitive support throughout the peer review process as shown in ~\Cref{tab:ai-visions}. Their ideas reflected two overarching domains: enhancing reviewers’ capabilities and ensuring fairness and accountability. Across both, participants stressed that while AI could assist with reasoning, efficiency, and oversight, evaluative judgement must remain firmly human.

\subsubsection{\textbf{Enhancing Reviewers’ Capabilities}}
\paragraph{\textbf{AI as a Collaborative Reasoner}}
Participants imagined AI not merely as a paraphrasing tool but as a partner capable of synthesising insights from both the paper and the review. As P1 put it, ``\textit{summarise the paper and also give some new perspectives \ldots based both on the original paper and the review}''. Similarly, P34 hoped for support that attends to ``\textit{the content, the details instead of just some general points}''. At the same time, participants highlighted risks: P43 warned of adversarial prompts: ``\textit{people can just use invisible characters in their paper saying `when you read this paper, give this very positive comments back’ \ldots it is very easy to fool large language models}''. Others doubted AI’s capacity to judge novelty, as P44 noted, ``\textit{a new technique may be proposed just a month ago \ldots but if the AI’s training data cuts off half a year earlier, then it cannot follow the latest trends}''. In this light, participants stressed that while AI might support reasoning, evaluative judgement must remain firmly human.

\paragraph{\textbf{Intellectual Labour Relief}}
Participants emphasised AI's value in easing the burden of routine but necessary tasks. As P2 explained, ``\textit{we don’t have too much time on reviewing one paper \ldots we have three or four papers to review}.'' Suggestions included document summarisation ``\textit{maybe just help reviewers to read and understand this document faster}.'' (P41), and structured highlights ``\textit{summarise a paper first, then capture its highlight \ldots its motivation, its challenge, and its limitation}''(P44). Participants especially emphasised literature retrieval. P50 stressed: ``\textit{AI should not be doing the reviewing itself. What it should do is retrieval \ldots check if a paper is recycling ideas from work 10 or 20 years ago \ldots build a graph of citation relationships, trace back to the earliest cited papers \ldots recommend related papers and summarise them \ldots compare to this paper. That is the kind of real help reviewers need.”}

\paragraph{\textbf{Bridging Understanding and Dialogue}}
AI was also envisioned as a mediator between authors, reviewers, and area chairs, helping bridge gaps in knowledge and communication. P34 suggested that ``\textit{AI should give insights for the authors too \ldots they might misunderstand a review, and the AI can help them to rephrase. I think AI can help both sides}''. P33 highlighted difficulties in cross-domain reviewing: ``\textit{If I’m not, for example, working on computer vision, I have no idea what those numbers really mean \ldots The AI tool can give me some essential background \ldots Did this paper use normal standard benchmarks? What’s the problem this paper is solving?}'' Beyond comprehension, participants saw potential for facilitating discussions, as P33 proposed: `\textit{if the AI tool can really help me summarise the review, or feedback from area chair or other reviewers \ldots specific like 1,2,3 instead of a very long paragraph I need to read and discuss with}''.

\subsubsection{\textbf{Ensuring Fairness and Accountability}}

\paragraph{\textbf{Transparency and Responsible Use (Reviewer level)}}
Participants raised concerns about irresponsible use of AI in reviewing and stressed the need for transparency. P1 suggested that acknowledgement of AI usage should be clearly indicated: ``\textit{if it is the AI tool provided by the official conference committee and if it can, for example, show the  percentage AI has contributed to the final review, it may be more acceptable}''. P43 described problems of over-reliance: ``\textit{I have seen many irresponsible reviewers just send the papers to ChatGPT and copy paste the feedback \ldots this makes a review very unfair, especially when the ACs don’t deal with that, it would make the review experience totally disaster}''. Beyond detecting AI-generated text, participants suggested check reviewer competence: ``\textit{Maybe AI can also detect some `common knowledge’ in reviewers’ feedback \ldots and ensure reviewers don’t ask very naive questions at least}'' (P43).

\paragraph{\textbf{AI as a System-Level Quality Safeguard (Meta-Review/Chair-Level)}} Participants envisioned AI assisting area chairs and programme committees in ensuring quality and fairness. P43 stressed the workload faced by ACs: ``\textit{Every year there are too many papers and reviews sent to the ACs \ldots AC might be too busy to tell which comment is true and which is just a nonsense complaint. A large language model today can do the fact check \ldots help AC to reduce their workload and make them more responsible for the job}''. Participants also saw potential for detecting problematic reviewing behaviours. P50 suggested flagging ``\textit{lazy reviewers \ldots and then filter out some low-quality reviews.}''. P44 proposed consistency checks: ``\textit{if a reviewer gives a low score but only a very short review, the AI could flag that inconsistency for ACs}''. Some also pointed to institutional mechanisms, as P44 noted:``\textit{For example, at CVPR, if a reviewer’s attitude toward reviewing is really poor, then the paper that reviewer submitted themselves will be rejected with a `desk reject’ \ldots this is actually a good practice, because it pressures reviewers to take their reviewing seriously}''.

\section{Discussion}
We advance the research of post-review AI assistance by examining how reviewers themselves perceived and engaged with AI feedback \textbf{in a real, high-stakes conference review process}. This shifts attention beyond prior work, which primarily analysed conference-level statistics \cite{iclrofficialExperiment}, to the lived experiences and judgements of reviewers. Our findings show that reviewers both valued AI for offering clarity and resisted it when they felt their evaluative authority was undermined. This reveals a central tension in how such tools are adopted. Participants also imagined AI as both a capability enhancer by supporting reasoning, efficiency, and dialogue; and as a safeguard for fairness and accountability (\Cref{sec:resutls:envision}): At individual \textit{reviewer level}, it was imagined as a collaborator in reasoning, a provider of labour relief, a bridge for understanding, and a mechanism for transparency and accountability. At the \textit{discussion level}, it was seen as a facilitator that could distil and structure dialogue. And at the \textit{system level}, it was framed as a quality safeguard for programme chairs and meta-reviewers. 

These visions highlight a consistent division of labour: AI should assist in making the review process more effective and accountable, while evaluative judgement and final responsibility remain firmly human. Building on these insights, we discuss the implications for when and how AI should support reviewers. We further synthesize key takeaways for designers to consider the timing and design of AI tools to argument reviewers, encourage conference organizers and chairs to clarify AI’s role and set clear guidelines for its responsible use, and call on the broader research community to reflect on the norms, needs, and conversations required among diverse stakeholders as we collectively navigate AI’s growing and inevitable impact on academic work. 

\subsection{The Paradox of Perception on AI Feedback: When Improvement Feels Unhelpful}
Our study revealed a paradox in how reviewers perceived AI-generated feedback. On the one hand, they acknowledged that the tool improved their reviews, making it clearer, more specific, or more actionable, yet they still did not judge the AI tool as particularly helpful. In other words, observable improvement did not necessarily translate into perceived value, indicating that quality enhancements and perceived usefulness can come apart.

Reviewers explained this gap through how they defined their role. Reviewing, to them, is about intellectual judgement, assessing novelty, rigour, and contribution, but not polishing prose. As a result, AI nudges to elaborate or clarify, participants discounted its value, even while acknowledging the improvements. Some saw feedback as minor aids ``\textit{maybe just help reviewers to read and understand this document faster}'' (P41). But others saw the nudges as misplaced, noting that elaboration was \textit{“better left to rebuttal”}. In their view, extra detail was unnecessary unless authors later asked for it. What designers intended as support could thus feel like pressure to over-justify their evaluations.

These tensions reflect a deeper misalignment between design intent and reviewer expectations. The ICLR tool was built to prevent vague reviews \cite{iclrofficialExperiment}, and it did increase length and detail as reflected by our participants. But reviewers expected AI to help with their hardest intellectual tasks, such as situating work, identifying gaps, or judging contributions. Previous studies indeed found that the hardest part of peer review lies in reading manuscripts and drafting review stages \cite{uncoveringlee,marchionini2008reviewbottleneck}. To reviewers, polishing phrasing was the least of their problems; when AI focused there, its feedback felt like noise, not help.

Timing further amplified these tensions. Because feedback arrived only after submission, many felt their intellectual work was already “done”, making revisions appear costly and unnecessary. This also clarifies why our study and ICLR’s field deployment diverged: nearly half of our participants revised their text, but in ICLR’s real-world deployment, uptake was much lower \cite{iclrofficialExperiment}. In our study, participants who joined the study may already tend to be more engaged in the peer review were primed to reflect (see \Cref{sec:limitation}); in practice, reviewers under time pressure often ignored the tool altogether. Thus, reviewers could agree with suggestions yet still dismiss them as unnecessary.

Finally, feedback is not always welcome. Prior research shows that people often seek confirmation of adequacy rather than critique \cite{hattie2007power}. Our data echoed this. Some participants read the AI’s comments as criticism, for example, {\itshape ``I don’t like this AI. It’s like a critic''} (P41). Reviewers felt the tool implied their reviews were not good enough, so they dismissed the suggestions. Compared with the current AI feedback, participants wanted comments that engaged more directly with the paper and offered actionable guidance that prompted reflection on substantive aspects of their evaluation, not surface-level phrasing. This aligns with prior work \cite{ngoon2018improvingfeedback} showing that static, generic suggestions offer little benefit for improving people’s feedback on others’ work, while adaptive and interactive guidance is more effective.

Overall, the lesson is clear: in peer review, improvement is not the same as usefulness. Designing effective AI support requires aligning with reviewers’ understanding of their role, focusing on their intellectual challenges rather than surface expression, and delivering feedback at moments when they are most open to it.

\subsection{Rethinking Review Objectives and AI's Role in Peer Review}
Prior discussions of AI in peer review have centred on automated reviewers, raising concerns about bias, accountability, and substitution of human expertise \cite{lin2021automated},during-review assistants such as ChatGPT, which are framed as reducing workload and supporting decision-making \cite{chen2025envisioning, naddaf2025ai}. Our study contributes a different perspective: when AI enters after a review is written, it does not alter evaluative content but comments on how reviewers express judgements they have already made.

This created a kind of role reversal. In most AI-supported scholarly tasks, such as drafting text, coding, or searching literature, humans initiate prompts and the AI generates responses. Here, the roles were reversed: the AI prompted reviewers to elaborate, clarify, or justify their own comments. Some participants welcomed these nudges as reminders to provide clearer and more actionable feedback to authors. Others, however, resisted, perceiving the tool as prescriptive, redundant, or even patronising. Prior work on technology-mediated nudging shows that system side prompts can influence users’ behaviour and sense of agency \cite{caraban201923, bergram2022digital}. Studies in education report similar tensions, with instructors feeling constrained by AI supported teaching aids \cite{lau2023ban}. Recent HCI research shows that shifting agency from human to system can trigger discomfort and resistance \cite{parsons2025teaching, lau2023ban}. These reactions also appear when systems begin to take initiative rather than simply assist \cite{levy2021assessing}. This echoes broader HCI findings that systems framed as instructors rather than collaborators often provoke resistance, even when the suggestions are valid \cite{lau2023ban, parsons2025teaching}.

Our findings therefore suggest that subtle shifts in control can reshape reviewers' perceptions. Reviewers consistently positioned themselves as authoritative evaluators, not as respondents to algorithmic directives. By repositioning AI as the entity issuing prompts, the system redefined the reviewer–AI relationship: not as assistant or evaluator, but as meta-commentator. This role did not alter review decisions, but it influenced what reviewers considered a “good” review, nudging them toward greater elaboration and clarity. In this sense, the contribution of such tools may lie less in changing outcomes and more in reshaping the rhetorical norms of scholarly communication. Similar concerns appear in prior work on AI-mediated peer review \cite{sun2024reviewflow} and algorithmic influence on scholarly paper review practice \cite{lin2021automated}. For HCI, this raises a provocative question: what happens when AI systems are positioned not to answer human prompts, but to issue them? And how might repeated rhetorical nudges at scale gradually redefine expectations of peer review as a professional practice?

These insights also invite reconsideration of what should count as “better” peer review in the age of AI. At ICLR, longer reviews were treated as evidence of improved quality \cite{iclrofficialExperiment}, and within CHI, authors have similarly reported appreciating lengthy reviews as a sign that their work has been read carefully and taken seriously \cite{jansen2016authorsvalue}. Our findings complicate this view: participants often questioned whether additional elaboration actually benefited evaluation, or whether it simply increased reviewer workload. Some argued that AI prompts led to unnecessary detail, raising expectations for exhaustive reviews without improving substantive critique. This tension highlights that review quality is not singular but differently valued by stakeholders: what authors view as thoroughness may be experienced by reviewers as redundant or burdensome. The paradox of improvement and perceived usefulness points to the fact that AI technologies are prompting communities to redefine the objectives of peer review itself. Should AI encourage maximal elaboration, or help reviewers balance clarity with efficiency? Addressing this question will be central to designing supportive AI that strengthens, rather than distorts, the norms of scholarly evaluation.

\subsection{Design Implication for Reviewer-Oriented AI Support}
The reviewer needs in our study converged on two forms of support. First, reviewers wanted scaffolding that helps them express clearer and more structured critique, especially when articulating methodological concerns or linking comments to evidence. Second, they wanted feedback that adapts to their workflow, offering light guidance during sense-making, drafting and revision without challenging their evaluative decisions. These expectations outline where reviewers welcome assistance and where they draw boundaries. Effective support therefore needs to sit within the act of reviewing rather than being delivered purely post hoc.

Early-stage support can advance these roles by handling light, verifiable forms of information organisation. Tasks such as summarising key contributions, checking references, or surfacing related work \cite{wadden2022scifactopen} help reviewers form an initial understanding without adding cognitive burden. Retrieval-augmented or knowledge-graph approaches \cite{lewis2020retrieval} may further assist navigation in large or rapidly shifting literatures, even if models cannot always provide fully up-to-date contextual insight \cite{farber2025comparing, wagner2022artificial}. Positioned early in the workflow, such low-level assistance supports reviewers’ sense-making while leaving their evaluative judgement untouched.

During drafting, reviewer-oriented AI can more directly scaffold the production of clearer critique. Optional checklists, reminders, or cues tied to methodological reporting or evidence–claim alignment can structure how reviewers articulate their assessments. Participants valued support that strengthened clarity and expression without constraining their judgement, echoing broader findings that scaffolded suggestions improve writing when authorial agency is preserved \cite{reza2025co}. Mid-stage prompts therefore contribute to the first role by shaping the form and coherence of critique while respecting reviewers’ autonomy.

Once a review draft is complete, AI feedback can make refinement more adaptive by focusing on articulation rather than reassessment. Reviewers in our study treated the ICLR tool’s comments as prompts for elaboration rather than substantive revision, consistent with evidence that experts resist suggestions that contradict established judgement \cite{reza2025co}. Light-touch critique at this stage, such as highlighting unclear reasoning or gaps in explanation, can support the second role by enabling efficient, end-stage revision without disrupting evaluative decisions that reviewers consider settled.

\subsection{Sustainable Peer Review with AI}
Peer review relies on volunteer labour, and rising submission volumes place increasing stress on reviewers and area chairs yearly \cite{hanson2024strain, kim2025position}. A sustainable review process is therefore needed to protect reviewer wellbeing \cite{aczel2025present, nobarany2014rethinking, shah2021overview}, support sound judgement, and treat authors fairly \cite{snell2005reviewersperception, naddaf2025ai}. AI is already entering reviewing, formally and informally, which makes it crucial for the community to decide how technology could assist peer review \cite{naddaf2025ai, kim2025position}. 

In this section, we discuss directions for AI-enabled sustainable peer review. First, we broaden the lens beyond individual reviewers to show where AI can support multiple stakeholders across the review stage (\Cref{sec:dis:acrossstakeholders}). Second, we highlight that sustainability requires governance, through clearer policy and venue-led systems that keep AI use visible and bounded (\Cref{sec:dis:regulate}). Third, we call for open, community-level conversations and empirical studies on how the community wants AI to support peer review practices (\Cref{sec:dis:howaireview}).

\subsubsection{\textbf{Positioning AI support across stakeholders}}\label{sec:dis:acrossstakeholders} In this subsection, we shift from the individual reviewer to the wider peer-review ecosystem, considering the broader stakeholders who shape the review process. Participants’ reflections point to opportunities for AI to strengthen the system by \textbf{improving reviewer–paper matching}, \textbf{supporting authors before submission}, and \textbf{mediating communication} among reviewers, authors, and chairs.

Improving reviewer–paper matching is one such structural lever. Prior work demonstrates that models incorporating semantic similarity, expertise estimation, and fairness constraints can reduce mismatches and rebalance reviewer expertise \cite{stelmakh2019peerreview4all, kobren2019paper, aziz2023group, kuznetsov2024can}. In platforms such as OpenReview, matching scores already inform area-chair decision-making, with chairs applying domain knowledge to refine assignments. More accurate matching can reduce the need for later corrective feedback, shaping clearer interaction across subsequent stages of review.

AI can also provide upstream support by assisting authors before submission. Because peer review aims to help authors improve their work \cite{jansen2016authorsvalue}, tools that surface potential weaknesses, such as systems that provide personalised and on-demand feedback \cite{benharrak2024writerAIpersonas} or manuscript-level critique generators like Agentic Reviewer \cite{jiangtech}, may help authors address issues earlier and reduce unnecessary burden on reviewers. Conferences could adapt such tools to reflect venue-specific standards, creating clearer expectations and more equitable access to formative feedback. Understanding how authors interpret and incorporate such pre-submission feedback remains an important area for future study.

AI may further support area chairs and programme chairs by mediating communication across stakeholders. Tools that synthesise reviewer comments into coherent messages, highlight inconsistencies between evaluations, or flag vague or insufficiently substantiated claims could help chairs maintain clarity and fairness across the process. Recent work already demonstrates the feasibility of identifying problematic rebuttal exchanges \cite{kennard2022disapere} or detecting low-effort reviews \cite{guo2023automatic, purkayastha2025lazyreview}. Positioned in this coordinating role, AI can act as a communication mediator and quality safeguard, helping chairs anticipate issues and improving the transparency of interactions among authors and reviewers.

\subsubsection{\textbf{Regulating AI use in peer review through policy and venue-led systems}}\label{sec:dis:regulate}
In this subsection, we focus on governance. AI use in peer review is no longer hypothetical: researchers already employ AI tools during manuscript evaluation, largely within reviewers’ private workflows and beyond venues’ ability to monitor or intervene. Sustainability therefore depends on \textbf{clearer policies and venue-led systems that make AI involvement legible, constrain over-delegation, and support intervention when misuse occurs}. Our participants anticipated even more capable AI assistance (\Cref{sec:resutls:envision}), such as helping interpret or verify complex arguments, to reduce reviewers’ cognitive burden. Yet expanding AI’s role raises concerns for responsibility and research integrity: delegating evaluative work to AI can create “moral distancing” \cite{kobis2025delegation}, and recent attempts by authors to embed hidden prompts that manipulate AI-assisted reviewers \cite{Lin2025HiddenPrompts} show that such vulnerabilities are already being exploited.

Current venue-level policies have struggled to keep pace with emerging patterns of AI use in peer review. In HCI, many venues allow limited forms of AI assistance but rely heavily on reviewer self-regulation, offering few mechanisms to detect or challenge inappropriate AI involvement. This pattern mirrors trends across across major scholarly publishers \cite{mollaki2024aipublishingpolicies}. By contrast, Open-review platforms such as ICLR have recently \cite{ICLR2026LLMresponse2025} introduced procedures that apply LLM-detection tools to submitted reviews and allow authors to flag and report suspected AI-generated content. Reviewers who submit LLM-generated low-quality reviews may even face desk rejection of their own papers. These measures provide a degree of post-review visibility that confidential review systems do not readily support.

Looking ahead, HCI venues may benefit from a consistent framework that clarifies appropriate forms of AI assistance and outlines procedural steps for responding to potential misuse. A complementary direction is to provide AI support through venue-led review platforms rather than relying solely on individual discretion. Integrating a small set of transparent, tightly scoped AI functions within official review systems would allow venues to maintain visibility into AI’s involvement while limiting over-delegation. Such platform-level guardrails, combined with lightweight checks that surface AI-dominant or low-effort reviews, would enable area chairs to intervene when needed without discouraging permitted, light-touch AI use.

\subsubsection{\textbf{Calling for community discussion on values in AI-assisted peer review}} \label{sec:dis:howaireview}
As the scale and complexity of peer review continue to grow in HCI, which mirrors trends in neighbouring CS communities, \textbf{there is an urgent need for open conversations and empirical studies on how our community wants AI to support review practices}. High submission volumes, broad participation, and rapidly evolving research areas have already pushed fields like AI/ML to explore new approaches. Beyond ICLR, other conferences are also now piloting AI in peer review. For example, at the Association for the Advancement of Artificial Intelligence conference (AAAI‑26), LLMs were used to generate supplementary reviews at the initial stage alongside human reviewers, and served as discussion summarisation assistants for the Senior Program Committee, helping highlight reviewer agreement and disagreement\cite{AAAI2025online}. 

The HCI community has long voiced concerns about the sustainability of peer review \cite{bernstein2012reject,nobarany2014rethinking,vitak2024CSCWPanel}, leading to multiple rounds of procedural change \cite{chi2016change,chi2019change, chi2018change}. The move to a revise-and-resubmit model \cite{chi2022change} introduces greater reviewer workload than a single accept-or-reject cycle, in exchange for improved fairness and paper quality. Our findings suggest that AI assistance such as the feedback tool tested at ICLR and other emerging designs, including AI as a review partner \cite{chen2025envisioning}, a review scaffold \cite{sun2024reviewflow}, or a meta-writer \cite{sun2024metawriter}, could help alleviate growing pressures, though current tools still require meaningful improvement and validation in real conference settings. 

Importantly, practices from AI/ML venues cannot be directly transplanted into HCI. Review cultures differ: CS researchers often possess deeper familiarity with AI systems, operate within norms centred on technical innovation, and review via platforms such as OpenReview that support transparency and rapid experimentation with AI. We are not advocating that HCI adopt these models wholesale. Instead, we argue that the community now needs open and structured deliberation - through workshops, panels, and formal studies - to understand members’ attitudes, acceptance, and concerns regarding the benefits and risks of AI in peer review. Research such as \cite{hadan2024greatAIwitchHunt} has begun to investigate reviewers’ perceptions of AI assisted manuscripts. More work is needed to examine how AI assisted review aligns with or challenges HCI values, and to trial such AI responsibly in real review settings before the review system reaches a breaking point.

\subsection{Limitations and Future Work}\label{sec:limitation}
This study offers an early window into how reviewers perceive AI feedback in peer review,  but several limitations remain.

First, Our sample was skewed toward male, early-career researchers. This mirrors broader imbalances in the machine learning community \cite{simonite2018aiwomen,fortunato2018science,mohammad2020gendergap,ding2023voices}, but it limits the range of perspectives represented. Self-selection bias is also possible: reviewers skeptical of AI or less inclined to experiment may have opted out. Future work should recruit more diverse and senior participants to capture how experiences with AI feedback vary across career stage, gender, and disciplinary background.

Second, our analysis relied on self-reported perceptions and intentions rather than direct measures of review quality, such as independent expert ratings of the reviews themselves or text-based assessments of their accuracy, depth, or constructiveness. Ethical constraints prevented us from examining confidential reviews, but self-reports nonetheless reveal how reviewers interpreted and acted on AI support. These perspectives are valuable for understanding adoption, trust, and integration into professional practice, even if they cannot substitute for objective quality metrics.  

Finally, our findings are specific to the ICLR 2025 experiment, where AI feedback was delivered via OpenReview only after initial reviews were submitted. This sequencing preserved reviewers’ independence but also reduced uptake compared to more flexible, opt-in, or real-time designs. Moreover, the ICLR review model includes multi-week discussions, whereas venues such as CHI involve only a single review and rebuttal cycle. Comparative research across communities with different review structures is needed to understand how integration strategies and timing shape the usefulness and reception of AI feedback.

These limitations highlight that our study should be read as a first step in a longer conversation. AI in peer review remains an emerging space with many possible directions, and this work provides an early perspective grounded in real reviewing conditions. We hope it encourages broader discussion on how AI should be integrated into peer reviews and how future systems can better align with reviewers’ needs and professional values. As AI tools become more prevalent in peer review, HCI researchers are well-positioned to investigate how diverse communities define “quality”, how AI can align with reviewers’ professional identities, and what long-term effects such tools may have on scholarly practice.

\section{Conclusion}
This study explored what happens when reviewers receive AI feedback on their reviews in a live deployment during the ICLR 2025 review process. We show how reviewers made sense of AI feedback, how they chose to act on it, and how it intersected with their sense of expertise, ownership, and responsibility.

We identify a paradox of perception: reviewers often acknowledged the ICLR AI feedback tool improved clarity and specificity, yet still judged it as “not useful” because it targeted surface expression, such as polishing, elaboration, and tone, while the intellectual labour reviewers value centres on evaluation and judgement. This mismatch helps explain why reviewers may adopt suggested edits while remaining unconvinced of the tool’s value, and why uptake varies even when measurable text-level improvements occur.

Reviewers’ responses point to clear expectations about where AI support should be positioned peer review. Designing effective AI support therefore requires careful attention to \textbf{timing, framing, and role boundaries}. Light-touch, verifiable support earlier in the review process may be better received than post-hoc critique. Participants envisioned reviewer-level support that reduces cognitive load without substituting judgement, doing the “boring hard work” of sensemaking, such as structured summaries and literature retrieval. They also called for support for chair to oversee review quality and fairness at scale, for example fact-checking claims, flagging low-effort or internally inconsistent reviews. 

Our findings also show that AI in peer review is not only a design problem, but a governance one. Reviewers’ concerns extend beyond individual tools to how AI is introduced, constrained, and made accountable within peer review process. This highlights the need for venue-led efforts to clarify AI’s role, make AI involvement visible, and enable action when misuse occurs.

This work advances a broader agenda for HCI research on human-centred AI in peer review. As AI becomes embedded in review workflows, HCI is well positioned to study how AI involvement shape reviewers’ labour, authority, and accountability in practice. Advancing this requires empirical work in live review settings, as well as community discussion about the governance and values that should steer AI in peer review.

\begin{acks}
We would like to thank all participants for generously contributing their time to this research. We also extend our sincere gratitude to the reviewers for their constructive comments and suggestions which guided the refinement of this paper.
\end{acks}

\section*{Acknowledgements of the Use of AI}
The authors used a large language model in a limited capacity, specifically for language editing and assistance with \LaTeX{} formatting. All research ideas, analyses, and substantive content were conceived, developed, and validated solely by the authors.

\bibliographystyle{ACM-Reference-Format}



\begin{thebibliography}{117}


\ifx \showCODEN    \undefined \def \showCODEN     #1{\unskip}     \fi
\ifx \showISBNx    \undefined \def \showISBNx     #1{\unskip}     \fi
\ifx \showISBNxiii \undefined \def \showISBNxiii  #1{\unskip}     \fi
\ifx \showISSN     \undefined \def \showISSN      #1{\unskip}     \fi
\ifx \showLCCN     \undefined \def \showLCCN      #1{\unskip}     \fi
\ifx \shownote     \undefined \def \shownote      #1{#1}          \fi
\ifx \showarticletitle \undefined \def \showarticletitle #1{#1}   \fi
\ifx \showURL      \undefined \def \showURL       {\relax}        \fi
\providecommand\bibfield[2]{#2}
\providecommand\bibinfo[2]{#2}
\providecommand\natexlab[1]{#1}
\providecommand\showeprint[2][]{arXiv:#2}

\bibitem[{}(2018)]%
        {chi2019change}
\bibfield{author}{\bibinfo{person}{{}}.} \bibinfo{year}{2018}\natexlab{}.
\newblock \bibinfo{title}{Recent Changes to the CHI Submission and Review
  Process}.
\newblock
\urldef\tempurl%
\url{https://chi2019.acm.org/authors/papers/papers-review-process/recent-changes-to-the-chi-submission-and-review-process/}
\showURL{%
\tempurl}
\newblock
\shownote{[Online; accessed 2025-11-02]}.


\bibitem[{}(2021)]%
        {chi2022change}
\bibfield{author}{\bibinfo{person}{{}}.} \bibinfo{year}{2021}\natexlab{}.
\newblock \bibinfo{title}{Moving to a Revise and Resubmit Review Process for
  CHI 2022}.
\newblock
\urldef\tempurl%
\url{https://chi2022.acm.org/2021/05/26/moving-to-a-revise-and-resubmit-review-process-for-chi-2022/}
\showURL{%
\tempurl}
\newblock
\shownote{[Online; accessed 2025-11-23]}.


\bibitem[ICL(2023)]%
        {ICLRStat13:online}
 \bibinfo{year}{2023}\natexlab{}.
\newblock \bibinfo{title}{ICLR Statistics - Paper Copilot}.
\newblock
\urldef\tempurl%
\url{https://papercopilot.com/statistics/iclr-statistics/}
\showURL{%
\tempurl}
\newblock
\shownote{[Online; accessed 2025-12-02]}.


\bibitem[AAAI(2025)]%
        {AAAI2025online}
\bibfield{author}{\bibinfo{person}{AAAI}.} \bibinfo{year}{2025}\natexlab{}.
\newblock \bibinfo{title}{AAAI Launches AI-Powered Peer Review Assessment
  System: Overview of the AI Review System}.
\newblock
\urldef\tempurl%
\url{https://aaai.org/wp-content/uploads/2025/08/FAQ-for-the-AI-Assisted-Peer-Review-Process-Pilot-Program.pdf}
\showURL{%
\tempurl}
\newblock
\shownote{[Online; accessed 2025-11-11]}.


\bibitem[Aczel et~al\mbox{.}(2025)]%
        {aczel2025present}
\bibfield{author}{\bibinfo{person}{Balazs Aczel}, \bibinfo{person}{Ann-Sophie
  Barwich}, \bibinfo{person}{Amanda~B Diekman}, \bibinfo{person}{Ayelet
  Fishbach}, \bibinfo{person}{Robert~L Goldstone}, \bibinfo{person}{Pablo
  Gomez}, \bibinfo{person}{Odd~Erik Gundersen}, \bibinfo{person}{Paul~T von
  Hippel}, \bibinfo{person}{Alex~O Holcombe}, \bibinfo{person}{Stephan
  Lewandowsky}, {et~al\mbox{.}}} \bibinfo{year}{2025}\natexlab{}.
\newblock \showarticletitle{The present and future of peer review: Ideas,
  interventions, and evidence}.
\newblock \bibinfo{journal}{\emph{Proceedings of the National Academy of
  Sciences}} \bibinfo{volume}{122}, \bibinfo{number}{5} (\bibinfo{year}{2025}),
  \bibinfo{pages}{e2401232121}.
\newblock
\href{https://doi.org/10.1073/pnas.2401232121}{doi:\nolinkurl{10.1073/pnas.2401232121}}


\bibitem[Aziz et~al\mbox{.}(2023)]%
        {aziz2023group}
\bibfield{author}{\bibinfo{person}{Haris Aziz}, \bibinfo{person}{Evi Micha},
  {and} \bibinfo{person}{Nisarg Shah}.} \bibinfo{year}{2023}\natexlab{}.
\newblock \showarticletitle{Group Fairness in Peer Review}.
\newblock   \bibinfo{volume}{36} (\bibinfo{year}{2023}),
  \bibinfo{pages}{64885--64895}.
\newblock
\urldef\tempurl%
\url{https://proceedings.neurips.cc/paper_files/paper/2023/file/ccba10dd4e80e7276054222bb95d467c-Paper-Conference.pdf}
\showURL{%
\tempurl}


\bibitem[Barke et~al\mbox{.}(2023)]%
        {barke2022grounded}
\bibfield{author}{\bibinfo{person}{Shraddha Barke}, \bibinfo{person}{Michael~B.
  James}, {and} \bibinfo{person}{Nadia Polikarpova}.}
  \bibinfo{year}{2023}\natexlab{}.
\newblock \showarticletitle{Grounded Copilot: How Programmers Interact with
  Code-Generating Models}.
\newblock \bibinfo{journal}{\emph{Proc. ACM Program. Lang.}}
  \bibinfo{volume}{7}, \bibinfo{number}{OOPSLA1}, Article
  \bibinfo{articleno}{78} (\bibinfo{date}{April} \bibinfo{year}{2023}),
  \bibinfo{numpages}{27}~pages.
\newblock
\href{https://doi.org/10.1145/3586030}{doi:\nolinkurl{10.1145/3586030}}


\bibitem[Ba{\v{s}}i{\'c} et~al\mbox{.}(2023)]%
        {bavsic2023chatgpt}
\bibfield{author}{\bibinfo{person}{{\v{Z}}eljana Ba{\v{s}}i{\'c}},
  \bibinfo{person}{Ana Banovac}, \bibinfo{person}{Ivana Kru{\v{z}}i{\'c}},
  {and} \bibinfo{person}{Ivan Jerkovi{\'c}}.} \bibinfo{year}{2023}\natexlab{}.
\newblock \showarticletitle{ChatGPT-3.5 as writing assistance in students’
  essays}.
\newblock \bibinfo{journal}{\emph{Humanities and social sciences
  communications}} \bibinfo{volume}{10}, \bibinfo{number}{1}
  (\bibinfo{year}{2023}), \bibinfo{pages}{1--5}.
\newblock
\href{https://doi.org/10.1057/s41599-023-02269-7}{doi:\nolinkurl{10.1057/s41599-023-02269-7}}


\bibitem[Benharrak et~al\mbox{.}(2024)]%
        {benharrak2024writerAIpersonas}
\bibfield{author}{\bibinfo{person}{Karim Benharrak}, \bibinfo{person}{Tim
  Zindulka}, \bibinfo{person}{Florian Lehmann}, \bibinfo{person}{Hendrik
  Heuer}, {and} \bibinfo{person}{Daniel Buschek}.}
  \bibinfo{year}{2024}\natexlab{}.
\newblock \showarticletitle{Writer-Defined AI Personas for On-Demand Feedback
  Generation}. In \bibinfo{booktitle}{\emph{Proceedings of the 2024 CHI
  Conference on Human Factors in Computing Systems}} (Honolulu, HI, USA)
  \emph{(\bibinfo{series}{CHI '24})}. \bibinfo{publisher}{Association for
  Computing Machinery}, \bibinfo{address}{New York, NY, USA}, Article
  \bibinfo{articleno}{1049}, \bibinfo{numpages}{18}~pages.
\newblock
\showISBNx{9798400703300}
\href{https://doi.org/10.1145/3613904.3642406}{doi:\nolinkurl{10.1145/3613904.3642406}}


\bibitem[Bergram et~al\mbox{.}(2022)]%
        {bergram2022digital}
\bibfield{author}{\bibinfo{person}{Kristoffer Bergram}, \bibinfo{person}{Marija
  Djokovic}, \bibinfo{person}{Val\'{e}ry Bezen\c{c}on}, {and}
  \bibinfo{person}{Adrian Holzer}.} \bibinfo{year}{2022}\natexlab{}.
\newblock \showarticletitle{The Digital Landscape of Nudging: A Systematic
  Literature Review of Empirical Research on Digital Nudges}. In
  \bibinfo{booktitle}{\emph{Proceedings of the 2022 CHI Conference on Human
  Factors in Computing Systems}} (New Orleans, LA, USA)
  \emph{(\bibinfo{series}{CHI '22})}. \bibinfo{publisher}{Association for
  Computing Machinery}, \bibinfo{address}{New York, NY, USA}, Article
  \bibinfo{articleno}{62}, \bibinfo{numpages}{16}~pages.
\newblock
\showISBNx{9781450391573}
\href{https://doi.org/10.1145/3491102.3517638}{doi:\nolinkurl{10.1145/3491102.3517638}}


\bibitem[Bernstein et~al\mbox{.}(2012)]%
        {bernstein2012reject}
\bibfield{author}{\bibinfo{person}{Michael Bernstein}, \bibinfo{person}{Dan
  Cosley}, \bibinfo{person}{Carl DiSalvo}, \bibinfo{person}{Sanjay Kairam},
  \bibinfo{person}{David Karger}, \bibinfo{person}{Travis Kriplean},
  \bibinfo{person}{Cliff Lampe}, \bibinfo{person}{Wendy Mackay},
  \bibinfo{person}{Loren Terveen}, \bibinfo{person}{Jacob Wobbrock}, {and}
  \bibinfo{person}{Sarita Yardi}.} \bibinfo{year}{2012}\natexlab{}.
\newblock \showarticletitle{Reject me: peer review and SIGCHI}. In
  \bibinfo{booktitle}{\emph{CHI '12 Extended Abstracts on Human Factors in
  Computing Systems}} (Austin, Texas, USA) \emph{(\bibinfo{series}{CHI EA
  '12})}. \bibinfo{publisher}{Association for Computing Machinery},
  \bibinfo{address}{New York, NY, USA}, \bibinfo{pages}{1197–1200}.
\newblock
\showISBNx{9781450310161}
\href{https://doi.org/10.1145/2212776.2212422}{doi:\nolinkurl{10.1145/2212776.2212422}}


\bibitem[Bhat et~al\mbox{.}(2023)]%
        {bhat2023interacting}
\bibfield{author}{\bibinfo{person}{Advait Bhat}, \bibinfo{person}{Saaket
  Agashe}, \bibinfo{person}{Parth Oberoi}, \bibinfo{person}{Niharika Mohile},
  \bibinfo{person}{Ravi Jangir}, {and} \bibinfo{person}{Anirudha Joshi}.}
  \bibinfo{year}{2023}\natexlab{}.
\newblock \showarticletitle{Interacting with Next-Phrase Suggestions: How
  Suggestion Systems Aid and Influence the Cognitive Processes of Writing}. In
  \bibinfo{booktitle}{\emph{Proceedings of the 28th International Conference on
  Intelligent User Interfaces}} (Sydney, NSW, Australia)
  \emph{(\bibinfo{series}{IUI '23})}. \bibinfo{publisher}{Association for
  Computing Machinery}, \bibinfo{address}{New York, NY, USA},
  \bibinfo{pages}{436–452}.
\newblock
\showISBNx{9798400701061}
\href{https://doi.org/10.1145/3581641.3584060}{doi:\nolinkurl{10.1145/3581641.3584060}}


\bibitem[Brachman et~al\mbox{.}(2024)]%
        {LLMframework}
\bibfield{author}{\bibinfo{person}{Michelle Brachman}, \bibinfo{person}{Amina
  El-Ashry}, \bibinfo{person}{Casey Dugan}, {and} \bibinfo{person}{Werner
  Geyer}.} \bibinfo{year}{2024}\natexlab{}.
\newblock \showarticletitle{How Knowledge Workers Use and Want to Use LLMs in
  an Enterprise Context}. In \bibinfo{booktitle}{\emph{Extended Abstracts of
  the CHI Conference on Human Factors in Computing Systems}} (Honolulu, HI,
  USA) \emph{(\bibinfo{series}{CHI EA '24})}. \bibinfo{publisher}{Association
  for Computing Machinery}, \bibinfo{address}{New York, NY, USA}, Article
  \bibinfo{articleno}{189}, \bibinfo{numpages}{8}~pages.
\newblock
\showISBNx{9798400703317}
\href{https://doi.org/10.1145/3613905.3650841}{doi:\nolinkurl{10.1145/3613905.3650841}}


\bibitem[Braun and Clarke(2006)]%
        {braun2024thematic}
\bibfield{author}{\bibinfo{person}{Virginia Braun} {and}
  \bibinfo{person}{Victoria Clarke}.} \bibinfo{year}{2006}\natexlab{}.
\newblock \showarticletitle{Using thematic analysis in psychology}.
\newblock \bibinfo{journal}{\emph{Qualitative Research in Psychology}}
  \bibinfo{volume}{3}, \bibinfo{number}{2} (\bibinfo{year}{2006}),
  \bibinfo{pages}{77--101}.
\newblock
\href{https://doi.org/10.1191/1478088706qp063oa}{doi:\nolinkurl{10.1191/1478088706qp063oa}}


\bibitem[Buschek et~al\mbox{.}(2021)]%
        {buschek2021impact}
\bibfield{author}{\bibinfo{person}{Daniel Buschek}, \bibinfo{person}{Martin
  Z\"{u}rn}, {and} \bibinfo{person}{Malin Eiband}.}
  \bibinfo{year}{2021}\natexlab{}.
\newblock \showarticletitle{The Impact of Multiple Parallel Phrase Suggestions
  on Email Input and Composition Behaviour of Native and Non-Native English
  Writers}. In \bibinfo{booktitle}{\emph{Proceedings of the 2021 CHI Conference
  on Human Factors in Computing Systems}} (Yokohama, Japan)
  \emph{(\bibinfo{series}{CHI '21})}. \bibinfo{publisher}{Association for
  Computing Machinery}, \bibinfo{address}{New York, NY, USA}, Article
  \bibinfo{articleno}{732}, \bibinfo{numpages}{13}~pages.
\newblock
\showISBNx{9781450380966}
\href{https://doi.org/10.1145/3411764.3445372}{doi:\nolinkurl{10.1145/3411764.3445372}}


\bibitem[Caraban et~al\mbox{.}(2019)]%
        {caraban201923}
\bibfield{author}{\bibinfo{person}{Ana Caraban}, \bibinfo{person}{Evangelos
  Karapanos}, \bibinfo{person}{Daniel Gon\c{c}alves}, {and}
  \bibinfo{person}{Pedro Campos}.} \bibinfo{year}{2019}\natexlab{}.
\newblock \showarticletitle{23 Ways to Nudge: A Review of Technology-Mediated
  Nudging in Human-Computer Interaction}. In
  \bibinfo{booktitle}{\emph{Proceedings of the 2019 CHI Conference on Human
  Factors in Computing Systems}} (Glasgow, Scotland Uk)
  \emph{(\bibinfo{series}{CHI '19})}. \bibinfo{publisher}{Association for
  Computing Machinery}, \bibinfo{address}{New York, NY, USA},
  \bibinfo{pages}{1–15}.
\newblock
\showISBNx{9781450359702}
\href{https://doi.org/10.1145/3290605.3300733}{doi:\nolinkurl{10.1145/3290605.3300733}}


\bibitem[Chakrabarty et~al\mbox{.}(2022)]%
        {chakrabarty2022help}
\bibfield{author}{\bibinfo{person}{Tuhin Chakrabarty}, \bibinfo{person}{Vishakh
  Padmakumar}, {and} \bibinfo{person}{He He}.} \bibinfo{year}{2022}\natexlab{}.
\newblock \showarticletitle{Help me write a poem: Instruction Tuning as a
  Vehicle for Collaborative Poetry Writing}. In
  \bibinfo{booktitle}{\emph{Proceedings of the 2022 Conference on Empirical
  Methods in Natural Language Processing}}. \bibinfo{publisher}{Association for
  Computational Linguistics}, \bibinfo{address}{Abu Dhabi, United Arab
  Emirates}, \bibinfo{pages}{6848--6863}.
\newblock
\href{https://doi.org/10.18653/v1/2022.emnlp-main.460}{doi:\nolinkurl{10.18653/v1/2022.emnlp-main.460}}


\bibitem[Checco et~al\mbox{.}(2021)]%
        {checco2021ai}
\bibfield{author}{\bibinfo{person}{Alessandro Checco}, \bibinfo{person}{Lorenzo
  Bracciale}, \bibinfo{person}{Pierpaolo Loreti}, \bibinfo{person}{Stephen
  Pinfield}, {and} \bibinfo{person}{Giuseppe Bianchi}.}
  \bibinfo{year}{2021}\natexlab{}.
\newblock \showarticletitle{AI-assisted peer review}.
\newblock \bibinfo{journal}{\emph{Humanities and social sciences
  communications}} \bibinfo{volume}{8}, \bibinfo{number}{1}
  (\bibinfo{year}{2021}), \bibinfo{pages}{1--11}.
\newblock
\href{https://doi.org/10.1057/s41599-020-00703-8}{doi:\nolinkurl{10.1057/s41599-020-00703-8}}


\bibitem[Chen et~al\mbox{.}(2025)]%
        {chen2025envisioning}
\bibfield{author}{\bibinfo{person}{Shiping Chen}, \bibinfo{person}{Duncan
  Brumby}, {and} \bibinfo{person}{Anna Cox}.} \bibinfo{year}{2025}\natexlab{}.
\newblock \showarticletitle{Envisioning the Future of Peer Review:
  Investigating LLM-Assisted Reviewing Using ChatGPT as a Case Study}. In
  \bibinfo{booktitle}{\emph{Proceedings of the 4th Annual Symposium on
  Human-Computer Interaction for Work}} \emph{(\bibinfo{series}{CHIWORK '25})}.
  \bibinfo{publisher}{Association for Computing Machinery},
  \bibinfo{address}{New York, NY, USA}, Article \bibinfo{articleno}{8},
  \bibinfo{numpages}{18}~pages.
\newblock
\showISBNx{9798400713842}
\href{https://doi.org/10.1145/3729176.3729196}{doi:\nolinkurl{10.1145/3729176.3729196}}


\bibitem[Chew et~al\mbox{.}(2023)]%
        {chew2023llm}
\bibfield{author}{\bibinfo{person}{Robert Chew}, \bibinfo{person}{John
  Bollenbacher}, \bibinfo{person}{Michael Wenger}, \bibinfo{person}{Jessica
  Speer}, {and} \bibinfo{person}{Annice Kim}.} \bibinfo{year}{2023}\natexlab{}.
\newblock \showarticletitle{LLM-assisted content analysis: Using large language
  models to support deductive coding}.
\newblock \bibinfo{journal}{\emph{arXiv preprint arXiv:2306.14924}}
  (\bibinfo{year}{2023}).
\newblock
\urldef\tempurl%
\url{https://arxiv.org/abs/2306.14924}
\showURL{%
\tempurl}


\bibitem[CHI(2017)]%
        {chi2018change}
\bibfield{author}{\bibinfo{person}{CHI}.} \bibinfo{year}{2017}\natexlab{}.
\newblock \bibinfo{title}{CHI 2018 -- Engage with CHI}.
\newblock
\urldef\tempurl%
\url{https://chi2018.acm.org/2017/06/27/changes-to-the-papers-reviewing-and-submission-process/}
\showURL{%
\tempurl}
\shownote{[Online; accessed 2025-11-01]}.


%



\bibitem[Chubb et~al\mbox{.}(2022)]%
        {chubb2022speeding}
\bibfield{author}{\bibinfo{person}{Jennifer Chubb}, \bibinfo{person}{Peter
  Cowling}, {and} \bibinfo{person}{Darren Reed}.}
  \bibinfo{year}{2022}\natexlab{}.
\newblock \showarticletitle{Speeding up to keep up: exploring the use of AI in
  the research process}.
\newblock \bibinfo{journal}{\emph{AI \& society}} \bibinfo{volume}{37},
  \bibinfo{number}{4} (\bibinfo{year}{2022}), \bibinfo{pages}{1439--1457}.
\newblock
\href{https://doi.org/10.1007/s00146-021-01259-0}{doi:\nolinkurl{10.1007/s00146-021-01259-0}}


\bibitem[Clark et~al\mbox{.}(2018)]%
        {clark2018creative}
\bibfield{author}{\bibinfo{person}{Elizabeth Clark},
  \bibinfo{person}{Anne~Spencer Ross}, \bibinfo{person}{Chenhao Tan},
  \bibinfo{person}{Yangfeng Ji}, {and} \bibinfo{person}{Noah~A. Smith}.}
  \bibinfo{year}{2018}\natexlab{}.
\newblock \showarticletitle{Creative Writing with a Machine in the Loop: Case
  Studies on Slogans and Stories}. In \bibinfo{booktitle}{\emph{Proceedings of
  the 23rd International Conference on Intelligent User Interfaces}} (Tokyo,
  Japan) \emph{(\bibinfo{series}{IUI '18})}. \bibinfo{publisher}{Association
  for Computing Machinery}, \bibinfo{address}{New York, NY, USA},
  \bibinfo{pages}{329–340}.
\newblock
\showISBNx{9781450349451}
\href{https://doi.org/10.1145/3172944.3172983}{doi:\nolinkurl{10.1145/3172944.3172983}}


\bibitem[CVPR({[n.\,d.]})]%
        {CVPR202556online}
\bibfield{author}{\bibinfo{person}{CVPR}.} \bibinfo{year}{[n.\,d.]}\natexlab{}.
\newblock \bibinfo{title}{CVPR 2025 Changes}.
\newblock
\urldef\tempurl%
\url{https://cvpr.thecvf.com/Conferences/2025/CVPRChanges}
\showURL{%
\tempurl}
\newblock
\shownote{[Online; accessed 2025-08-30]}.


\bibitem[Dai et~al\mbox{.}(2023)]%
        {dai2023llm}
\bibfield{author}{\bibinfo{person}{Shih-Chieh Dai}, \bibinfo{person}{Aiping
  Xiong}, {and} \bibinfo{person}{Lun-Wei Ku}.} \bibinfo{year}{2023}\natexlab{}.
\newblock \showarticletitle{LLM-in-the-loop: Leveraging Large Language Model
  for Thematic Analysis}. In \bibinfo{booktitle}{\emph{Findings of the
  Association for Computational Linguistics: EMNLP 2023}}.
  \bibinfo{publisher}{Association for Computational Linguistics},
  \bibinfo{pages}{9993--10001}.
\newblock
\href{https://doi.org/10.18653/v1/2023.findings-emnlp.669}{doi:\nolinkurl{10.18653/v1/2023.findings-emnlp.669}}


\bibitem[Dhillon et~al\mbox{.}(2024)]%
        {Dhillon2024ScaffoldingCoWriting}
\bibfield{author}{\bibinfo{person}{Paramveer~S. Dhillon},
  \bibinfo{person}{Somayeh Molaei}, \bibinfo{person}{Jiaqi Li},
  \bibinfo{person}{Maximilian Golub}, \bibinfo{person}{Shaochun Zheng}, {and}
  \bibinfo{person}{Lionel~Peter Robert}.} \bibinfo{year}{2024}\natexlab{}.
\newblock \showarticletitle{Shaping Human-AI Collaboration: Varied Scaffolding
  Levels in Co-writing with Language Models}. In
  \bibinfo{booktitle}{\emph{Proceedings of the 2024 CHI Conference on Human
  Factors in Computing Systems}} (Honolulu, HI, USA)
  \emph{(\bibinfo{series}{CHI '24})}. \bibinfo{publisher}{Association for
  Computing Machinery}, \bibinfo{address}{New York, NY, USA}, Article
  \bibinfo{articleno}{1044}, \bibinfo{numpages}{18}~pages.
\newblock
\showISBNx{9798400703300}
\href{https://doi.org/10.1145/3613904.3642134}{doi:\nolinkurl{10.1145/3613904.3642134}}


\bibitem[Ding et~al\mbox{.}(2025)]%
        {ding2023voices}
\bibfield{author}{\bibinfo{person}{Yiwen Ding}, \bibinfo{person}{Jiarui Liu},
  \bibinfo{person}{Zhiheng Lyu}, \bibinfo{person}{Kun Zhang},
  \bibinfo{person}{Bernhard Sch{\"o}lkopf}, \bibinfo{person}{Zhijing Jin},
  {and} \bibinfo{person}{Rada Mihalcea}.} \bibinfo{year}{2025}\natexlab{}.
\newblock \showarticletitle{Voices of Her: Analyzing Gender Differences in the
  {AI} Publication World}. In \bibinfo{booktitle}{\emph{Proceedings of the
  Fourth Workshop on NLP for Positive Impact (NLP4PI)}}.
  \bibinfo{publisher}{Association for Computational Linguistics},
  \bibinfo{address}{Vienna, Austria}, \bibinfo{pages}{196--214}.
\newblock
\showISBNx{978-1-959429-19-7}
\href{https://doi.org/10.18653/v1/2025.nlp4pi-1.17}{doi:\nolinkurl{10.18653/v1/2025.nlp4pi-1.17}}


\bibitem[Farber(2025)]%
        {farber2025comparing}
\bibfield{author}{\bibinfo{person}{Shai Farber}.}
  \bibinfo{year}{2025}\natexlab{}.
\newblock \showarticletitle{Comparing human and AI expertise in the academic
  peer review process: towards a hybrid approach}.
\newblock \bibinfo{journal}{\emph{Higher Education Research \& Development}}
  \bibinfo{volume}{44}, \bibinfo{number}{4} (\bibinfo{year}{2025}),
  \bibinfo{pages}{871--885}.
\newblock
\href{https://doi.org/10.1080/07294360.2024.2445575}{doi:\nolinkurl{10.1080/07294360.2024.2445575}}
\showeprint{https://doi.org/10.1080/07294360.2024.2445575}


\bibitem[Fortunato et~al\mbox{.}(2018)]%
        {fortunato2018science}
\bibfield{author}{\bibinfo{person}{Santo Fortunato}, \bibinfo{person}{Carl~T.
  Bergstrom}, \bibinfo{person}{Katy Börner}, \bibinfo{person}{James~A. Evans},
  \bibinfo{person}{Dirk Helbing}, \bibinfo{person}{Staša Milojević},
  \bibinfo{person}{Alexander~M. Petersen}, \bibinfo{person}{Filippo Radicchi},
  \bibinfo{person}{Roberta Sinatra}, \bibinfo{person}{Brian Uzzi},
  \bibinfo{person}{Alessandro Vespignani}, \bibinfo{person}{Dashun Wang}, {and}
  \bibinfo{person}{Albert-László Barabási}.}
  \bibinfo{year}{2018}\natexlab{}.
\newblock \showarticletitle{Science of science}.
\newblock \bibinfo{journal}{\emph{Science}} \bibinfo{volume}{359},
  \bibinfo{number}{6379} (\bibinfo{year}{2018}), \bibinfo{pages}{eaao0185}.
\newblock
\href{https://doi.org/10.1126/science.aao0185}{doi:\nolinkurl{10.1126/science.aao0185}}


\bibitem[Gero et~al\mbox{.}(2023)]%
        {gero2023social}
\bibfield{author}{\bibinfo{person}{Katy~Ilonka Gero}, \bibinfo{person}{Tao
  Long}, {and} \bibinfo{person}{Lydia~B Chilton}.}
  \bibinfo{year}{2023}\natexlab{}.
\newblock \showarticletitle{Social Dynamics of AI Support in Creative Writing}.
  In \bibinfo{booktitle}{\emph{Proceedings of the 2023 CHI Conference on Human
  Factors in Computing Systems}} (Hamburg, Germany) \emph{(\bibinfo{series}{CHI
  '23})}. \bibinfo{publisher}{Association for Computing Machinery},
  \bibinfo{address}{New York, NY, USA}, Article \bibinfo{articleno}{245},
  \bibinfo{numpages}{15}~pages.
\newblock
\showISBNx{9781450394215}
\href{https://doi.org/10.1145/3544548.3580782}{doi:\nolinkurl{10.1145/3544548.3580782}}


\bibitem[Goldberg et~al\mbox{.}(2024)]%
        {goldberg2024usefulness}
\bibfield{author}{\bibinfo{person}{Alexander Goldberg}, \bibinfo{person}{Ihsan
  Ullah}, \bibinfo{person}{Thanh Gia~Hieu Khuong},
  \bibinfo{person}{Benedictus~Kent Rachmat}, \bibinfo{person}{Zhen Xu},
  \bibinfo{person}{Isabelle Guyon}, {and} \bibinfo{person}{Nihar~B Shah}.}
  \bibinfo{year}{2024}\natexlab{}.
\newblock \showarticletitle{Usefulness of LLMs as an Author Checklist Assistant
  for Scientific Papers: NeurIPS'24 Experiment}.
\newblock \bibinfo{journal}{\emph{arXiv preprint arXiv:2411.03417}}
  (\bibinfo{year}{2024}).
\newblock
\urldef\tempurl%
\url{https://arxiv.org/abs/2411.03417}
\showURL{%
\tempurl}


\bibitem[Gruda(2025)]%
        {UseAIforReviewNature}
\bibfield{author}{\bibinfo{person}{Dritjon Gruda}.}
  \bibinfo{year}{2025}\natexlab{}.
\newblock \showarticletitle{Three AI-powered steps to faster, smarter peer
  review}.
\newblock \bibinfo{journal}{\emph{Nature}} (\bibinfo{year}{2025}).
\newblock
\urldef\tempurl%
\url{https://www.nature.com/articles/d41586-025-00526-0}
\showURL{%
\tempurl}


\bibitem[Guo et~al\mbox{.}(2023)]%
        {guo2023automatic}
\bibfield{author}{\bibinfo{person}{Yanzhu Guo}, \bibinfo{person}{Guokan Shang},
  \bibinfo{person}{Virgile Rennard}, \bibinfo{person}{Michalis Vazirgiannis},
  {and} \bibinfo{person}{Chlo{\'e} Clavel}.} \bibinfo{year}{2023}\natexlab{}.
\newblock \showarticletitle{Automatic analysis of substantiation in scientific
  peer reviews}. In \bibinfo{booktitle}{\emph{Findings of the Association for
  Computational Linguistics: EMNLP 2023}}. \bibinfo{publisher}{Association for
  Computational Linguistics}, \bibinfo{address}{Singapore},
  \bibinfo{pages}{10198--10216}.
\newblock
\href{https://doi.org/10.18653/v1/2023.findings-emnlp.684}{doi:\nolinkurl{10.18653/v1/2023.findings-emnlp.684}}


\bibitem[Hadan et~al\mbox{.}(2024)]%
        {hadan2024greatAIwitchHunt}
\bibfield{author}{\bibinfo{person}{Hilda Hadan}, \bibinfo{person}{Derrick~M
  Wang}, \bibinfo{person}{Reza~Hadi Mogavi}, \bibinfo{person}{Joseph Tu},
  \bibinfo{person}{Leah Zhang-Kennedy}, {and} \bibinfo{person}{Lennart~E
  Nacke}.} \bibinfo{year}{2024}\natexlab{}.
\newblock \showarticletitle{The great AI witch hunt: Reviewers’ perception
  and (Mis) conception of generative AI in research writing}.
\newblock \bibinfo{journal}{\emph{Computers in Human Behavior: Artificial
  Humans}} \bibinfo{volume}{2}, \bibinfo{number}{2} (\bibinfo{year}{2024}),
  \bibinfo{pages}{100095}.
\newblock
\showISSN{2949-8821}
\href{https://doi.org/10.1016/j.chbah.2024.100095}{doi:\nolinkurl{10.1016/j.chbah.2024.100095}}


\bibitem[Hanson et~al\mbox{.}(2024)]%
        {hanson2024strain}
\bibfield{author}{\bibinfo{person}{Mark~A Hanson},
  \bibinfo{person}{Pablo~G{\'o}mez Barreiro}, \bibinfo{person}{Paolo Crosetto},
  {and} \bibinfo{person}{Dan Brockington}.} \bibinfo{year}{2024}\natexlab{}.
\newblock \showarticletitle{The strain on scientific publishing}.
\newblock \bibinfo{journal}{\emph{Quantitative Science Studies}}
  \bibinfo{volume}{5}, \bibinfo{number}{4} (\bibinfo{date}{11}
  \bibinfo{year}{2024}), \bibinfo{pages}{823--843}.
\newblock
\showISSN{2641-3337}
\href{https://doi.org/10.1162/qss_a_00327}{doi:\nolinkurl{10.1162/qss_a_00327}}


\bibitem[Hattie and Timperley(2007)]%
        {hattie2007power}
\bibfield{author}{\bibinfo{person}{John Hattie} {and} \bibinfo{person}{Helen
  Timperley}.} \bibinfo{year}{2007}\natexlab{}.
\newblock \showarticletitle{The power of feedback}.
\newblock \bibinfo{journal}{\emph{Review of educational research}}
  \bibinfo{volume}{77}, \bibinfo{number}{1} (\bibinfo{year}{2007}),
  \bibinfo{pages}{81--112}.
\newblock
\href{https://doi.org/10.3102/003465430298487}{doi:\nolinkurl{10.3102/003465430298487}}


\bibitem[Hicks et~al\mbox{.}(2016)]%
        {hicks2016framing}
\bibfield{author}{\bibinfo{person}{Catherine~M. Hicks}, \bibinfo{person}{Vineet
  Pandey}, \bibinfo{person}{C.~Ailie Fraser}, {and} \bibinfo{person}{Scott
  Klemmer}.} \bibinfo{year}{2016}\natexlab{}.
\newblock \showarticletitle{Framing Feedback: Choosing Review Environment
  Features that Support High Quality Peer Assessment}. In
  \bibinfo{booktitle}{\emph{Proceedings of the 2016 CHI Conference on Human
  Factors in Computing Systems}} (San Jose, California, USA)
  \emph{(\bibinfo{series}{CHI '16})}. \bibinfo{publisher}{Association for
  Computing Machinery}, \bibinfo{address}{New York, NY, USA},
  \bibinfo{pages}{458–469}.
\newblock
\showISBNx{9781450333627}
\href{https://doi.org/10.1145/2858036.2858195}{doi:\nolinkurl{10.1145/2858036.2858195}}


\bibitem[Hoque et~al\mbox{.}(2024)]%
        {Hoque2024HaLLMark}
\bibfield{author}{\bibinfo{person}{Md~Naimul Hoque}, \bibinfo{person}{Tasfia
  Mashiat}, \bibinfo{person}{Bhavya Ghai}, \bibinfo{person}{Cecilia~D.
  Shelton}, \bibinfo{person}{Fanny Chevalier}, \bibinfo{person}{Kari Kraus},
  {and} \bibinfo{person}{Niklas Elmqvist}.} \bibinfo{year}{2024}\natexlab{}.
\newblock \showarticletitle{The HaLLMark Effect: Supporting Provenance and
  Transparent Use of Large Language Models in Writing with Interactive
  Visualization}. In \bibinfo{booktitle}{\emph{Proceedings of the 2024 CHI
  Conference on Human Factors in Computing Systems}} (Honolulu, HI, USA)
  \emph{(\bibinfo{series}{CHI '24})}. \bibinfo{publisher}{Association for
  Computing Machinery}, \bibinfo{address}{New York, NY, USA}, Article
  \bibinfo{articleno}{1045}, \bibinfo{numpages}{15}~pages.
\newblock
\showISBNx{9798400703300}
\href{https://doi.org/10.1145/3613904.3641895}{doi:\nolinkurl{10.1145/3613904.3641895}}


\bibitem[Horta and Jung(2024)]%
        {horta2024crisis}
\bibfield{author}{\bibinfo{person}{Hugo Horta} {and} \bibinfo{person}{Jisun
  Jung}.} \bibinfo{year}{2024}\natexlab{}.
\newblock \showarticletitle{The crisis of peer review: Part of the evolution of
  science}.
\newblock \bibinfo{journal}{\emph{Higher Education Quarterly}}
  \bibinfo{volume}{78}, \bibinfo{number}{4} (\bibinfo{year}{2024}),
  \bibinfo{pages}{e12511}.
\newblock
\urldef\tempurl%
\url{https://onlinelibrary.wiley.com/doi/10.1111/hequ.12511}
\showURL{%
\tempurl}


\bibitem[Hou and Jung(2021)]%
        {hou2021expert}
\bibfield{author}{\bibinfo{person}{Yoyo Tsung-Yu Hou} {and}
  \bibinfo{person}{Malte~F. Jung}.} \bibinfo{year}{2021}\natexlab{}.
\newblock \showarticletitle{Who is the Expert? Reconciling Algorithm Aversion
  and Algorithm Appreciation in AI-Supported Decision Making}.
\newblock \bibinfo{journal}{\emph{Proc. ACM Hum.-Comput. Interact.}}
  \bibinfo{volume}{5}, \bibinfo{number}{CSCW2}, Article
  \bibinfo{articleno}{477} (\bibinfo{date}{Oct.} \bibinfo{year}{2021}),
  \bibinfo{numpages}{25}~pages.
\newblock
\href{https://doi.org/10.1145/3479864}{doi:\nolinkurl{10.1145/3479864}}


\bibitem[Hwang et~al\mbox{.}(2025a)]%
        {hwang2025human}
\bibfield{author}{\bibinfo{person}{Angel Hsing-Chi Hwang},
  \bibinfo{person}{Michael~S. Bernstein}, \bibinfo{person}{S.~Shyam Sundar},
  \bibinfo{person}{Renwen Zhang}, \bibinfo{person}{Manoel Horta~Ribeiro},
  \bibinfo{person}{Yingdan Lu}, \bibinfo{person}{Serina Chang},
  \bibinfo{person}{Tongshuang Wu}, \bibinfo{person}{Aimei Yang},
  \bibinfo{person}{Dmitri Williams}, \bibinfo{person}{Joon~Sung Park},
  \bibinfo{person}{Katherine Ognyanova}, \bibinfo{person}{Ziang Xiao},
  \bibinfo{person}{Aaron Shaw}, {and} \bibinfo{person}{David~A. Shamma}.}
  \bibinfo{year}{2025}\natexlab{a}.
\newblock \showarticletitle{Human Subjects Research in the Age of Generative
  AI: Opportunities and Challenges of Applying LLM-Simulated Data to HCI
  Studies}. In \bibinfo{booktitle}{\emph{Proceedings of the Extended Abstracts
  of the CHI Conference on Human Factors in Computing Systems}}
  \emph{(\bibinfo{series}{CHI EA '25})}. \bibinfo{publisher}{Association for
  Computing Machinery}, \bibinfo{address}{New York, NY, USA}, Article
  \bibinfo{articleno}{761}, \bibinfo{numpages}{7}~pages.
\newblock
\showISBNx{9798400713958}
\href{https://doi.org/10.1145/3706599.3716299}{doi:\nolinkurl{10.1145/3706599.3716299}}


\bibitem[Hwang et~al\mbox{.}(2025b)]%
        {Hwang2025Authenticity}
\bibfield{author}{\bibinfo{person}{Angel Hsing-Chi Hwang},
  \bibinfo{person}{Q.~Vera Liao}, \bibinfo{person}{Su~Lin Blodgett},
  \bibinfo{person}{Alexandra Olteanu}, {and} \bibinfo{person}{Adam Trischler}.}
  \bibinfo{year}{2025}\natexlab{b}.
\newblock \showarticletitle{'It was 80\% me, 20\% AI': Seeking Authenticity in
  Co-Writing with Large Language Models}.
\newblock \bibinfo{journal}{\emph{Proc. ACM Hum.-Comput. Interact.}}
  \bibinfo{volume}{9}, \bibinfo{number}{2}, Article
  \bibinfo{articleno}{CSCW122} (\bibinfo{date}{May} \bibinfo{year}{2025}),
  \bibinfo{numpages}{41}~pages.
\newblock
\href{https://doi.org/10.1145/3711020}{doi:\nolinkurl{10.1145/3711020}}


\bibitem[{ICLR 2026 Program Chairs}(2025)]%
        {ICLR2026LLMresponse2025}
\bibfield{author}{\bibinfo{person}{{ICLR 2026 Program Chairs}}.}
  \bibinfo{year}{2025}\natexlab{}.
\newblock \bibinfo{title}{{ICLR 2026 Response to LLM-Generated Papers and
  Reviews}}.
\newblock
\urldef\tempurl%
\url{https://blog.iclr.cc/2025/11/19/iclr-2026-response-to-llm-generated-papers-and-reviews/}
\showURL{%
\tempurl}
\newblock
\shownote{[Online; accessed 2025-11-19]}.


\bibitem[Ippolito et~al\mbox{.}(2022)]%
        {ippolito2022creative}
\bibfield{author}{\bibinfo{person}{Daphne Ippolito}, \bibinfo{person}{Ann
  Yuan}, \bibinfo{person}{Andy Coenen}, {and} \bibinfo{person}{Sehmon Burnam}.}
  \bibinfo{year}{2022}\natexlab{}.
\newblock \showarticletitle{Creative Writing with an AI-Powered Writing
  Assistant: Perspectives from Professional Writers}.
\newblock \bibinfo{journal}{\emph{arXiv preprint arXiv:2211.05030}}
  (\bibinfo{year}{2022}).
\newblock
\urldef\tempurl%
\url{https://arxiv.org/abs/2211.05030}
\showURL{%
\tempurl}


\bibitem[Jakesch et~al\mbox{.}(2023)]%
        {Jakesch2023OpinionatedLMs}
\bibfield{author}{\bibinfo{person}{Maurice Jakesch}, \bibinfo{person}{Advait
  Bhat}, \bibinfo{person}{Daniel Buschek}, \bibinfo{person}{Lior Zalmanson},
  {and} \bibinfo{person}{Mor Naaman}.} \bibinfo{year}{2023}\natexlab{}.
\newblock \showarticletitle{Co-Writing with Opinionated Language Models Affects
  Users’ Views}. In \bibinfo{booktitle}{\emph{Proceedings of the 2023 CHI
  Conference on Human Factors in Computing Systems}} (Hamburg, Germany)
  \emph{(\bibinfo{series}{CHI '23})}. \bibinfo{publisher}{Association for
  Computing Machinery}, \bibinfo{address}{New York, NY, USA}, Article
  \bibinfo{articleno}{111}, \bibinfo{numpages}{15}~pages.
\newblock
\showISBNx{9781450394215}
\href{https://doi.org/10.1145/3544548.3581196}{doi:\nolinkurl{10.1145/3544548.3581196}}


\bibitem[Jansen et~al\mbox{.}(2016)]%
        {jansen2016authorsvalue}
\bibfield{author}{\bibinfo{person}{Yvonne Jansen}, \bibinfo{person}{Kasper
  Hornb\ae{}k}, {and} \bibinfo{person}{Pierre Dragicevic}.}
  \bibinfo{year}{2016}\natexlab{}.
\newblock \showarticletitle{What Did Authors Value in the CHI'16 Reviews They
  Received?}. In \bibinfo{booktitle}{\emph{Proceedings of the 2016 CHI
  Conference Extended Abstracts on Human Factors in Computing Systems}} (San
  Jose, California, USA) \emph{(\bibinfo{series}{CHI EA '16})}.
  \bibinfo{publisher}{Association for Computing Machinery},
  \bibinfo{address}{New York, NY, USA}, \bibinfo{pages}{596–608}.
\newblock
\showISBNx{9781450340823}
\href{https://doi.org/10.1145/2851581.2892576}{doi:\nolinkurl{10.1145/2851581.2892576}}


\bibitem[Jelson and Lee(2024)]%
        {jelson2024empirical}
\bibfield{author}{\bibinfo{person}{Andrew Jelson} {and}
  \bibinfo{person}{Sang~Won Lee}.} \bibinfo{year}{2024}\natexlab{}.
\newblock \showarticletitle{An empirical study to understand how students use
  ChatGPT for writing essays and how it affects their ownership}. In
  \bibinfo{booktitle}{\emph{Proceedings of the Third Workshop on Intelligent
  and Interactive Writing Assistants}} (Honolulu, HI, USA)
  \emph{(\bibinfo{series}{In2Writing '24})}. \bibinfo{publisher}{Association
  for Computing Machinery}, \bibinfo{address}{New York, NY, USA},
  \bibinfo{pages}{26–30}.
\newblock
\showISBNx{9798400710315}
\href{https://doi.org/10.1145/3690712.3690720}{doi:\nolinkurl{10.1145/3690712.3690720}}


\bibitem[Jiang and Ng(2025)]%
        {jiangtech}
\bibfield{author}{\bibinfo{person}{Yixing Jiang} {and} \bibinfo{person}{Andrew
  Ng}.} \bibinfo{year}{2025}\natexlab{}.
\newblock \showarticletitle{Tech Overview - Stanford Agentic Reviewer}.
\newblock  (\bibinfo{year}{2025}).
\newblock
\urldef\tempurl%
\url{https://paperreview.ai/tech-overview}
\showURL{%
\tempurl}


\bibitem[Kapania et~al\mbox{.}(2025)]%
        {kapania2025m}
\bibfield{author}{\bibinfo{person}{Shivani Kapania}, \bibinfo{person}{Ruiyi
  Wang}, \bibinfo{person}{Toby Jia-Jun Li}, \bibinfo{person}{Tianshi Li}, {and}
  \bibinfo{person}{Hong Shen}.} \bibinfo{year}{2025}\natexlab{}.
\newblock \showarticletitle{'I'm Categorizing LLM as a Productivity Tool':
  Examining Ethics of LLM Use in HCI Research Practices}.
\newblock \bibinfo{journal}{\emph{Proc. ACM Hum.-Comput. Interact.}}
  \bibinfo{volume}{9}, \bibinfo{number}{2}, Article
  \bibinfo{articleno}{CSCW102} (\bibinfo{date}{May} \bibinfo{year}{2025}),
  \bibinfo{numpages}{26}~pages.
\newblock
\href{https://doi.org/10.1145/3711000}{doi:\nolinkurl{10.1145/3711000}}


\bibitem[Kennard et~al\mbox{.}(2022)]%
        {kennard2022disapere}
\bibfield{author}{\bibinfo{person}{Neha~Nayak Kennard}, \bibinfo{person}{Tim
  O’Gorman}, \bibinfo{person}{Rajarshi Das}, \bibinfo{person}{Akshay Sharma},
  \bibinfo{person}{Chhandak Bagchi}, \bibinfo{person}{Matthew Clinton},
  \bibinfo{person}{Pranay~Kumar Yelugam}, \bibinfo{person}{Hamed Zamani}, {and}
  \bibinfo{person}{Andrew McCallum}.} \bibinfo{year}{2022}\natexlab{}.
\newblock \showarticletitle{DISAPERE: A dataset for discourse structure in peer
  review discussions}. In \bibinfo{booktitle}{\emph{Proceedings of the 2022
  Conference of the North American Chapter of the Association for Computational
  Linguistics: Human Language Technologies}}. \bibinfo{publisher}{Association
  for Computational Linguistics}, \bibinfo{address}{Seattle, United States},
  \bibinfo{pages}{1234--1249}.
\newblock
\href{https://doi.org/10.18653/v1/2022.naacl-main.89}{doi:\nolinkurl{10.18653/v1/2022.naacl-main.89}}


\bibitem[Khalifa and Albadawy(2024)]%
        {khalifa2024using}
\bibfield{author}{\bibinfo{person}{Mohamed Khalifa} {and} \bibinfo{person}{Mona
  Albadawy}.} \bibinfo{year}{2024}\natexlab{}.
\newblock \showarticletitle{Using artificial intelligence in academic writing
  and research: An essential productivity tool}.
\newblock \bibinfo{journal}{\emph{Computer Methods and Programs in Biomedicine
  Update}}  \bibinfo{volume}{5} (\bibinfo{year}{2024}),
  \bibinfo{pages}{100145}.
\newblock
\showISSN{2666-9900}
\href{https://doi.org/10.1016/j.cmpbup.2024.100145}{doi:\nolinkurl{10.1016/j.cmpbup.2024.100145}}


\bibitem[Kim et~al\mbox{.}(2025)]%
        {kim2025position}
\bibfield{author}{\bibinfo{person}{Jaeho Kim}, \bibinfo{person}{Yunseok Lee},
  {and} \bibinfo{person}{Seulki Lee}.} \bibinfo{year}{2025}\natexlab{}.
\newblock \showarticletitle{Position: The {AI} Conference Peer Review Crisis
  Demands Author Feedback and Reviewer Rewards}. In
  \bibinfo{booktitle}{\emph{Forty-second International Conference on Machine
  Learning Position Paper Track}}.
\newblock
\urldef\tempurl%
\url{https://openreview.net/forum?id=l8QemUZaIA}
\showURL{%
\tempurl}


\bibitem[K{\"o}bis et~al\mbox{.}(2025)]%
        {kobis2025delegation}
\bibfield{author}{\bibinfo{person}{Nils K{\"o}bis}, \bibinfo{person}{Zoe
  Rahwan}, \bibinfo{person}{Raluca Rilla}, \bibinfo{person}{Bramantyo~Ibrahim
  Supriyatno}, \bibinfo{person}{Clara Bersch}, \bibinfo{person}{Tamer Ajaj},
  \bibinfo{person}{Jean-Fran{\c{c}}ois Bonnefon}, {and} \bibinfo{person}{Iyad
  Rahwan}.} \bibinfo{year}{2025}\natexlab{}.
\newblock \showarticletitle{Delegation to artificial intelligence can increase
  dishonest behaviour}.
\newblock \bibinfo{journal}{\emph{Nature}}  \bibinfo{volume}{646}
  (\bibinfo{year}{2025}), \bibinfo{pages}{126--134}.
\newblock
\href{https://doi.org/10.1038/s41586-025-09505-x}{doi:\nolinkurl{10.1038/s41586-025-09505-x}}


\bibitem[Kobren et~al\mbox{.}(2019)]%
        {kobren2019paper}
\bibfield{author}{\bibinfo{person}{Ari Kobren}, \bibinfo{person}{Barna Saha},
  {and} \bibinfo{person}{Andrew McCallum}.} \bibinfo{year}{2019}\natexlab{}.
\newblock \showarticletitle{Paper Matching with Local Fairness Constraints}. In
  \bibinfo{booktitle}{\emph{Proceedings of the 25th ACM SIGKDD International
  Conference on Knowledge Discovery \& Data Mining}} (Anchorage, AK, USA)
  \emph{(\bibinfo{series}{KDD '19})}. \bibinfo{publisher}{Association for
  Computing Machinery}, \bibinfo{address}{New York, NY, USA},
  \bibinfo{pages}{1247–1257}.
\newblock
\showISBNx{9781450362016}
\href{https://doi.org/10.1145/3292500.3330899}{doi:\nolinkurl{10.1145/3292500.3330899}}


\bibitem[Kuznetsov et~al\mbox{.}(2024)]%
        {kuznetsov2024can}
\bibfield{author}{\bibinfo{person}{Ilia Kuznetsov},
  \bibinfo{person}{Osama~Mohammed Afzal}, \bibinfo{person}{Koen Dercksen},
  \bibinfo{person}{Nils Dycke}, \bibinfo{person}{Alexander Goldberg},
  \bibinfo{person}{Tom Hope}, \bibinfo{person}{Dirk Hovy},
  \bibinfo{person}{Jonathan~K Kummerfeld}, \bibinfo{person}{Anne Lauscher},
  \bibinfo{person}{Kevin Leyton-Brown}, {et~al\mbox{.}}}
  \bibinfo{year}{2024}\natexlab{}.
\newblock \showarticletitle{What can natural language processing do for peer
  review?}
\newblock \bibinfo{journal}{\emph{arXiv preprint arXiv:2405.06563}}
  (\bibinfo{year}{2024}).
\newblock
\urldef\tempurl%
\url{https://arxiv.org/abs/2405.06563}
\showURL{%
\tempurl}


\bibitem[Lau and Guo(2023)]%
        {lau2023ban}
\bibfield{author}{\bibinfo{person}{Sam Lau} {and} \bibinfo{person}{Philip
  Guo}.} \bibinfo{year}{2023}\natexlab{}.
\newblock \showarticletitle{From "Ban It Till We Understand It" to "Resistance
  is Futile": How University Programming Instructors Plan to Adapt as More
  Students Use AI Code Generation and Explanation Tools such as ChatGPT and
  GitHub Copilot}. In \bibinfo{booktitle}{\emph{Proceedings of the 2023 ACM
  Conference on International Computing Education Research - Volume 1}}
  (Chicago, IL, USA) \emph{(\bibinfo{series}{ICER '23})}.
  \bibinfo{publisher}{Association for Computing Machinery},
  \bibinfo{address}{New York, NY, USA}, \bibinfo{pages}{106–121}.
\newblock
\showISBNx{9781450399760}
\href{https://doi.org/10.1145/3568813.3600138}{doi:\nolinkurl{10.1145/3568813.3600138}}


\bibitem[Lee et~al\mbox{.}(2024)]%
        {lee2024design}
\bibfield{author}{\bibinfo{person}{Mina Lee}, \bibinfo{person}{Katy~Ilonka
  Gero}, \bibinfo{person}{John Joon~Young Chung},
  \bibinfo{person}{Simon~Buckingham Shum}, \bibinfo{person}{Vipul Raheja},
  \bibinfo{person}{Hua Shen}, \bibinfo{person}{Subhashini Venugopalan},
  \bibinfo{person}{Thiemo Wambsganss}, \bibinfo{person}{David Zhou},
  \bibinfo{person}{Emad~A. Alghamdi}, \bibinfo{person}{Tal August},
  \bibinfo{person}{Avinash Bhat}, \bibinfo{person}{Madiha~Zahrah Choksi},
  \bibinfo{person}{Senjuti Dutta}, \bibinfo{person}{Jin~L.C. Guo},
  \bibinfo{person}{Md~Naimul Hoque}, \bibinfo{person}{Yewon Kim},
  \bibinfo{person}{Simon Knight}, \bibinfo{person}{Seyed~Parsa Neshaei},
  \bibinfo{person}{Antonette Shibani}, \bibinfo{person}{Disha Shrivastava},
  \bibinfo{person}{Lila Shroff}, \bibinfo{person}{Agnia Sergeyuk},
  \bibinfo{person}{Jessi Stark}, \bibinfo{person}{Sarah Sterman},
  \bibinfo{person}{Sitong Wang}, \bibinfo{person}{Antoine Bosselut},
  \bibinfo{person}{Daniel Buschek}, \bibinfo{person}{Joseph~Chee Chang},
  \bibinfo{person}{Sherol Chen}, \bibinfo{person}{Max Kreminski},
  \bibinfo{person}{Joonsuk Park}, \bibinfo{person}{Roy Pea},
  \bibinfo{person}{Eugenia Ha~Rim Rho}, \bibinfo{person}{Zejiang Shen}, {and}
  \bibinfo{person}{Pao Siangliulue}.} \bibinfo{year}{2024}\natexlab{}.
\newblock \showarticletitle{A Design Space for Intelligent and Interactive
  Writing Assistants}. In \bibinfo{booktitle}{\emph{Proceedings of the 2024 CHI
  Conference on Human Factors in Computing Systems}} (Honolulu, HI, USA)
  \emph{(\bibinfo{series}{CHI '24})}. \bibinfo{publisher}{Association for
  Computing Machinery}, \bibinfo{address}{New York, NY, USA}, Article
  \bibinfo{articleno}{1054}, \bibinfo{numpages}{35}~pages.
\newblock
\showISBNx{9798400703300}
\href{https://doi.org/10.1145/3613904.3642697}{doi:\nolinkurl{10.1145/3613904.3642697}}


\bibitem[Lee et~al\mbox{.}(2022)]%
        {lee2022coauthor}
\bibfield{author}{\bibinfo{person}{Mina Lee}, \bibinfo{person}{Percy Liang},
  {and} \bibinfo{person}{Qian Yang}.} \bibinfo{year}{2022}\natexlab{}.
\newblock \showarticletitle{CoAuthor: Designing a Human-AI Collaborative
  Writing Dataset for Exploring Language Model Capabilities}. In
  \bibinfo{booktitle}{\emph{Proceedings of the 2022 CHI Conference on Human
  Factors in Computing Systems}} (New Orleans, LA, USA)
  \emph{(\bibinfo{series}{CHI '22})}. \bibinfo{publisher}{Association for
  Computing Machinery}, \bibinfo{address}{New York, NY, USA}, Article
  \bibinfo{articleno}{388}, \bibinfo{numpages}{19}~pages.
\newblock
\showISBNx{9781450391573}
\href{https://doi.org/10.1145/3491102.3502030}{doi:\nolinkurl{10.1145/3491102.3502030}}


\bibitem[Lee et~al\mbox{.}(2023)]%
        {lee2022AIHumanEvaluation}
\bibfield{author}{\bibinfo{person}{Mina Lee}, \bibinfo{person}{Megha
  Srivastava}, \bibinfo{person}{Amelia Hardy}, \bibinfo{person}{John
  Thickstun}, \bibinfo{person}{Esin Durmus}, \bibinfo{person}{Ashwin
  Paranjape}, \bibinfo{person}{Ines Gerard-Ursin}, \bibinfo{person}{Xiang~Lisa
  Li}, \bibinfo{person}{Faisal Ladhak}, \bibinfo{person}{Frieda Rong},
  {et~al\mbox{.}}} \bibinfo{year}{2023}\natexlab{}.
\newblock \showarticletitle{Evaluating human-language model interaction}.
\newblock \bibinfo{journal}{\emph{Transactions on Machine Learning Research}}
  (\bibinfo{year}{2023}).
\newblock
\showISSN{2835-8856}
\urldef\tempurl%
\url{https://openreview.net/forum?id=hjDYJUn9l1}
\showURL{%
\tempurl}


\bibitem[Lee et~al\mbox{.}(2020)]%
        {uncoveringlee}
\bibfield{author}{\bibinfo{person}{Wanhae Lee}, \bibinfo{person}{Minji Kwon},
  \bibinfo{person}{Yewon Hyun}, \bibinfo{person}{Jihyun Lee},
  \bibinfo{person}{Joonho Gwon}, {and} \bibinfo{person}{Hyunggu Jung}.}
  \bibinfo{year}{2020}\natexlab{}.
\newblock \showarticletitle{Uncovering CHI Reviewers Needs and Barriers}. In
  \bibinfo{booktitle}{\emph{Proceedings of the 2020 Symposium on Emerging
  Research from Asia and on Asian Contexts and Cultures}} (Honolulu, HI, USA)
  \emph{(\bibinfo{series}{AsianCHI '20})}. \bibinfo{publisher}{Association for
  Computing Machinery}, \bibinfo{address}{New York, NY, USA},
  \bibinfo{pages}{57–60}.
\newblock
\showISBNx{9781450387682}
\href{https://doi.org/10.1145/3391203.3391218}{doi:\nolinkurl{10.1145/3391203.3391218}}


\bibitem[Levy et~al\mbox{.}(2021)]%
        {levy2021assessing}
\bibfield{author}{\bibinfo{person}{Ariel Levy}, \bibinfo{person}{Monica
  Agrawal}, \bibinfo{person}{Arvind Satyanarayan}, {and} \bibinfo{person}{David
  Sontag}.} \bibinfo{year}{2021}\natexlab{}.
\newblock \showarticletitle{Assessing the Impact of Automated Suggestions on
  Decision Making: Domain Experts Mediate Model Errors but Take Less
  Initiative}. In \bibinfo{booktitle}{\emph{Proceedings of the 2021 CHI
  Conference on Human Factors in Computing Systems}} (Yokohama, Japan)
  \emph{(\bibinfo{series}{CHI '21})}. \bibinfo{publisher}{Association for
  Computing Machinery}, \bibinfo{address}{New York, NY, USA}, Article
  \bibinfo{articleno}{72}, \bibinfo{numpages}{13}~pages.
\newblock
\showISBNx{9781450380966}
\href{https://doi.org/10.1145/3411764.3445522}{doi:\nolinkurl{10.1145/3411764.3445522}}


\bibitem[Lewis et~al\mbox{.}(2020)]%
        {lewis2020retrieval}
\bibfield{author}{\bibinfo{person}{Patrick Lewis}, \bibinfo{person}{Ethan
  Perez}, \bibinfo{person}{Aleksandra Piktus}, \bibinfo{person}{Fabio Petroni},
  \bibinfo{person}{Vladimir Karpukhin}, \bibinfo{person}{Naman Goyal},
  \bibinfo{person}{Heinrich K{\"u}ttler}, \bibinfo{person}{Mike Lewis},
  \bibinfo{person}{Wen-tau Yih}, \bibinfo{person}{Tim Rockt{\"a}schel},
  {et~al\mbox{.}}} \bibinfo{year}{2020}\natexlab{}.
\newblock \showarticletitle{Retrieval-augmented generation for
  knowledge-intensive nlp tasks}.
\newblock \bibinfo{journal}{\emph{Advances in neural information processing
  systems}}  \bibinfo{volume}{33} (\bibinfo{year}{2020}),
  \bibinfo{pages}{9459--9474}.
\newblock
\urldef\tempurl%
\url{https://proceedings.neurips.cc/paper_files/paper/2020/file/6b493230205f780e1bc26945df7481e5-Paper.pdf}
\showURL{%
\tempurl}


\bibitem[Li et~al\mbox{.}(2024)]%
        {Li2024ValueBenefitsConcerns}
\bibfield{author}{\bibinfo{person}{Zhuoyan Li}, \bibinfo{person}{Chen Liang},
  \bibinfo{person}{Jing Peng}, {and} \bibinfo{person}{Ming Yin}.}
  \bibinfo{year}{2024}\natexlab{}.
\newblock \showarticletitle{The Value, Benefits, and Concerns of Generative
  AI-Powered Assistance in Writing}. In \bibinfo{booktitle}{\emph{Proceedings
  of the 2024 CHI Conference on Human Factors in Computing Systems}} (Honolulu,
  HI, USA) \emph{(\bibinfo{series}{CHI '24})}. \bibinfo{publisher}{Association
  for Computing Machinery}, \bibinfo{address}{New York, NY, USA}, Article
  \bibinfo{articleno}{1048}, \bibinfo{numpages}{25}~pages.
\newblock
\showISBNx{9798400703300}
\href{https://doi.org/10.1145/3613904.3642625}{doi:\nolinkurl{10.1145/3613904.3642625}}


\bibitem[Liang et~al\mbox{.}(2024a)]%
        {liang2024LargeScale}
\bibfield{author}{\bibinfo{person}{Weixin Liang}, \bibinfo{person}{Yuhui
  Zhang}, \bibinfo{person}{Hancheng Cao}, \bibinfo{person}{Binglu Wang},
  \bibinfo{person}{Daisy~Yi Ding}, \bibinfo{person}{Xinyu Yang},
  \bibinfo{person}{Kailas Vodrahalli}, \bibinfo{person}{Siyu He},
  \bibinfo{person}{Daniel~Scott Smith}, \bibinfo{person}{Yian Yin},
  {et~al\mbox{.}}} \bibinfo{year}{2024}\natexlab{a}.
\newblock \showarticletitle{Can large language models provide useful feedback
  on research papers? A large-scale empirical analysis}.
\newblock \bibinfo{journal}{\emph{NEJM AI}} \bibinfo{volume}{1},
  \bibinfo{number}{8} (\bibinfo{year}{2024}), \bibinfo{pages}{AIoa2400196}.
\newblock
\href{https://doi.org/10.1056/AIoa2400196}{doi:\nolinkurl{10.1056/AIoa2400196}}


\bibitem[Liang et~al\mbox{.}(2024b)]%
        {liang2024mapping}
\bibfield{author}{\bibinfo{person}{Weixin Liang}, \bibinfo{person}{Yaohui
  Zhang}, \bibinfo{person}{Zhengxuan Wu}, \bibinfo{person}{Haley Lepp},
  \bibinfo{person}{Wenlong Ji}, \bibinfo{person}{Xuandong Zhao},
  \bibinfo{person}{Hancheng Cao}, \bibinfo{person}{Sheng Liu},
  \bibinfo{person}{Siyu He}, \bibinfo{person}{Zhi Huang}, \bibinfo{person}{Diyi
  Yang}, \bibinfo{person}{Christopher Potts}, \bibinfo{person}{Christopher~D
  Manning}, {and} \bibinfo{person}{James~Y. Zou}.}
  \bibinfo{year}{2024}\natexlab{b}.
\newblock \showarticletitle{Mapping the Increasing Use of {LLM}s in Scientific
  Papers}. In \bibinfo{booktitle}{\emph{First Conference on Language
  Modeling}}.
\newblock
\urldef\tempurl%
\url{https://openreview.net/forum?id=YX7QnhxESU}
\showURL{%
\tempurl}


\bibitem[Lin et~al\mbox{.}(2023)]%
        {lin2021automated}
\bibfield{author}{\bibinfo{person}{Jialiang Lin}, \bibinfo{person}{Jiaxin
  Song}, \bibinfo{person}{Zhangping Zhou}, \bibinfo{person}{Yidong Chen}, {and}
  \bibinfo{person}{Xiaodong Shi}.} \bibinfo{year}{2023}\natexlab{}.
\newblock \showarticletitle{Automated scholarly paper review: Concepts,
  technologies, and challenges}.
\newblock \bibinfo{journal}{\emph{Information Fusion}}  \bibinfo{volume}{98}
  (\bibinfo{year}{2023}), \bibinfo{pages}{101830}.
\newblock
\showISSN{1566-2535}
\href{https://doi.org/10.1016/j.inffus.2023.101830}{doi:\nolinkurl{10.1016/j.inffus.2023.101830}}


\bibitem[Lin and Yu(2025)]%
        {lin2025elucidating}
\bibfield{author}{\bibinfo{person}{Yupeng Lin} {and} \bibinfo{person}{Zhonggen
  Yu}.} \bibinfo{year}{2025}\natexlab{}.
\newblock \showarticletitle{Elucidating university students’ intentions to
  seek automated writing feedback from Grammarly: toward perceptual and
  systemic predictors}.
\newblock \bibinfo{journal}{\emph{Humanities and Social Sciences
  Communications}} \bibinfo{volume}{12}, \bibinfo{number}{7}
  (\bibinfo{year}{2025}).
\newblock
\href{https://doi.org/10.1057/s41599-024-03861-1}{doi:\nolinkurl{10.1057/s41599-024-03861-1}}


\bibitem[Lin(2025)]%
        {Lin2025HiddenPrompts}
\bibfield{author}{\bibinfo{person}{Zhicheng Lin}.}
  \bibinfo{year}{2025}\natexlab{}.
\newblock \showarticletitle{Hidden Prompts in Manuscripts Exploit AI-Assisted
  Peer Review}.
\newblock \bibinfo{journal}{\emph{arXiv preprint arXiv:2507.06185}}
  (\bibinfo{year}{2025}).
\newblock
\urldef\tempurl%
\url{https://arxiv.org/abs/2507.06185}
\showURL{%
\tempurl}
\newblock
\shownote{Preprint}.


\bibitem[Liu and Shah(2023)]%
        {liu2023reviewergpt}
\bibfield{author}{\bibinfo{person}{Ryan Liu} {and} \bibinfo{person}{Nihar~B
  Shah}.} \bibinfo{year}{2023}\natexlab{}.
\newblock \showarticletitle{Reviewergpt? an exploratory study on using large
  language models for paper reviewing}.
\newblock \bibinfo{journal}{\emph{arXiv preprint arXiv:2306.00622}}
  (\bibinfo{year}{2023}).
\newblock
\urldef\tempurl%
\url{https://arxiv.org/abs/2306.00622}
\showURL{%
\tempurl}


\bibitem[Marchionini(2008)]%
        {marchionini2008reviewbottleneck}
\bibfield{author}{\bibinfo{person}{Gary Marchionini}.}
  \bibinfo{year}{2008}\natexlab{}.
\newblock \showarticletitle{Editorial: Reviewer merits and review control in an
  age of electronic manuscript management systems}.
\newblock \bibinfo{journal}{\emph{ACM Trans. Inf. Syst.}} \bibinfo{volume}{26},
  \bibinfo{number}{4}, Article \bibinfo{articleno}{25} (\bibinfo{date}{Oct.}
  \bibinfo{year}{2008}), \bibinfo{numpages}{6}~pages.
\newblock
\showISSN{1046-8188}
\href{https://doi.org/10.1145/1402256.1402264}{doi:\nolinkurl{10.1145/1402256.1402264}}


\bibitem[Mason and Chong(2023)]%
        {mason2023bringing}
\bibfield{author}{\bibinfo{person}{Shannon Mason} {and}
  \bibinfo{person}{Sin~Wang Chong}.} \bibinfo{year}{2023}\natexlab{}.
\newblock \showarticletitle{Bringing light to a hidden genre: the peer review
  report}.
\newblock \bibinfo{journal}{\emph{Higher Education Research \& Development}}
  \bibinfo{volume}{42}, \bibinfo{number}{3} (\bibinfo{year}{2023}),
  \bibinfo{pages}{664--678}.
\newblock
\href{https://doi.org/10.1080/07294360.2022.2073976}{doi:\nolinkurl{10.1080/07294360.2022.2073976}}


\bibitem[Mastrianni et~al\mbox{.}(2025)]%
        {mastrianni2025aiGenetic}
\bibfield{author}{\bibinfo{person}{Angela Mastrianni}, \bibinfo{person}{Hope
  Twede}, \bibinfo{person}{Aleksandra Sarcevic}, \bibinfo{person}{Jeremiah
  Wander}, \bibinfo{person}{Christina Austin-Tse}, \bibinfo{person}{Scott
  Saponas}, \bibinfo{person}{Heidi Rehm}, \bibinfo{person}{Ashley~Mae Conard},
  {and} \bibinfo{person}{Amanda~K. Hall}.} \bibinfo{year}{2025}\natexlab{}.
\newblock \showarticletitle{AI-Enhanced Sensemaking: Exploring the Design of a
  Generative AI-Based Assistant to Support Genetic Professionals}.
\newblock \bibinfo{journal}{\emph{ACM Trans. Interact. Intell. Syst.}}
  \bibinfo{volume}{15}, \bibinfo{number}{4}, Article \bibinfo{articleno}{22}
  (\bibinfo{date}{Dec.} \bibinfo{year}{2025}), \bibinfo{numpages}{30}~pages.
\newblock
\showISSN{2160-6455}
\href{https://doi.org/10.1145/3756326}{doi:\nolinkurl{10.1145/3756326}}


\bibitem[Mayer(2025)]%
        {chi2026stats}
\bibfield{author}{\bibinfo{person}{Sven Mayer}.}
  \bibinfo{year}{2025}\natexlab{}.
\newblock \bibinfo{title}{Insights Into the Papers Track First Round Outcomes -
  ACM CHI 2026}.
\newblock
\urldef\tempurl%
\url{https://chi2026.acm.org/2025/12/12/insights-into-the-papers-track-first-round-outcomes/}
\showURL{%
\tempurl}


\bibitem[Mohammad(2020)]%
        {mohammad2020gendergap}
\bibfield{author}{\bibinfo{person}{Saif~M. Mohammad}.}
  \bibinfo{year}{2020}\natexlab{}.
\newblock \showarticletitle{Gender Gap in Natural Language Processing Research:
  Disparities in Authorship and Citations}. In
  \bibinfo{booktitle}{\emph{Proceedings of the 58th Annual Meeting of the
  Association for Computational Linguistics}}. \bibinfo{publisher}{Association
  for Computational Linguistics}, \bibinfo{address}{Online},
  \bibinfo{pages}{7860--7870}.
\newblock
\href{https://doi.org/10.18653/v1/2020.acl-main.702}{doi:\nolinkurl{10.18653/v1/2020.acl-main.702}}


\bibitem[Mollaki(2024)]%
        {mollaki2024aipublishingpolicies}
\bibfield{author}{\bibinfo{person}{Vasiliki Mollaki}.}
  \bibinfo{year}{2024}\natexlab{}.
\newblock \showarticletitle{Death of a reviewer or death of peer review
  integrity? The challenges of using AI tools in peer reviewing and the need to
  go beyond publishing policies}.
\newblock \bibinfo{journal}{\emph{Research Ethics}} \bibinfo{volume}{20},
  \bibinfo{number}{2} (\bibinfo{year}{2024}), \bibinfo{pages}{239--250}.
\newblock
\href{https://doi.org/10.1177/17470161231224552}{doi:\nolinkurl{10.1177/17470161231224552}}


\bibitem[Naddaf(2025)]%
        {naddaf2025ai}
\bibfield{author}{\bibinfo{person}{Miryam Naddaf}.}
  \bibinfo{year}{2025}\natexlab{}.
\newblock \showarticletitle{AI is transforming peer review—and many
  scientists are worried}.
\newblock \bibinfo{journal}{\emph{Nature}}  \bibinfo{volume}{639}
  (\bibinfo{year}{2025}), \bibinfo{pages}{852--854}.
\newblock
\urldef\tempurl%
\url{https://www.nature.com/articles/d41586-025-00894-7}
\showURL{%
\tempurl}


\bibitem[Ngoon et~al\mbox{.}(2018)]%
        {ngoon2018improvingfeedback}
\bibfield{author}{\bibinfo{person}{Tricia~J. Ngoon}, \bibinfo{person}{C.~Ailie
  Fraser}, \bibinfo{person}{Ariel~S. Weingarten}, \bibinfo{person}{Mira
  Dontcheva}, {and} \bibinfo{person}{Scott Klemmer}.}
  \bibinfo{year}{2018}\natexlab{}.
\newblock \showarticletitle{Interactive Guidance Techniques for Improving
  Creative Feedback}. In \bibinfo{booktitle}{\emph{Proceedings of the 2018 CHI
  Conference on Human Factors in Computing Systems}} (Montreal QC, Canada)
  \emph{(\bibinfo{series}{CHI '18})}. \bibinfo{publisher}{Association for
  Computing Machinery}, \bibinfo{address}{New York, NY, USA},
  \bibinfo{pages}{1–11}.
\newblock
\showISBNx{9781450356206}
\href{https://doi.org/10.1145/3173574.3173629}{doi:\nolinkurl{10.1145/3173574.3173629}}


\bibitem[Nobarany(2014)]%
        {nobarany2014rethinking}
\bibfield{author}{\bibinfo{person}{Syavash Nobarany}.}
  \bibinfo{year}{2014}\natexlab{}.
\newblock \showarticletitle{Rethinking the peer review process}. In
  \bibinfo{booktitle}{\emph{Proceedings of the Companion Publication of the
  17th ACM Conference on Computer Supported Cooperative Work \& Social
  Computing}} (Baltimore, Maryland, USA) \emph{(\bibinfo{series}{CSCW Companion
  '14})}. \bibinfo{publisher}{Association for Computing Machinery},
  \bibinfo{address}{New York, NY, USA}, \bibinfo{pages}{77–80}.
\newblock
\showISBNx{9781450325417}
\href{https://doi.org/10.1145/2556420.2556828}{doi:\nolinkurl{10.1145/2556420.2556828}}


\bibitem[O'Brien(2025)]%
        {OBrien2025SciUseLLMsProgram}
\bibfield{author}{\bibinfo{person}{Gabrielle O'Brien}.}
  \bibinfo{year}{2025}\natexlab{}.
\newblock \showarticletitle{How Scientists Use Large Language Models to
  Program}. In \bibinfo{booktitle}{\emph{Proceedings of the 2025 CHI Conference
  on Human Factors in Computing Systems}} \emph{(\bibinfo{series}{CHI '25})}.
  \bibinfo{publisher}{Association for Computing Machinery},
  \bibinfo{address}{New York, NY, USA}, Article \bibinfo{articleno}{876},
  \bibinfo{numpages}{16}~pages.
\newblock
\showISBNx{9798400713941}
\href{https://doi.org/10.1145/3706598.3713668}{doi:\nolinkurl{10.1145/3706598.3713668}}


\bibitem[Parsons and Toombs(2025)]%
        {parsons2025teaching}
\bibfield{author}{\bibinfo{person}{Paul~C. Parsons} {and}
  \bibinfo{person}{Austin~L. Toombs}.} \bibinfo{year}{2025}\natexlab{}.
\newblock \showarticletitle{Teaching to Fail (Before It Happens): Premortem as
  a Pedagogical Strategy in HCI Education}. In
  \bibinfo{booktitle}{\emph{Proceedings of the 7th Annual Symposium on HCI
  Education}} \emph{(\bibinfo{series}{EduCHI '25})}.
  \bibinfo{publisher}{Association for Computing Machinery},
  \bibinfo{address}{New York, NY, USA}, Article \bibinfo{articleno}{14},
  \bibinfo{numpages}{6}~pages.
\newblock
\showISBNx{9798400714634}
\href{https://doi.org/10.1145/3742901.3742908}{doi:\nolinkurl{10.1145/3742901.3742908}}


\bibitem[Pividori and Greene(2024)]%
        {pividori2023publishing}
\bibfield{author}{\bibinfo{person}{Milton Pividori} {and}
  \bibinfo{person}{Casey~S Greene}.} \bibinfo{year}{2024}\natexlab{}.
\newblock \showarticletitle{A publishing infrastructure for AI-assisted
  academic authoring}.
\newblock \bibinfo{journal}{\emph{Journal of the American Medical Informatics
  Association}} \bibinfo{volume}{31}, \bibinfo{number}{9} (\bibinfo{date}{06}
  \bibinfo{year}{2024}), \bibinfo{pages}{2103--2113}.
\newblock
\showISSN{1527-974X}
\href{https://doi.org/10.1093/jamia/ocae139}{doi:\nolinkurl{10.1093/jamia/ocae139}}


\bibitem[Posner and Fei-Fei(2020)]%
        {posner2020ai}
\bibfield{author}{\bibinfo{person}{Tess Posner} {and} \bibinfo{person}{Li
  Fei-Fei}.} \bibinfo{year}{2020}\natexlab{}.
\newblock \showarticletitle{AI will change the world, so it’s time to change
  AI}.
\newblock \bibinfo{journal}{\emph{Nature}}  \bibinfo{volume}{588}
  (\bibinfo{year}{2020}), \bibinfo{pages}{S118--S118}.
\newblock
\href{https://doi.org/10.1038/d41586-020-03412-z}{doi:\nolinkurl{10.1038/d41586-020-03412-z}}


\bibitem[Pradhan et~al\mbox{.}(2021)]%
        {pradhan2021claver}
\bibfield{author}{\bibinfo{person}{Tribikram Pradhan},
  \bibinfo{person}{Prashant Kumar}, {and} \bibinfo{person}{Sukomal Pal}.}
  \bibinfo{year}{2021}\natexlab{}.
\newblock \showarticletitle{CLAVER: An integrated framework of convolutional
  layer, bidirectional LSTM with attention mechanism based scholarly venue
  recommendation}.
\newblock \bibinfo{journal}{\emph{Information Sciences}}  \bibinfo{volume}{559}
  (\bibinfo{year}{2021}), \bibinfo{pages}{212--235}.
\newblock
\showISSN{0020-0255}
\href{https://doi.org/10.1016/j.ins.2020.12.024}{doi:\nolinkurl{10.1016/j.ins.2020.12.024}}


\bibitem[Pradhan and Pal(2020)]%
        {pradhan2020cnaver}
\bibfield{author}{\bibinfo{person}{Tribikram Pradhan} {and}
  \bibinfo{person}{Sukomal Pal}.} \bibinfo{year}{2020}\natexlab{}.
\newblock \showarticletitle{CNAVER: A content and network-based academic venue
  recommender system}.
\newblock \bibinfo{journal}{\emph{Knowledge-Based Systems}}
  \bibinfo{volume}{189} (\bibinfo{year}{2020}), \bibinfo{pages}{105092}.
\newblock
\showISSN{0950-7051}
\href{https://doi.org/10.1016/j.knosys.2019.105092}{doi:\nolinkurl{10.1016/j.knosys.2019.105092}}


\bibitem[Purkayastha et~al\mbox{.}(2025)]%
        {purkayastha2025lazyreview}
\bibfield{author}{\bibinfo{person}{Sukannya Purkayastha},
  \bibinfo{person}{Zhuang Li}, \bibinfo{person}{Anne Lauscher},
  \bibinfo{person}{Lizhen Qu}, {and} \bibinfo{person}{Iryna Gurevych}.}
  \bibinfo{year}{2025}\natexlab{}.
\newblock \showarticletitle{LazyReview A Dataset for Uncovering Lazy Thinking
  in NLP Peer Reviews}. In \bibinfo{booktitle}{\emph{Proceedings of the 63rd
  Annual Meeting of the Association for Computational Linguistics (Volume 1:
  Long Papers)}}. \bibinfo{publisher}{Association for Computational
  Linguistics}, \bibinfo{pages}{3280--3308}.
\newblock
\showISBNx{979-8-89176-251-0}
\href{https://doi.org/10.18653/v1/2025.acl-long.165}{doi:\nolinkurl{10.18653/v1/2025.acl-long.165}}


\bibitem[Rao et~al\mbox{.}(2025)]%
        {rao2025detecting}
\bibfield{author}{\bibinfo{person}{Vishisht~Srihari Rao},
  \bibinfo{person}{Aounon Kumar}, \bibinfo{person}{Himabindu Lakkaraju}, {and}
  \bibinfo{person}{Nihar~B Shah}.} \bibinfo{year}{2025}\natexlab{}.
\newblock \showarticletitle{Detecting LLM-generated peer reviews}.
\newblock \bibinfo{journal}{\emph{PLoS One}} \bibinfo{volume}{20},
  \bibinfo{number}{9} (\bibinfo{year}{2025}), \bibinfo{pages}{e0331871}.
\newblock
\href{https://doi.org/10.1371/journal.pone.0331871}{doi:\nolinkurl{10.1371/journal.pone.0331871}}


\bibitem[Reza et~al\mbox{.}(2024)]%
        {Reza2024ABScribe}
\bibfield{author}{\bibinfo{person}{Mohi Reza}, \bibinfo{person}{Nathan~M
  Laundry}, \bibinfo{person}{Ilya Musabirov}, \bibinfo{person}{Peter Dushniku},
  \bibinfo{person}{Zhi Yuan~“Michael” Yu}, \bibinfo{person}{Kashish
  Mittal}, \bibinfo{person}{Tovi Grossman}, \bibinfo{person}{Michael Liut},
  \bibinfo{person}{Anastasia Kuzminykh}, {and} \bibinfo{person}{Joseph~Jay
  Williams}.} \bibinfo{year}{2024}\natexlab{}.
\newblock \showarticletitle{ABScribe: Rapid Exploration \& Organization of
  Multiple Writing Variations in Human-AI Co-Writing Tasks using Large Language
  Models}. In \bibinfo{booktitle}{\emph{Proceedings of the 2024 CHI Conference
  on Human Factors in Computing Systems}} (Honolulu, HI, USA)
  \emph{(\bibinfo{series}{CHI '24})}. \bibinfo{publisher}{Association for
  Computing Machinery}, \bibinfo{address}{New York, NY, USA}, Article
  \bibinfo{articleno}{1042}, \bibinfo{numpages}{18}~pages.
\newblock
\showISBNx{9798400703300}
\href{https://doi.org/10.1145/3613904.3641899}{doi:\nolinkurl{10.1145/3613904.3641899}}


\bibitem[Reza et~al\mbox{.}(2025)]%
        {reza2025co}
\bibfield{author}{\bibinfo{person}{Mohi Reza}, \bibinfo{person}{Jeb
  Thomas-Mitchell}, \bibinfo{person}{Peter Dushniku}, \bibinfo{person}{Nathan
  Laundry}, \bibinfo{person}{Joseph~Jay Williams}, {and}
  \bibinfo{person}{Anastasia Kuzminykh}.} \bibinfo{year}{2025}\natexlab{}.
\newblock \showarticletitle{Co-Writing with AI, on Human Terms: Aligning
  Research with User Demands Across the Writing Process}.
\newblock \bibinfo{journal}{\emph{Proc. ACM Hum.-Comput. Interact.}}
  \bibinfo{volume}{9}, \bibinfo{number}{7}, Article
  \bibinfo{articleno}{CSCW385} (\bibinfo{date}{Oct.} \bibinfo{year}{2025}),
  \bibinfo{numpages}{37}~pages.
\newblock
\href{https://doi.org/10.1145/3757566}{doi:\nolinkurl{10.1145/3757566}}


\bibitem[Shah(2022)]%
        {shah2021overview}
\bibfield{author}{\bibinfo{person}{Nihar~B. Shah}.}
  \bibinfo{year}{2022}\natexlab{}.
\newblock \showarticletitle{Challenges, experiments, and computational
  solutions in peer review}.
\newblock \bibinfo{journal}{\emph{Commun. ACM}} \bibinfo{volume}{65},
  \bibinfo{number}{6} (\bibinfo{date}{May} \bibinfo{year}{2022}),
  \bibinfo{pages}{76–87}.
\newblock
\showISSN{0001-0782}
\href{https://doi.org/10.1145/3528086}{doi:\nolinkurl{10.1145/3528086}}


\bibitem[SIGCHI(2015)]%
        {chi2016change}
\bibfield{author}{\bibinfo{person}{SIGCHI}.} \bibinfo{year}{2015}\natexlab{}.
\newblock \bibinfo{title}{Changes to the Submission and Review Process for
  CHI2016}.
\newblock
\urldef\tempurl%
\url{https://sigchi.tumblr.com/post/108282241520/changes-to-the-submission-and-review-process-for}
\showURL{%
\tempurl}
\newblock
\shownote{[Online; accessed 2025-11-01]}.


\bibitem[Simonite(2018)]%
        {simonite2018aiwomen}
\bibfield{author}{\bibinfo{person}{Tom Simonite}.}
  \bibinfo{year}{2018}\natexlab{}.
\newblock \showarticletitle{AI Is the Future—But Where Are the Women?}
\newblock \bibinfo{journal}{\emph{Wired}} (\bibinfo{date}{Aug.}
  \bibinfo{year}{2018}).
\newblock
\urldef\tempurl%
\url{https://www.wired.com/story/artificial-intelligence-researchers-gender-imbalance}
\showURL{%
\tempurl}
\newblock
\shownote{Accessed: 2025-09-09}.


\bibitem[Snell and Spencer(2005)]%
        {snell2005reviewersperception}
\bibfield{author}{\bibinfo{person}{Linda Snell} {and} \bibinfo{person}{John
  Spencer}.} \bibinfo{year}{2005}\natexlab{}.
\newblock \showarticletitle{Reviewers' perceptions of the peer review process
  for a medical education journal}.
\newblock \bibinfo{journal}{\emph{Medical education}} \bibinfo{volume}{39},
  \bibinfo{number}{1} (\bibinfo{year}{2005}), \bibinfo{pages}{90--97}.
\newblock
\href{https://doi.org/10.1111/j.1365-2929.2004.02026.x}{doi:\nolinkurl{10.1111/j.1365-2929.2004.02026.x}}


\bibitem[Sperber et~al\mbox{.}(2025)]%
        {sperber2025ReerandAIreview}
\bibfield{author}{\bibinfo{person}{Lisa Sperber}, \bibinfo{person}{Marit
  MacArthur}, \bibinfo{person}{Sophia Minnillo}, \bibinfo{person}{Nicholas
  Stillman}, {and} \bibinfo{person}{Carl Whithaus}.}
  \bibinfo{year}{2025}\natexlab{}.
\newblock \showarticletitle{Peer and AI Review+ Reflection (PAIRR): A
  human-centered approach to formative assessment}.
\newblock \bibinfo{journal}{\emph{Computers and Composition}}
  \bibinfo{volume}{76} (\bibinfo{year}{2025}), \bibinfo{pages}{102921}.
\newblock
\showISSN{8755-4615}
\href{https://doi.org/10.1016/j.compcom.2025.102921}{doi:\nolinkurl{10.1016/j.compcom.2025.102921}}


\bibitem[Stelmakh et~al\mbox{.}(2019)]%
        {stelmakh2019peerreview4all}
\bibfield{author}{\bibinfo{person}{Ivan Stelmakh}, \bibinfo{person}{Nihar~B.
  Shah}, {and} \bibinfo{person}{Aarti Singh}.} \bibinfo{year}{2019}\natexlab{}.
\newblock \showarticletitle{PeerReview4All: Fair and accurate reviewer
  assignment in peer review}. In \bibinfo{booktitle}{\emph{Proceedings of the
  30th International Conference on Algorithmic Learning Theory}}
  \emph{(\bibinfo{series}{Proceedings of Machine Learning Research},
  Vol.~\bibinfo{volume}{98})}, \bibfield{editor}{\bibinfo{person}{Aurélien
  Garivier} {and} \bibinfo{person}{Satyen Kale}} (Eds.).
  \bibinfo{publisher}{PMLR}, \bibinfo{pages}{828--856}.
\newblock
\urldef\tempurl%
\url{https://proceedings.mlr.press/v98/stelmakh19a.html}
\showURL{%
\tempurl}


\bibitem[Sun et~al\mbox{.}(2024a)]%
        {sun2024reviewflow}
\bibfield{author}{\bibinfo{person}{Lu Sun}, \bibinfo{person}{Aaron Chan},
  \bibinfo{person}{Yun~Seo Chang}, {and} \bibinfo{person}{Steven~P. Dow}.}
  \bibinfo{year}{2024}\natexlab{a}.
\newblock \showarticletitle{ReviewFlow: Intelligent Scaffolding to Support
  Academic Peer Reviewing}. In \bibinfo{booktitle}{\emph{Proceedings of the
  29th International Conference on Intelligent User Interfaces}} (Greenville,
  SC, USA) \emph{(\bibinfo{series}{IUI '24})}. \bibinfo{publisher}{Association
  for Computing Machinery}, \bibinfo{address}{New York, NY, USA},
  \bibinfo{pages}{120–137}.
\newblock
\showISBNx{9798400705083}
\href{https://doi.org/10.1145/3640543.3645159}{doi:\nolinkurl{10.1145/3640543.3645159}}


\bibitem[Sun et~al\mbox{.}(2024b)]%
        {sun2024metawriter}
\bibfield{author}{\bibinfo{person}{Lu Sun}, \bibinfo{person}{Stone Tao},
  \bibinfo{person}{Junjie Hu}, {and} \bibinfo{person}{Steven~P. Dow}.}
  \bibinfo{year}{2024}\natexlab{b}.
\newblock \showarticletitle{MetaWriter: Exploring the Potential and Perils of
  AI Writing Support in Scientific Peer Review}.
\newblock \bibinfo{journal}{\emph{Proc. ACM Hum.-Comput. Interact.}}
  \bibinfo{volume}{8}, \bibinfo{number}{CSCW1}, Article \bibinfo{articleno}{94}
  (\bibinfo{date}{April} \bibinfo{year}{2024}), \bibinfo{numpages}{32}~pages.
\newblock
\href{https://doi.org/10.1145/3637371}{doi:\nolinkurl{10.1145/3637371}}


\bibitem[Sun and Kalar(2025)]%
        {GeminiAtWork_AI4knowledgework}
\bibfield{author}{\bibinfo{person}{Na Sun} {and} \bibinfo{person}{Donald
  Kalar}.} \bibinfo{year}{2025}\natexlab{}.
\newblock \showarticletitle{Gemini at Work: Knowledge Workers' Perceptions and
  Assessment of Productivity Gains}. In \bibinfo{booktitle}{\emph{Proceedings
  of the 2025 ACM Designing Interactive Systems Conference}}
  \emph{(\bibinfo{series}{DIS '25})}. \bibinfo{publisher}{Association for
  Computing Machinery}, \bibinfo{address}{New York, NY, USA},
  \bibinfo{pages}{3681–3695}.
\newblock
\showISBNx{9798400714856}
\href{https://doi.org/10.1145/3715336.3735679}{doi:\nolinkurl{10.1145/3715336.3735679}}


\bibitem[Thakkar et~al\mbox{.}(2025)]%
        {iclrofficialExperiment}
\bibfield{author}{\bibinfo{person}{Nitya Thakkar}, \bibinfo{person}{Mert
  Yuksekgonul}, \bibinfo{person}{Jake Silberg}, \bibinfo{person}{Animesh Garg},
  \bibinfo{person}{Nanyun Peng}, \bibinfo{person}{Fei Sha},
  \bibinfo{person}{Rose Yu}, \bibinfo{person}{Carl Vondrick}, {and}
  \bibinfo{person}{James Zou}.} \bibinfo{year}{2025}\natexlab{}.
\newblock \showarticletitle{Can llm feedback enhance review quality? a
  randomized study of 20k reviews at iclr 2025}.
\newblock \bibinfo{journal}{\emph{arXiv preprint arXiv:2504.09737}}
  (\bibinfo{year}{2025}).
\newblock
\urldef\tempurl%
\url{https://arxiv.org/abs/2504.09737}
\showURL{%
\tempurl}


\bibitem[Thaler and Sunstein(2021)]%
        {thaler2021nudge}
\bibfield{author}{\bibinfo{person}{Richard~H Thaler} {and}
  \bibinfo{person}{Cass~R Sunstein}.} \bibinfo{year}{2021}\natexlab{}.
\newblock \bibinfo{booktitle}{\emph{Nudge: The final edition}}.
\newblock \bibinfo{publisher}{Penguin}.
\newblock
\showISBNx{9780143137009}


\bibitem[Tu et~al\mbox{.}(2024)]%
        {tu_etal_2024_AI_Collaboration_with_Authors_academicwriting}
\bibfield{author}{\bibinfo{person}{Joseph Tu}, \bibinfo{person}{Hilda Hadan},
  \bibinfo{person}{Derrick~M Wang}, \bibinfo{person}{Sabrina~A Sgandurra},
  \bibinfo{person}{Reza~Hadi Mogavi}, {and} \bibinfo{person}{Lennart~E Nacke}.}
  \bibinfo{year}{2024}\natexlab{}.
\newblock \showarticletitle{Augmenting the author: Exploring the potential of
  AI collaboration in academic writing}.
\newblock \bibinfo{journal}{\emph{arXiv preprint arXiv:2404.16071}}
  (\bibinfo{year}{2024}).
\newblock
\urldef\tempurl%
\url{https://arxiv.org/abs/2404.16071}
\showURL{%
\tempurl}


\bibitem[Tversky and Kahneman(1981)]%
        {tversky1981framing}
\bibfield{author}{\bibinfo{person}{Amos Tversky} {and} \bibinfo{person}{Daniel
  Kahneman}.} \bibinfo{year}{1981}\natexlab{}.
\newblock \showarticletitle{The framing of decisions and the psychology of
  choice}.
\newblock \bibinfo{journal}{\emph{Science}} \bibinfo{volume}{211},
  \bibinfo{number}{4481} (\bibinfo{year}{1981}), \bibinfo{pages}{453--458}.
\newblock
\href{https://doi.org/10.1126/science.7455683}{doi:\nolinkurl{10.1126/science.7455683}}


\bibitem[Vaithilingam et~al\mbox{.}(2022)]%
        {vaithilingam2022expectation}
\bibfield{author}{\bibinfo{person}{Priyan Vaithilingam},
  \bibinfo{person}{Tianyi Zhang}, {and} \bibinfo{person}{Elena~L. Glassman}.}
  \bibinfo{year}{2022}\natexlab{}.
\newblock \showarticletitle{Expectation vs. Experience: Evaluating the
  Usability of Code Generation Tools Powered by Large Language Models}. In
  \bibinfo{booktitle}{\emph{Extended Abstracts of the 2022 CHI Conference on
  Human Factors in Computing Systems}} (New Orleans, LA, USA)
  \emph{(\bibinfo{series}{CHI EA '22})}. \bibinfo{publisher}{Association for
  Computing Machinery}, \bibinfo{address}{New York, NY, USA}, Article
  \bibinfo{articleno}{332}, \bibinfo{numpages}{7}~pages.
\newblock
\showISBNx{9781450391566}
\href{https://doi.org/10.1145/3491101.3519665}{doi:\nolinkurl{10.1145/3491101.3519665}}


\bibitem[Vitak et~al\mbox{.}(2024)]%
        {vitak2024CSCWPanel}
\bibfield{author}{\bibinfo{person}{Jessica Vitak}, \bibinfo{person}{Amy
  Bruckman}, \bibinfo{person}{Cliff Lampe}, \bibinfo{person}{Xinru Page}, {and}
  \bibinfo{person}{Marisol Wong-Villacr\'{e}s}.}
  \bibinfo{year}{2024}\natexlab{}.
\newblock \showarticletitle{Beyond "Reviewer 2" Problems: Responding to the
  Peer Review Crisis in Computing Research}. In
  \bibinfo{booktitle}{\emph{Companion Publication of the 2024 Conference on
  Computer-Supported Cooperative Work and Social Computing}} (San Jose, Costa
  Rica) \emph{(\bibinfo{series}{CSCW Companion '24})}.
  \bibinfo{publisher}{Association for Computing Machinery},
  \bibinfo{address}{New York, NY, USA}, \bibinfo{pages}{110–113}.
\newblock
\showISBNx{9798400711145}
\href{https://doi.org/10.1145/3678884.3689131}{doi:\nolinkurl{10.1145/3678884.3689131}}


\bibitem[Vondrick(2024)]%
        {Assistin40:online}
\bibfield{author}{\bibinfo{person}{Carl Vondrick}.}
  \bibinfo{year}{2024}\natexlab{}.
\newblock \bibinfo{title}{Assisting ICLR 2025 reviewers with feedback – ICLR
  Blog}.
\newblock
\urldef\tempurl%
\url{https://blog.iclr.cc/2024/10/09/iclr2025-assisting-reviewers/}
\showURL{%
\tempurl}
\newblock
\shownote{[Online; accessed 2025-11-11]}.


\bibitem[Vondrick(2025)]%
        {Leveragi63:online}
\bibfield{author}{\bibinfo{person}{Carl Vondrick}.}
  \bibinfo{year}{2025}\natexlab{}.
\newblock \bibinfo{title}{Leveraging LLM feedback to enhance review quality –
  ICLR Blog}.
\newblock
\urldef\tempurl%
\url{https://blog.iclr.cc/2025/04/15/leveraging-llm-feedback-to-enhance-review-quality/}
\showURL{%
\tempurl}
\newblock
\shownote{[Online; accessed 2025-11-22]}.


\bibitem[Wadden et~al\mbox{.}(2022)]%
        {wadden2022scifactopen}
\bibfield{author}{\bibinfo{person}{David Wadden}, \bibinfo{person}{Kyle Lo},
  \bibinfo{person}{Bailey Kuehl}, \bibinfo{person}{Arman Cohan},
  \bibinfo{person}{Iz Beltagy}, \bibinfo{person}{Lucy~Lu Wang}, {and}
  \bibinfo{person}{Hannaneh Hajishirzi}.} \bibinfo{year}{2022}\natexlab{}.
\newblock \showarticletitle{SciFact-Open: Towards Open-Domain Scientific Claim
  Verification}. In \bibinfo{booktitle}{\emph{Findings of the Association for
  Computational Linguistics: EMNLP 2022}},
  \bibfield{editor}{\bibinfo{person}{Yoav Goldberg}, \bibinfo{person}{Zornitsa
  Kozareva}, {and} \bibinfo{person}{Yue Zhang}} (Eds.).
  \bibinfo{publisher}{Association for Computational Linguistics},
  \bibinfo{address}{Abu Dhabi, United Arab Emirates},
  \bibinfo{pages}{4719--4734}.
\newblock
\href{https://doi.org/10.18653/v1/2022.findings-emnlp.349}{doi:\nolinkurl{10.18653/v1/2022.findings-emnlp.349}}


\bibitem[Wagner et~al\mbox{.}(2022)]%
        {wagner2022artificial}
\bibfield{author}{\bibinfo{person}{Gerit Wagner}, \bibinfo{person}{Roman
  Lukyanenko}, {and} \bibinfo{person}{Guy Par{\'e}}.}
  \bibinfo{year}{2022}\natexlab{}.
\newblock \showarticletitle{Artificial intelligence and the conduct of
  literature reviews}.
\newblock \bibinfo{journal}{\emph{Journal of Information Technology}}
  \bibinfo{volume}{37}, \bibinfo{number}{2} (\bibinfo{year}{2022}),
  \bibinfo{pages}{209--226}.
\newblock
\href{https://doi.org/10.1177/02683962211048201}{doi:\nolinkurl{10.1177/02683962211048201}}
\showeprint{https://doi.org/10.1177/02683962211048201}


\bibitem[Wan et~al\mbox{.}(2024)]%
        {WanHuZhang2024}
\bibfield{author}{\bibinfo{person}{Qian Wan}, \bibinfo{person}{Siying Hu},
  \bibinfo{person}{Yu Zhang}, \bibinfo{person}{Piaohong Wang},
  \bibinfo{person}{Bo Wen}, {and} \bibinfo{person}{Zhicong Lu}.}
  \bibinfo{year}{2024}\natexlab{}.
\newblock \showarticletitle{"It Felt Like Having a Second Mind": Investigating
  Human-AI Co-creativity in Prewriting with Large Language Models}.
\newblock \bibinfo{journal}{\emph{Proc. ACM Hum.-Comput. Interact.}}
  \bibinfo{volume}{8}, \bibinfo{number}{CSCW1}, Article \bibinfo{articleno}{84}
  (\bibinfo{date}{April} \bibinfo{year}{2024}), \bibinfo{numpages}{26}~pages.
\newblock
\href{https://doi.org/10.1145/3637361}{doi:\nolinkurl{10.1145/3637361}}


\bibitem[Wang et~al\mbox{.}(2019)]%
        {wang2019human}
\bibfield{author}{\bibinfo{person}{Dakuo Wang}, \bibinfo{person}{Justin~D.
  Weisz}, \bibinfo{person}{Michael Muller}, \bibinfo{person}{Parikshit Ram},
  \bibinfo{person}{Werner Geyer}, \bibinfo{person}{Casey Dugan},
  \bibinfo{person}{Yla Tausczik}, \bibinfo{person}{Horst Samulowitz}, {and}
  \bibinfo{person}{Alexander Gray}.} \bibinfo{year}{2019}\natexlab{}.
\newblock \showarticletitle{Human-AI Collaboration in Data Science: Exploring
  Data Scientists' Perceptions of Automated AI}.
\newblock \bibinfo{journal}{\emph{Proc. ACM Hum.-Comput. Interact.}}
  \bibinfo{volume}{3}, \bibinfo{number}{CSCW}, Article \bibinfo{articleno}{211}
  (\bibinfo{date}{Nov.} \bibinfo{year}{2019}), \bibinfo{numpages}{24}~pages.
\newblock
\href{https://doi.org/10.1145/3359313}{doi:\nolinkurl{10.1145/3359313}}


\bibitem[Wang et~al\mbox{.}(2020)]%
        {wang-etal-2020-reviewrobot}
\bibfield{author}{\bibinfo{person}{Qingyun Wang}, \bibinfo{person}{Qi Zeng},
  \bibinfo{person}{Lifu Huang}, \bibinfo{person}{Kevin Knight},
  \bibinfo{person}{Heng Ji}, {and} \bibinfo{person}{Nazneen~Fatema Rajani}.}
  \bibinfo{year}{2020}\natexlab{}.
\newblock \showarticletitle{{R}eview{R}obot: Explainable Paper Review
  Generation based on Knowledge Synthesis}. In
  \bibinfo{booktitle}{\emph{Proceedings of the 13th International Conference on
  Natural Language Generation}}. \bibinfo{publisher}{Association for
  Computational Linguistics}, \bibinfo{address}{Dublin, Ireland},
  \bibinfo{pages}{384--397}.
\newblock
\href{https://doi.org/10.18653/v1/2020.inlg-1.44}{doi:\nolinkurl{10.18653/v1/2020.inlg-1.44}}


\bibitem[Woodruff et~al\mbox{.}(2024)]%
        {woodruff2024knowledgeworkerAItransformIndustry}
\bibfield{author}{\bibinfo{person}{Allison Woodruff}, \bibinfo{person}{Renee
  Shelby}, \bibinfo{person}{Patrick~Gage Kelley}, \bibinfo{person}{Steven
  Rousso-Schindler}, \bibinfo{person}{Jamila Smith-Loud}, {and}
  \bibinfo{person}{Lauren Wilcox}.} \bibinfo{year}{2024}\natexlab{}.
\newblock \showarticletitle{How Knowledge Workers Think Generative AI Will
  (Not) Transform Their Industries}. In \bibinfo{booktitle}{\emph{Proceedings
  of the 2024 CHI Conference on Human Factors in Computing Systems}} (Honolulu,
  HI, USA) \emph{(\bibinfo{series}{CHI '24})}. \bibinfo{publisher}{Association
  for Computing Machinery}, \bibinfo{address}{New York, NY, USA}, Article
  \bibinfo{articleno}{641}, \bibinfo{numpages}{26}~pages.
\newblock
\showISBNx{9798400703300}
\href{https://doi.org/10.1145/3613904.3642700}{doi:\nolinkurl{10.1145/3613904.3642700}}


\bibitem[Ye et~al\mbox{.}(2025)]%
        {Ye2025}
\bibfield{author}{\bibinfo{person}{Runlong Ye}, \bibinfo{person}{Patrick Lee},
  \bibinfo{person}{Matthew Varona}, \bibinfo{person}{Oliver Huang}, {and}
  \bibinfo{person}{Carolina Nobre}.} \bibinfo{year}{2025}\natexlab{}.
\newblock \showarticletitle{ScholarMate: A Mixed-Initiative Tool for
  Qualitative Knowledge Work and Information Sensemaking}. In
  \bibinfo{booktitle}{\emph{Adjunct Proceedings of the 4th Annual Symposium on
  Human-Computer Interaction for Work}} \emph{(\bibinfo{series}{CHIWORK '25
  Adjunct})}. \bibinfo{publisher}{Association for Computing Machinery},
  \bibinfo{address}{New York, NY, USA}, Article \bibinfo{articleno}{7},
  \bibinfo{numpages}{7}~pages.
\newblock
\showISBNx{9798400713972}
\href{https://doi.org/10.1145/3707640.3731913}{doi:\nolinkurl{10.1145/3707640.3731913}}


\bibitem[Yuan and Liu(2022)]%
        {yuan-liu-2022-kidreview}
\bibfield{author}{\bibinfo{person}{Weizhe Yuan} {and} \bibinfo{person}{Pengfei
  Liu}.} \bibinfo{year}{2022}\natexlab{}.
\newblock \showarticletitle{KID-Review: Knowledge-Guided Scientific Review
  Generation with Oracle Pre-training}. In
  \bibinfo{booktitle}{\emph{Proceedings of the Thirty-Sixth AAAI Conference on
  Artificial Intelligence (AAAI-22)}}, Vol.~\bibinfo{volume}{36}.
  \bibinfo{pages}{11639--11647}.
\newblock
\urldef\tempurl%
\url{https://ojs.aaai.org/index.php/AAAI/article/view/21418}
\showURL{%
\tempurl}


\bibitem[Yuan et~al\mbox{.}(2022)]%
        {yuan2022automate}
\bibfield{author}{\bibinfo{person}{Weizhe Yuan}, \bibinfo{person}{Pengfei Liu},
  {and} \bibinfo{person}{Graham Neubig}.} \bibinfo{year}{2022}\natexlab{}.
\newblock \showarticletitle{Can We Automate Scientific Reviewing?}
\newblock \bibinfo{journal}{\emph{Journal of Artificial Intelligence Research}}
   \bibinfo{volume}{75} (\bibinfo{year}{2022}), \bibinfo{pages}{171--212}.
\newblock
\href{https://doi.org/10.1613/jair.1.12862}{doi:\nolinkurl{10.1613/jair.1.12862}}


\bibitem[Yun et~al\mbox{.}(2025)]%
        {Yun2025Yodeai}
\bibfield{author}{\bibinfo{person}{Bhada Yun}, \bibinfo{person}{Dana Feng},
  \bibinfo{person}{Ace~S. Chen}, \bibinfo{person}{Afshin Nikzad}, {and}
  \bibinfo{person}{Niloufar Salehi}.} \bibinfo{year}{2025}\natexlab{}.
\newblock \showarticletitle{Generative AI in Knowledge Work: Design
  Implications for Data Navigation and Decision-Making}. In
  \bibinfo{booktitle}{\emph{Proceedings of the 2025 CHI Conference on Human
  Factors in Computing Systems}} \emph{(\bibinfo{series}{CHI '25})}.
  \bibinfo{publisher}{Association for Computing Machinery},
  \bibinfo{address}{New York, NY, USA}, Article \bibinfo{articleno}{634},
  \bibinfo{numpages}{19}~pages.
\newblock
\showISBNx{9798400713941}
\href{https://doi.org/10.1145/3706598.3713337}{doi:\nolinkurl{10.1145/3706598.3713337}}


\bibitem[Zeng et~al\mbox{.}(2024)]%
        {zeng2024orsum}
\bibfield{author}{\bibinfo{person}{Qi Zeng}, \bibinfo{person}{Mankeerat Sidhu},
  \bibinfo{person}{Ansel Blume}, \bibinfo{person}{Hou~Pong Chan},
  \bibinfo{person}{Lu Wang}, {and} \bibinfo{person}{Heng Ji}.}
  \bibinfo{year}{2024}\natexlab{}.
\newblock \bibinfo{title}{Scientific Opinion Summarization: Paper Meta-review
  Generation Dataset, Methods, and Evaluation}.
\newblock
\href{https://doi.org/10.1007/978-981-97-9536-9_2}{doi:\nolinkurl{10.1007/978-981-97-9536-9_2}}
\showeprint[arxiv]{2305.14647}~[cs.CL]


\bibitem[Zhou et~al\mbox{.}(2024)]%
        {zhou-etal-2024-llm}
\bibfield{author}{\bibinfo{person}{Ruiyang Zhou}, \bibinfo{person}{Lu Chen},
  {and} \bibinfo{person}{Kai Yu}.} \bibinfo{year}{2024}\natexlab{}.
\newblock \showarticletitle{Is {LLM} a Reliable Reviewer? A Comprehensive
  Evaluation of {LLM} on Automatic Paper Reviewing Tasks}. In
  \bibinfo{booktitle}{\emph{Proceedings of the 2024 Joint International
  Conference on Computational Linguistics, Language Resources and Evaluation
  (LREC-COLING 2024)}}. \bibinfo{publisher}{ELRA and ICCL},
  \bibinfo{address}{Torino, Italia}, \bibinfo{pages}{9340--9351}.
\newblock
\urldef\tempurl%
\url{https://aclanthology.org/2024.lrec-main.816/}
\showURL{%
\tempurl}


\end{thebibliography}
\end{document}